\documentclass{aa}  

\usepackage{graphicx}
\usepackage[varg]{txfonts}
\usepackage{natbib}
\bibpunct{(}{)}{;}{a}{}{,} 
\usepackage[colorlinks=true,linkcolor=blue,citecolor=blue]{hyperref}%
\usepackage{orcidlink}

\newcommand{\jofour}{\textrm{J}0458\mbox{--}5202}
\newcommand{\Msun}{\hbox{$\rm\thinspace M_{\odot}$}}
\newcommand{\ecgs}{erg~cm$^{-2}$~s$^{-1}$}
\newcommand{\ecgsA}{erg~cm$^{-2}$~s$^{-1}$~\AA$^{-1}$}

\newcommand{\aox}{$\alpha_{\rm OX}$}

\newcommand{\lboledd}{$L_{\rm Bol}/L_{\rm Edd}$ }

\begin{document}

   \title{eROSITA detection of a cloud obscuration event in the Seyfert AGN EC~04570$-$5206}

\author{
Alex Markowitz\orcidlink{0000-0002-2173-0673},\inst{1,2}
Mirko Krumpe,\inst{3}  
David Homan,\inst{3}
Mariusz Gromadzki\orcidlink{0000-0002-1650-1518},\inst{4}
Malte Schramm,\inst{5}
Thomas Boller,\orcidlink{0000-0001-5874-9362}\inst{6}
Saikruba Krishnan,\inst{1}
Tathagata Saha,\inst{1}
Joern Wilms,\inst{7}
Andrea Gokus,\orcidlink{0000-0002-5726-5216}\inst{7,8}
Steven Haemmerich\orcidlink{0000-0002-1113-0041},\inst{7}
Hartmut Winkler,\inst{9} 
Johannes Buchner\orcidlink{0000-0003-0426-6634},\inst{6} 
David A.\ H.\ Buckley\orcidlink{0000-0002-7004-9956}\inst{10,11,12},
Roisin Brogan\inst{3}, and
Daniel E.\ Reichart\inst{13}
}

   \institute{$^1$Nicolaus Copernicus Astronomical Center, Polish Academy of Sciences, ul.\ Bartycka 18, 00-716 Warsaw, Poland\\
$^2$University of California, San Diego, Center for Astrophysics and Space
     Sciences, MC 0424, La Jolla, CA, 92093-0424, USA\\
$^3$Leibniz-Institut f\"ur Astrophysik Potsdam (AIP), Potsdam, Germany \\
$^4$Astronomical Observatory, University of Warsaw, Al.\ Ujazdowskie 4, 00-478 Warsaw, Poland \\
$^5$Graduate School of Science and Engineering, Saitama Univ., 255
Shimo-Okubo, Sakura-ku, Saitama City, Saitama 338-8570, Japan \\
$^6$Max-Planck-Institut f\"ur Extraterrestrische Physik, Giessenbachstr.\ 1, 85748 Garching, Germany\\
$^7$Dr.\ Karl Remeis-Observatory and Erlangen Centre for Astroparticle
Physics, Friedrich-Alexander Universit\"at Erlangen-N\"urnberg,
Sternwartstr.\ 7, 96049 Bamberg, Germany \\
$^8$Department of Physics \& McDonnell Center for the Space Sciences, Washington University in St.~Louis, One Brookings Drive, St.~Louis, MO 63130, USA \\
$^9$Department of Physics, University of Johannesburg, PO Box 524, 2006 Auckland Park, South Africa \\
$^{10}$South African Astronomical Observatory, PO Box 9, Observatory, Cape Town 7935, South Africa\\
$^{11}$Department of Astronomy, University of Cape Town, Private Bag X3, Rondebosch 7701, South Africa\\
$^{12}$Department of Physics, University of the Free State, PO Box 339, Bloemfontein 9300, South Africa\\
$^{13}$Department of Physics and Astronomy, University of North Carolina at Chapel Hill, Campus Box 3255, Chapel Hill, NC 27599-3255, USA \\
             }  

   \date{Received August 17, 2023; accepted --- }

 
  \abstract
{Recent years have seen broad observational support for the presence of 
a clumpy
component within the circumnuclear gas around supermassive black holes (SMBHs).  In the X-ray band, individual clouds can manifest
themselves when they transit the line of sight to the X-ray corona,
temporarily obscuring the X-ray continuum and thereby indicating the characteristics and
location of these clouds.}
{X-ray flux monitoring with \textit{Spectrum Roentgen Gamma} 
extended ROentgen Survey with an Imaging Telescope Array (\textit{SRG}/eROSITA)      
has revealed that in the Seyfert 1 active galactic nucleus (AGN)
EC~04570$-$5206, the soft X-ray flux dipped abruptly for about 10--18
months over 2020--2021, only to recover and then drop a second time by
early 2022.  Here, we investigate whether these flux dips and recoveries could be associated with
cloud occultation events.}
{We complemented the eROSITA scans with multiwavelength follow-up
observations, including X-ray/UV observations with \textit{Swift}, 
\textit{XMM-Newton}, and NICER, along with ground-based optical photometric and spectroscopic
observations to investigate the spectral and flux variability.}
  {\textit{XMM-Newton} spectra confirm that
  the soft X-ray flux dips were caused by partial-covering obscuration
  by two separate clouds.  The 2020-2021 event was caused by a cloud
  with column density near $1 \times 10^{22}$~cm$^{-2}$ and a
  covering fraction of roughly 60 percent.  The cloud in the 2022 event had a
  column density near $ 3 \times10^{23}$~cm$^{-2}$ and a
  covering fraction near 80 percent.  The optical/UV continuum
  flux varied minimally and the optical emission line
  spectra showed no variability in Balmer profiles or intensity.}
  {    The transiting gas clouds are neutral or lowly-ionized, while the lower limits
on their radial distances are commensurate with the dust sublimation
zone (cloud 1) or the optical broad line region (cloud 2).  One
possible explanation is a dust-free, outflowing wind with embedded X-ray
clumps.
These events are the first cloud obscuration events detected in a
Seyfert galaxy using eROSITA's X-ray monitoring capabilities.      }

\authorrunning{A.\ Markowitz et al.}
\titlerunning{Cloud obscuration in EC~04570$-$5206}
   \maketitle
\nolinenumbers
%
%

\section{Introduction}   \label{sec:intro}   

Active galactic nuclei (AGNs) are generally thought to be powered by the
accretion of gas onto a supermassive ($10^{6-9} \Msun$) black hole.
However, open questions remain regarding the precise morphology and mechanics of
some of the various accreting and outflowing components. 
 
Circumnuclear X-ray-obscuring 
gas, colloquially called the ``torus,'' likely supplies
gas for the accretion disk that feeds the supermassive black hole,
while obscuring certain lines of sight to the central engine.
Though its exact morphology and nature remain unclear,
the torus likely plays a role in radiatively driven
outflows and AGN feedback \citep*[e.g.,][]{Murray05, Hoenig19}.
Its morphology governs the fractions of unobscured,
Compton thin-obscured, and Compton thick-obscured AGNs, thereby
impacting their relative fractions in comprising the cosmic X-ray
background \citep[e.g.,][]{Comastri95, Gilli07}.
The morphology of the X-ray-obscuring torus 
is generally accepted to be preferentially distributed 
toward the equatorial plane, but morphological
parameters such as the global covering factor may depend on luminosity
\citep{Burlon11, Ricci17, Ananna22}.


In nearby Seyferts, mid-IR and sub-mm interferometry, with the
Atacama Large Millimeter Array (ALMA) or Very Large Telescope (VLT)
MIDI \citep{Kishimoto09, Tristram09, Pott10, GarciaBurillo16,
  Tristram22}, for instance, and reverberation mapping of the thermal emission from
warm dust \citep{Suganuma06} have been revealing the presence of dusty components 
to the torus, residing at radial scales of parsecs down to tenths of a parsec.  
This dusty component has been invoked in
orientation-dependent ``unification'' schemes for Seyferts, in which
"type 1" or "type 2" objects either display or lack (respectively) highly
Doppler-broadened (FWHM on the order of thousands km s$^{-1}$)
emission lines from the compact optical broad line region (BLR).
As noted by \citet{Netzer93}, \citet{Elitzur07}, and
\citet{Gaskell08}, BLR structures and dusty-torus structures may
comprise one continuous, radially extended structure straddling both
inside and outside the dust sublimation zone, as dust embedded in
dense clouds gas can suppresses optical/UV line emission.


There is mounting observational evidence that dusty and non-dusty
circumnuclear gas in nearby AGNs contain components comprised of clumps
or filaments;
consequently, the AGN community tends to use the term ``torus'' to
denote circumnuclear gas, although it might not necessarily form a simple
axisymmetric “donut” shape.
In X-rays, the community has been accumulating observations of
variations in the line-of-sight column density $N_{\rm H}$ on
timescales of days to years in more than roughly 20 Seyferts
\citep[e.g.,][]
{Mushotzky78, Reichert86, REN02, Risaliti09, Risaliti11, MKN14, Ricci16,
Zaino20}; a review is given in \citet{Ricci23}.
Column densities are
typically observed to vary by factors of a few to ten or more; many
variations are associated with structures that are neutral or
lowly-ionized, and are Compton-thin (on the order of $10^{22-23}$~cm$^{-2}$)
or Compton-thick (on the order of $10^{24}$~cm$^{-2}$).  Column density
variations have been observed both in Seyferts that usually lack
significant amounts of neutral or lowly-ionized gas along their line
of sight, as well as in perpetually obscured Seyferts
\citep[e.g.,][]{REN02, Bianchi09,Rivers11,Laha20}, with some objects
in this latter group transitioning between Compton-thin and
Compton-thick obscured states
\citep[e.g.,][]{Bianchi09,Ricci16,Marchesi22}.  Variations in $N_{\rm
H}$ have been attributed to both full-covering and partial-covering
obscurers, with covering fractions of partial-covering obscurers
varying in some cases \citep[e.g.,][]{Puccetti07, Turner08,
Sanfrutos13}.  

Such observations support the notion that discrete, dynamic structures
temporarily transit the line of sight to the X-ray corona.  
Typically, eclipse events that occur on timescales of roughly a day
and shorter are inferred to be associated with structures
commensurate with the optical BLR
\citep[e.g.,][]{Risaliti11,Sanfrutos13,Gallo21}, based on constraints
from ionization and event duration, inferred number densities, and/or
from obscuring structures' being partial-covering as opposed to
full-covering.  Meanwhile, eclipses or transitions that take 
months to years to complete are typically inferred to be commensurate with the
outer BLR  or inner dusty torus \citep{MKN14,Beuchert15}.


In parallel, high-spatial resolution IR/sub-mm imaging
\citep{Raban09,Izumi18,GarciaBurillo19} and IR SED modeling of thermal
dust emission \citep{RamosAlmeida11, RamosAlmeida14} dust also
resolve, or support, clumpy or filamentary structures.
All these observations have motivated the development of models wherein
the obscuring gas is composed of discrete clouds \citep{Hoenig13}.  In
such models, the total observed line-of-sight obscuration is a probability
that depends on the number of clouds lying along the line of sight,
which, in turn, depends on both viewing angle and the angular and
radial extents of the cloud distribution; however, in most models (e.g.,
\citealt{Nenkova08} and \citealt{LiuLi14} for IR and X-ray spectra,
respectively), clouds are preferentially distributed toward the
equatorial plane.

However, outflowing clumpy winds, likely launched from the inner
accretion disk, can also produce temporary X-ray obscuration events,
as observed in a handful of nearby Seyferts.
Such obscurers tend to be partial covering in the X-rays, moderately ionized,
and have total columns on the order of $10^{22-23}$~cm$^{-2}$.
Such winds are often (but not always) accompanied by 
UV absorption lines that are blueshifted by an order of a few thousand 
km~s$^{-1}$.
Events can last up to a month or longer (e.g., NGC3783,
\citealt{Mehdipour17}; Mkn~817, \citealt{Kara21}), and even up to a
decade \citep[NGC5548][]{Kaastra14,Mehdipour22}. 
Correlations between the properties of the X-ray- and UV-obscuring
components \citep{Mehdipour22} hint at X-ray obscuration being due to
dense clumps embedded in UV-absorbing, outflowing gas.
Wind locations are generally inferred to be on the order light-days to light-weeks
from the black hole, and such winds may act as a "filter,"
obscuring and modifying the ionization continuum that is
intercepted by the BLR \citep{Dehghanian19}. 


Accumulating statistics on the properties of individual
X-ray-obscuring structures -- locations, dust content, ionization structure,
connections to both relativistically and
sub-relativistically outflowing ionized winds -- is needed to
constrain their morphology and possible origins.  However,  the detection of new obscuration events poses a challenge. With some
exceptions such as NGC~1365, in which discrete eclipses or rapid
transitions between Compton-thin and -thick states are frequently
observed \citep{Ricci23}, obscuration events in Seyferts tend to occur
very rarely on a per-object basis. 
\citet{MKN14} and \citet{TorricelliCiamponi14} 
demonstrated the power of long-term X-ray monitoring of
a large starting sample of Seyferts to aggregate a sample of
obscuration events. However, those studies were archival and
had assessed events that had occurred prior. An ideal situation is
to monitor a sample of Seyferts and catch obscuration events
as they are occurring and to immediately apply follow-up
observations to obtain optimal constraints on the properties
of obscurers.

The extended ROentgen Survey with an Imaging Telescope Array
\citep[eROSITA;][]{Predehl21} is the soft X-ray instrument  
aboard \textit{Spectrum Roentgen Gamma}
\citep[\textit{SRG};][]{Sunyaev21}.  Starting in December\ 2019, eROSITA
began conducting deep 0.2--10 keV all-sky surveys (eRASS), scanning
the entire sky once every six months.  Its repeated scans enable time
domain astronomy studies on a variety of variable-emission sources,
including AGNs.  By monitoring a starting sample on the order of $10^6$
AGNs and visiting each source every six months, eROSITA can amplify
small numbers of rare transient AGN events.

Our team has been monitoring AGNs for major changes in soft X-ray flux
between successive eRASS scans.  We identified a Seyfert 1 at redshift
$z$=0.276 whose soft X-ray flux drop from eRASS1 to eRASS2 ranked it
among the most significant (11$\sigma$) drops among all targets
monitored during eRASS1 and 2.  Continued tracking of this object
revealed that this object's 0.5--2~keV flux varied drastically
(factors $\gtrsim$10) on timescales of months and longer.
Specifically, its 0.5--2 keV flux dropped by $\sim$11 from eRASS1
(February 2020) to eRASS2 (August 2020), then recovered, increasing by a
factor of $\sim$15 from eRASS3 (February 2021) to eRASS4 (August 2021)
and then dropped a second time, by a factor of $\sim$9, by eRASS5
(February 2022).

Immediately after we detected each of these major X-ray flux
variations,  we triggered
target-of-opportunity observations encompassing X-ray spectroscopy and
photometry, space-based UV/optical photometry, and ground-based
optical photometry and spectroscopy.  In total, the eRASS
scans and these multiwavelength follow-up observations spanned almost
three years.  We conclude that the two low soft X-ray states -- one
lasting from before August\ 2020 until early 2021, with the second starting
by February\ 2022 and lasting through at least mid-2023 -- are associated
with two discrete structures that temporarily transited the line of
sight to (and partially covered) the X-ray corona.  Meanwhile, the
high soft-X-ray flux states were associated with lack of obscuration.
Our results thus represent the first two major X-ray obscuration
events to be detected with eROSITA.  Moreover, this object is the only
candidate identified by comparing eRASS1 to eRASS2 flux changes where
follow-up observations supported a changing-obscuration event.

The remainder of this paper is organized as follows: In
Sect.~\ref{sec:section2}, we discuss the source's counterpart, the
observing campaign, and the data reduction, and we summarize
the sources' multiband continuum variability characteristics.  In
Sect.~\ref{sec:Xspecfits}, we discuss the spectral fits to the X-ray
data to characterize the obscurers.  In
Sect.~\ref{sec:OUVSED} and Sect.~\ref{sec:optspec}, we fit the optical/UV SED 
and the optical spectra, respectively, to check for signs of reddening by dust.  
Section~\ref{sec:Discussion} presents a discussion, where we infer the locations of the obscuring structures
and discuss their nature.  A summary of our results is given in
Sect.~\ref{sec:Conclusions}.

%
%

\section{Counterpart; follow-up observations and data reduction}     \label{sec:section2}   

\subsection{Identification and counterpart}

The primary goals of eROSITA are to conduct all-sky X-ray scans and to map hot
gas in $\sim$$10^5$ galaxy clusters and intercluster filaments out to
redshifts $\sim$1.3, thus tracing evolution of large-scale structures
across cosmic time, and to detect on order of a million AGNs.
\textit{SRG} orbits the Earth-Sun L2 point. During eROSITA's all-sky
scan mode, \textit{SRG} rotates once every four hours such that
eROSITA, with its $\sim$$1^{\circ}$ field of view, traces a narrow
ring on the sky, following a great circle.  \textit{SRG}'s orbit
precesses by roughly one degree per day, tracking the Earth-Sun line,
so that the entire sky is mapped every six months (each eRASS).

In August\ 2020, we compared X-ray fluxes between eRASS1 and
eRASS2 and identified an X-ray point source displaying a drop in
0.5--2.0~keV flux by a factor of $11.1^{+15.8}_{-4.2}$ between
eRASS1 (February 2020; $F_{0.5-2}$ = (5.0$\pm$0.4) $\times$ $10^{-13}$ \ecgs) to
eRASS2 (August 2020; $F_{0.5-2} = (4.5^{+2.2}_{-2.5}) \times 10^{-14}$
\ecgs). This source was located
at coordinates $\alpha = 04^{\rm h} 58^{\rm m} 15{\fs}72$, $\delta=
-52^{\circ} 02\arcmin 00{\farcs}7$, with a positional uncertainty of
$1{\farcs}0$\footnote{Includes statistical and systematic uncertainties given
the right ascension and declination for this source.}, and named eRASSt~J045815$-$520200
(henceforth \jofour).  The most likely counterpart is the infrared
source WISEA~J045815.63$-$520202.5, located 1{\farcs}7 
away at $\alpha = 04^{\rm h}
58^{\rm m} 15{\fs}64$, $\delta= -52^{\circ} 02\arcmin 02{\farcs}57$,
and cross-listed as EC~04570$-$5206 \citep[Edinburgh-Cape Blue Object
Survey;][]{Stobie97}.

There were no archival optical spectra available and, thus, its redshift was not known previously. 
Our first follow-up optical spectra, discussed in further detail below, 
yielded a redshift of $0.276$.  
In this paper, we assume a flat cosmology,
$H_0=70$~km~s$^{-1}$~Mpc$^{-1}$, $\Omega_{\rm M} = 0.3$, and
$\Omega_{\rm vac}=0.7$, which yields
a luminosity distance of 1400 Mpc 
\citep{Wright06}\footnote{Ned Wright's Cosmology Calculator, at \url{http://www.astro.ucla.edu/~wright/CosmoCalc.html}} for this best-fit redshift value.
Its infrared color as measured by \textit{WISE} in the AllWISE survey \citep{Cutri13} is
W1 $-$ W2 = 1.0 mag, suggesting the 
presence of an AGN following \citet{Stern12} and \citet{Assef18}. 

\subsection{X-ray observations and data reduction}      

Our follow-up campaign included X-ray data from 
eROSITA, \textit{XMM-Newton}, \textit{Swift}, and NICER. 
In Table~\ref{tab:Xobs}, we present a log of the X-ray observations.

\begin{table*}[h]
\caption[]{X-ray observation log of \jofour}
        \centering
\label{tab:Xobs}
        \begin{tabular}{lccccc} \hline\hline
Telescope     & ObsID            & Date                  & Date    & Exposure &  Abbr. \\
                   &             &                       & (MJD)   & (ks)     &        \\ \hline
eROSITA/eRASS1     &             & 18--19 \& 21--25 Feb.\ 2020 & 58902.22 & 0.9         &  eR1\\
eROSITA/eRASS2     &             & 18--23 Aug.\ 2020 & 59081.91 & 1.0  &  eR2\\
\textit{Swift} XRT & 00013747001 & 8--9 Oct.\ 2020   & 59130.89 & 1.9  &  Sw1\\
\textit{Swift} XRT & 00013747002 & 15 Oct.\ 2020     & 59137.49 & 0.8  &  Sw2 \\
\textrm{NICER}     & 3505010101  & 30 Oct.\ 2020     & 59152.93 & 0.5  & N1  \\
\textrm{NICER}     & 3505010102  & 31 Oct.\ 2020     & 59153.45 & 0.8  & N2 \\
\textrm{NICER}     & 3505010103  & 1 Nov.\ 2020      & 59154.86 & 0.6  & N3 \\
\textrm{NICER}     & 3505010104  & 2 Nov.\ 2020      & 59155.57 & 7.8  & N4 \\
\textrm{NICER}     & 3505010105  & 3 Nov.\ 2020      & 59156.51 & 8.3  & N5 \\
\textrm{NICER}     & 3505010106  & 3--4 Nov.\ 2020   & 59157.25 & 2.8  & N6\\
\textit{XMM} EPIC  & 0862770601  & 26--27 Dec.\ 2020 & 59209.62 & 53.7, 66.1, 65.4 & XM1\\ 
\textit{XMM} EPIC  & 0872391601  & 28 Jan.\ 2021     & 59242.44 & 22.5, 31.0, 32.5 & XM2 \\ 
eROSITA/eRASS3     &             & 5--9 Feb.\ 2021   & 59252.53 & 0.7  & eR3\\ 
\textit{Swift} XRT & 00014393001 & 25--26 Jun.\ 2021 & 59390.76 & 6.1  &  Sw3\\
eROSITA/eRASS4     &             & 12--17 Aug.\ 2021 & 59441.16 & 0.9  & eR4  \\
\textit{XMM} EPIC  & 0891801601  & 8 Sep.\ 2021      & 59465.40 & 19.3, 25.6, 26.6 &  XM3   \\ 
eROSITA/eRASS5     &             & 11--15 Feb.\ 2022 & 59623.99 & 0.8  &  eR5 \\
\textit{Swift} XRT & 00015090001 & 24 Mar.\ 2022     & 59661.58 & 5.8  &    Sw4\\
\textit{XMM} EPIC  & 0862771101  & 25--26 Apr.\ 2022 & 59695.01 & 56.4, 64.0, 64.0 &   XM4 \\
\textit{Swift} XRT & 00089398001 & 3 Jun.\ 2022      & 59733.06 & 1.7  &  Sw5 \\
\textit{Swift} XRT & 00015276001 & 23 Jul.\ 2022     & 59783.45 & 4.8  &  Sw6 \\
\textit{Swift} XRT & 00015382001 & 21 Oct.\ 2022     & 59873.53 & 5.1  & Sw7 \\
\textit{Swift} XRT & 00015880001 & 17 Feb.\ 2023     & 59992.20 & 4.9  & Sw8 \\
\hline \end{tabular}
\tablefoot{All MJD dates refer to the midpoint of the observation.  
Exposure refers to good time after
screening. For \textit{XMM-Newton}, the three exposure values refer to
pn, MOS1, and MOS2, respectively.}
\end{table*}


\subsubsection{\textit{SRG}/eROSITA}      

eROSITA completed four full scans and part of a fifth scan (eRASS5)
before the instrument was placed into safe mode in February 2022;
$\jofour$ was visited in each of eRASS1--5, and observations are
referred to as eR1, ..., eR5, henceforth. Within each eRASS, objects at
the orbital equator receive six epochs, each with a $\sim$40~s
exposure, separated by four hours, in each eRASS scan. However,
objects closer to the orbital poles receive a larger number of
successive epochs. $\jofour$ lies very roughly $15^\circ$ from
\textit{SRG}'s orbital poles within each eRASS scan.  Within each of the
five eRASS scans, eROSITA scanned the target 24--30 passes, once every
four hours, over the course of 92--112 hours (the number of passes
varied slightly due to small orbital changes between eRASS scans).

\begin{figure*}[h]\centering
  \includegraphics[width=0.4\columnwidth]{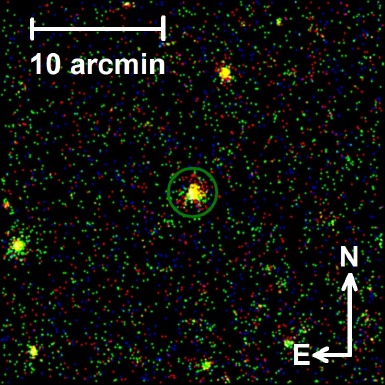}
  \includegraphics[width=0.4\columnwidth]{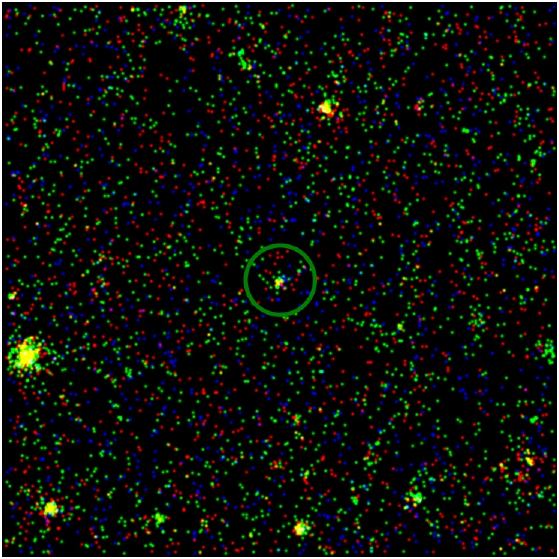}
  \includegraphics[width=0.4\columnwidth]{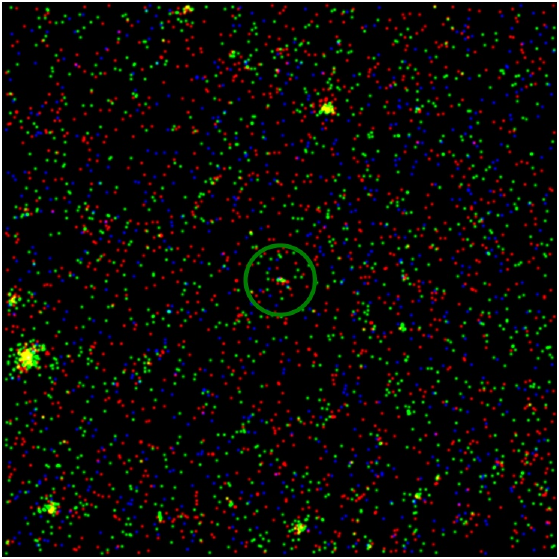}
  \includegraphics[width=0.4\columnwidth]{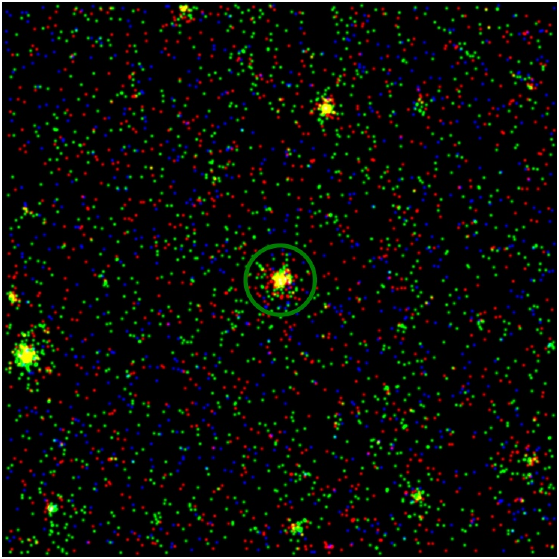}
  \includegraphics[width=0.4\columnwidth]{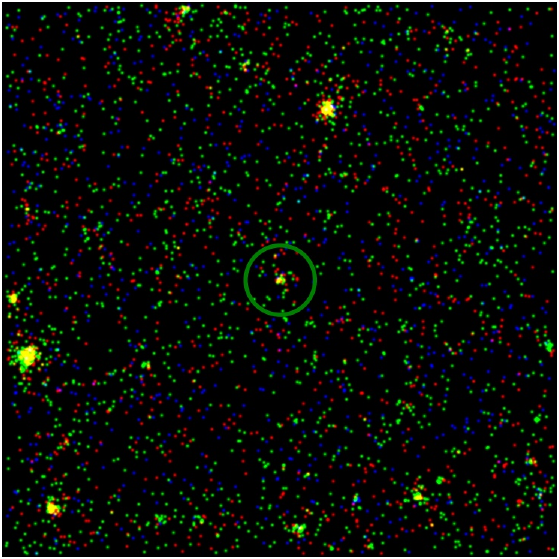}
    \caption{eRASS1--5 X-ray images of the field around $\jofour$. 
Gaussian smoothing is applied. The different colors correspond to the
different energies of the photons (red channel: 0.2--0.6 keV, green
channel: 0.6--2.3 keV, blue channel: 2.3--5.0 keV). In the eRASS1 image,
we show the scale and the orientation of the image; this is the same
for all images. }
\label{fig:erimages}
\end{figure*}

Data were extracted using event processing version c020. We used eSASS               
version 21121\_0\_4 \citep{Brunner22} and HEASOFT version 6.30.1.  We combined     
data from all seven Telescope Modules. We extracted the
source using a circular extraction region with the radius scaled to
the 0.2--2.3~keV maximum likelihood (ML) count rate from the eRASS source catalog;
the extraction regions therefore differ somewhat from one eRASS to the
next, with larger radii corresponding to higher count rates.
ML count rates take into account the time when the source was in the
field of view, and with corrections for vignetting effects applied.
Similarly, background regions were extracted using annuli whose inner
and outer radii depend on ML count rate. Extraction radii and
ML count rates are listed in Table~\ref{tab:erassreduction}.  Point sources detected in the
background extraction regions were excised, again using circular
regions whose radii depended on ML count rate.  Good exposure times
after screening, also listed in Table~\ref{tab:erassreduction}, were
in the range 724--964~s.  Images of $\jofour$ from eR1--5 are displayed
in Fig.~\ref{fig:erimages} to help visualize the strong soft X-ray
variability.

\begin{table*}[h]
\caption[]{eRASS extraction regions, exposure times, and ML count rates} 
        \centering
\label{tab:erassreduction}
        \begin{tabular}{ccccc} \hline\hline
eRASS    & Source & Background &  Exposure & ML count rate  \\
         & Radius ($\arcsec$) & Annulus Radii ($\arcsec$) & (s) & (ct s$^{-1}$; 0.2--2.3~keV)     \\ \hline
1        & 87.8   & 186.3, 1092.7 &   929 & $1.090\pm0.052$    \\  
2        & 40.2   &  95.7, 503.0 &   964 &  $0.063\pm0.012$   \\      
3        & 36.6   &  88.2, 457.4 &   724 & $0.050\pm0.013$\\  
4        & 75.0   & 162.7, 933.6 &   855 & $0.622\pm0.040$ \\  
5        & 40.3   &  95.8, 503.6 &   819 & $0.067\pm0.014$\\  \hline
\end{tabular}
\tablefoot{Exposure refers to good time after screening.
The ML count rate \citep{Merloni24}
denotes the maximum likelihood count rate taking into
account the time when the source was in the field of view and with
corrections applied for vignetting effects.}
\end{table*}

\subsubsection{\textit{XMM-Newton} EPIC}

The four \textit{XMM-Newton} \citep{Jansen01} observations occurred on 26 December 2020,
28 January 2021, 8 September 2021, and 25 April 2022, 
henceforth referred to
as XM1, XM2, XM3, and XM4, respectively.
All observations used both the EPIC \citep{Strueder01} pn and MOS cameras.
XM1 and XM2 each used the large-window mode for all three cameras, while
XM3 and XM4 each used full-frame mode for all three cameras.
The medium optical blocking filter was used for all EPIC cameras in all observations.

We reduced the data using XMM Science Analysis Software (XMMSAS) version 19.1.0
and HEASOFT version 6.28, following standard extraction procedures for point sources.
Source spectra were extracted from circles 40$\arcsec$ in radius;
background spectra were extracted from source-free regions with the same size
located a few arcminutes away and on the same CCD chip.
We screened data against strong, time-localized background flares due
to proton flux by visually inspecting $>$5~keV background light
curves. For the pn, we selected data from pattern 0 and pattern 1--4 separately
(henceforth pn0 and pn14). 
We checked for pileup using the XMMSAS task \texttt{epatplot} but found no evidence for any pileup.
Good exposure times after screening are listed in Table~\ref{tab:Xobs}.

\subsubsection{\textit{Swift} X-ray Telescope (XRT)}    

  The \textit{Neil Gehrels Swift Observatory} \citep[\textit{Swift};][]{Gehrels04}
observed $\jofour$ eight times between October 2020 and February 2023 (Sw1, ..., Sw8).
Each XRT \citep{Burrows05} observation was in photon counting (PC) mode.
Raw event files were reprocessed using
\texttt{xrtpipeline} version 0.13.5 in HEASOFT version 6.28.
and the latest XRT calibration files.
We extracted source spectra using circular regions of radius 20~pixels (47$\arcsec$);
background spectra were extracted from annular regions of inner radius
60~pixels (141$\arcsec$) and outer radius
65~pixels (153$\arcsec$), and confirmed to be free of background sources.
We generated ancillary response files using \texttt{xrtmkarf}, and we selected
the PC mode response files from the calibration database.
Good exposure times after screening are listed in Table~\ref{tab:Xobs}.

\subsubsection{NICER}

The Neutron Star Interior Composition Explorer Mission \citep[NICER;][]{Gendreau16}, aboard
the International Space Station (ISS), observed $\jofour$ six times between
30 October 2020 and 4 November 2020, as listed in Table~\ref{tab:Xobs}.
We used NICERDAS version 10 software and followed standard procedures 
to screen data, produce cleaned event files, and extract spectra.
We discarded data from detectors 14 and 34, which are prone to excessive noise.

We rejected time intervals when the detector undershoot
rate\footnote{Detector undershoots are reset events that occur when
incoming photons trigger a cascade of accumulate charge.} exceeded $150$ ct~s$^{-1}$ per module.  We also
screened out time intervals when the detector overshoot rate (caused
when high-energy particles deposit excess charge) exceeded
1.5~ct~s$^{-1}$.  Given the source faintness, we discarded data taken during the ISS' passage
through the "SAA" South Atlantic Anomaly boundary, which is defined to
be more conservative and cover a slightly larger area than for the standard
"NICERSAA" boundary. We used the 3C50 background estimation method,
screening out times where the background rate in the hard band
(13--15~keV) exceeded 0.5~ct~s$^{-1}$ in the hard band.

Good exposure times after screening are listed in
Table~\ref{tab:Xobs}.  Given the danger of underestimated optical
loading impacting the softest energies, and given the faintness of the
source, we discarded data below 0.4~keV and above 10~keV.

However, in each observation, as well as in spectrum summed from all
six observations, the source was not detected, as we obtained negative
net count rates after background subtraction. For the summed spectrum,
we estimate upper limits of $\sim$3~$\times10^{-13}$ 
and $\sim$2~$\times10^{-13}$ erg
cm$^{-2}$ s$^{-1}$ for the 0.2--5 and 0.5--2~keV bands, respectively.


\begin{table*}
\caption[]{Optical/UV photometric observations of \jofour}
        \centering
\label{tab:OUVobs}
        \begin{tabular}{lccc} \hline\hline
Observation \& & Date     & Filter & Exposure   \\
Instrument     & (MJD)    &        & (ks)       \\ \hline
Sw1 UVOT       & 59130.95 & B      & 1.9 \\
Sw2 UVOT       & 59137.49 & B      & 0.7 \\
XM1 OM         & 59209.82 & M2     & 60.8  \\
XM2 OM         & 59242.45 & M2     & 43.3  \\
Sw3 UVOT       & 59391.00 & V, B, U, W1, M2, W2 &  1.4, 0.5, 0.7, 0.7, 0.7, 2.0  \\
XM3 OM         & 59465.41 & M2 & 28.3 \\
Sw4 UVOT       & 59662.59 & V, B, U, W1, M2, W2 &  0.5, 0.5, 0.5, 0.9, 1.3, 1.9  \\
XM4 OM         & 59694.64  & B  & 8.0  \\
               & 59695.04  & M2 & 56.0 \\   
Sw5 UVOT       & 59733.06  & V, B, U, W1, M2, W2 &  0.1, 0.1, 0.1, 0.3, 0.4, 0.5 \\
Sw6 UVOT       & 59783.49  & V, B, U, W1, M2, W2 &  0.3, 0.3, 0.3, 0.5, 0.7, 1.1 \\ 
Sw7 UVOT       & 59873.53  & V, B, U, W1, M2, W2 &  0.4, 0.4, 0.4, 0.8, 1.2, 1.7 \\ 
Sw8 UVOT       & 59992.20  & V, B, U, W1, M2, W2 &  0.4, 0.4, 0.4, 0.9, 1.1, 1.7 \\ 
\hline
\end{tabular}
\tablefoot{MJD date refers to the midpoint of the observation (time average of all exposures).}
\end{table*}

\subsection{Optical and UV photometric observations and data reduction}    

\subsubsection{  {\it XMM-Newton} Optical Monitor} \label{sec:OMred}

The \textit{XMM-Newton} Optical Monitor \citep[OM;][]{Mason01} observed $\jofour$ with the
UVM2 filter (effective wavelength: 231~nm) simultaneously to each of
the four EPIC observations; XM4 additionally used the B filter
(450~nm). Dates, start-stop times, and total good exposure times (sum
of all images) are listed in Table~\ref{tab:OUVobs}.  The numbers and
lengths of individual exposures are listed in
Table~\ref{tab:XMMOMlog}.
We reduced the data using the XMM\_SAS routines \texttt{omichain} and
\texttt{omfchain} for the image and fast modes, respectively.  These
routines apply flat-fielding, source detection, and aperture
photometry for each exposure, and they combine all exposure images
into a mosaiced image, and perform source detection and aperture
photometry on the mosaiced image.  The source extraction radius was 12
pixels = 5$\farcs$7.  These routines also correct fluxes for detector
dead time.  We verified that the source was well detected within each
exposure, that there were no obvious imaging artifacts in any
exposure, and that the source was not too close to the edge of the window
in fast mode.  AB magnitudes were converted to Vega magnitudes
following the zeropoints listed in Sect.~3.5.3 of the XMM-Newton User's
Handbook, issue
2.20\footnote{\url{https://xmm-tools.cosmos.esa.int/external/xmm_user_support/documentation/uhb/omfilters.html}}
(add 0.185 mag for the B band; subtract 1.640 mag for M2).
The resulting Vega and AB magnitudes are listed in
Table~\ref{tab:OUVfluxes}.  Table~\ref{tab:OUVfluxes} also contains
the corresponding flux densities, corrected for 
Galactic extinction. We used $R$=3.1 and the Galactic
extinction curve of \citet{Cardelli89}, which for $E(B-V)=0.009$
\citep*{Schlegel98} yielded increases of $A_{\rm M2}$ = 0.07
magnitudes and $A_{\rm B}$ = 0.03 magnitudes in M2 and B,
respectively.

\begin{table}
\caption[]{\textit{XMM-Newton} Optical Monitor log}
        \centering
\label{tab:XMMOMlog}
        \begin{tabular}{lcccc} \hline\hline
Obs.\  &  Modes   &  Filter  &  Num.\ of   &  Duration \\
       &          &          & exposures   & of each expo. \\ \hline
XM1 & Image \& Fast & M2 & 14 & 4.3 ks \\ 
XM2 & Image \& Fast & M2 & 11 &   4.0 ks\tablefootmark{a} \\
XM3 & Image only    & M2 &  7 &   4.0 ks\tablefootmark{b} \\
XM4 & Image \& Fast & B  &  2 &   4.0 ks \\
    & Image \& Fast & M2 & 14 &  4.0 ks \\ \hline \end{tabular}
\tablefoot{ \\
\tablefoottext{a}{The last exposure lasted 3.3 ks.}
\tablefoottext{b}{The first exposure lasted 4.3 ks.}}
\end{table}


\subsubsection{\textit{Swift} Ultraviolet/Optical Telescope (UVOT)}

\textit{Swift's} Ultraviolet/Optical Telescope
\citep[UVOT;][]{Roming05} observed $\jofour$ concurrently with the six
XRT observations.  During observations Sw3--8, the UVOT rotated
through all six filters (effective wavelengths: V: 540~nm; B: 433~nm;
U: 350~nm; W1: 263~nm; M2: 223~nm; and W2: 203~nm.  During Sw1 and
Sw2, UVOT used only the B filter.  Dates, start and stop times, and
total good exposure times (summed over all images) are listed in
Table~\ref{tab:OUVobs}.

Aperture photometry for each UVOT filter was performed using the ASI
Space Science Data Center Multi-Mission Interactive Archive online
data analysis
tool\footnote{https://www.ssdc.asi.it/mmia/index.php?mission=swiftmastr}.
It sums all exposure fractions and performs source extraction using
\texttt{uvotdetect} and the latest CALDB.  Source counts were
extracted from a circle of radius 5$\arcsec$; background was extracted
from an annulus of inner and outer radii 27$\farcs$5 and 35$\arcsec$,
respectively.  Aperture-corrected, background-subtracted, and Galactic
extinction-corrected magnitudes and flux densities were derived; they
are listed in Table~\ref{tab:OUVfluxes}. The uncertainties on these
magnitudes are statistical only.  As detailed in
Appendix~\ref{sec:appdxoptmag}, we also extracted the magnitudes for
two stars nearby in the field of view to estimate systematic
uncertainties.  However, statistical uncertainties dominate, so we
neglect these systematic uncertainties in all plots below.

\begin{figure}
\includegraphics[width=0.99\columnwidth]{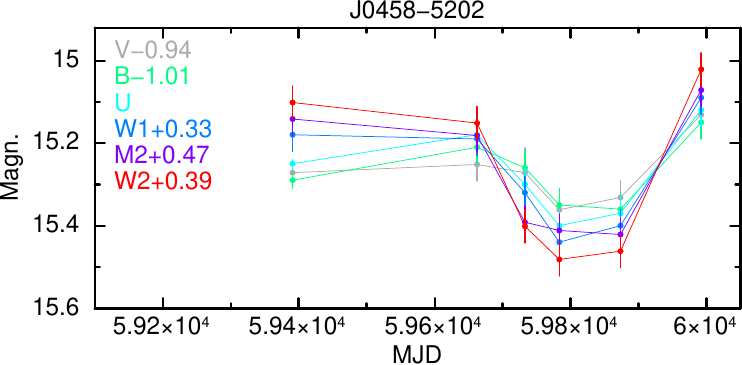}
\caption{Optical/UV light curves of $\jofour$ for observations Sw3--8, in which
\textit{Swift} UVOT observed with all six filters.  For this overplot, the V, B, W1, M2, and W2-band light curves have been offset
by the magnitude values indicated in the top left corner such that their mean magnitudes across all six observations
match the mean of the U-band light curve. This action 
helps to illustrate the optical/UV spectral variability as a function of flux;
as demonstrated in Sect.~\ref{sec:OUVSED}, such spectral variability is consistent with bluer-when-brighter behavior,
and not a product of variable extinction.}
\label{fig:lcs_sw38_J0458}
\end{figure}

In Fig.~\ref{fig:lcs_sw38_J0458}, we display the V, B, U, W1, M2, and
W2-band light curves obtained with \textit{Swift} UVOT during
observations Sw3--8.  Through Sw5--7, as the source flux decreases,
the relatively higher energy bands become fainter faster,
although the drop is only on the order of 0.3 mag in the far-UV.
By Sw8, the source flux recovers, becoming slightly brighter than during
Sw3--4. As we demonstrate in Sect.~\ref{sec:OUVSED} from broadband spectral fitting,
such spectral variability is consistent with an intrinsic bluer-when-brighter trend,
and not consistent with being due to a significant degree of variable extinction.

Finally, we used the ftools \texttt{uvotimsum} and \texttt{uvot2pha}
to generate single-channel spectral files for each filter for the
purpose of SED fitting.  We used standard UVOT response files from the
calibration database.  The time-resolved SED fits to the UVOT spectra
will be discussed in Sect.~\ref{sec:OUVSED}.

\subsubsection{Ground-based optical photometry}

We obtained 40 observations of $\jofour$ at the 0.4-m PROMPT6 telescope
at Cerro Tololo Inter-American Observatory, operated as part of
the Skynet Robotic Telescope Network. The first 33 observations (MJD 59164--59308) used the Johnson
B filter; the final seven (MJD 59389--59541) used Johnson V.  We
reduced all images in a standard manner, including bias correction and
flat-fielding.  We performed aperture photometry by extracting source-centered
circles and background annuli.
Observed (not corrected for Galactic reddening) Vega magnitudes are
listed in Table~\ref{tab:Bphotlog}.


\subsection{Optical spectroscopic observations}

We obtained ten longslit spectra of $\jofour$ between September 2020 and
October 2022, as listed in Table~\ref{tab:optspeclog}.  These
observations included the South African Large Telescope (SALT)
longslit Robert Stobie Spectrograph
\citep[RSS;][]{Burgh03, Kobulnicky03}, the FORS2 spectrograph
\citep{Appenzeller98} on the 8.2~m Very Large Telescope Array’s (VLT)
UT1 at Cerro Paranal, and the SpUpNIC spectrograph \citep{Crause19} at
the South African Astronomical Observatory (SAAO) 1.9~m telescope.  Spectra
\#1--7 were taken between September 2020 and April 2021, during the first
soft X-ray flux dip; spectrum \#8 was coincident with the high soft
X-ray flux state; spectra \#9 and 10 were taken in April 2022 and 
October 2022, during the second soft X-ray flux dip.

\begin{table*}[h]
\caption[]{Optical spectroscopic observations of $\jofour$}
        \centering
\label{tab:optspeclog}
        \begin{tabular}{lllcc} \hline\hline
  \#   & Telescope     & Date       & MJD              & Total Exposure\   \\
       & \& Instrument &            & MJD              & (s)        \\ \hline
 1 & SALT RSS      & 27 Sep.\ 2020 & 59119 & 150\\
 2 & SALT RSS      & 27 Oct.\ 2020 & 59149 & 150 \\
 3 & SALT RSS      &  8 Dec.\ 2020 & 59191 & 130 \\
 4 & VLT FORS2     & 27 Dec.\ 2020 & 59210 & 450+450+1000 \\
 5 & SALT RSS      &  2 Feb.\ 2021 & 59247 & 240+240 \\
 6 & SALT RSS      &  2 Mar.\ 2021 & 59275 & 240+240 \\
 7 & SALT RSS      &  3 Apr.\ 2021 & 59307 & 240+240 \\
 8 & SAAO 1.9 m    & 10 Sep.\ 2021 & 59467 & 1200+1200+1200\\
 9 & VLT FORS2     & 27 Apr.\ 2022 & 59696 & 1000+1000 \\
10 & SALT RSS      &  5 Oct.\ 2022 & 59857 & 600 \\
 \hline  \end{tabular}
\tablefoot{For spectra \#5, 6, and 7 at SALT RSS, exposures were taken
  with two RSS setups, hence, two exposure times are listed. 
For spectrum \#4 at VLT FORS2, the three values listed refer to exposures using the
G300V, G300I, and G1400V gratings, respectively.
For spectrum \#9 at VLT FORS2, the two values listed refer to exposures using
the G300V and G300I gratings.
}
\end{table*}

For all SALT RSS observations, the slit width was 1$\farcs$5. 
All observations used the pg0900 gratings (to achieve spectral
resolution $R \sim$ 800--1200).  For spectra \#5, 6, and 7, 
we took 4$\times$60~s exposures using a grating angle of 19.25$^{\circ}$ (camera angle
38.5$^{\circ}$), and 4$\times$60~s exposures using a grating angle of
13.25$^{\circ}$ (camera angle 26.5$^{\circ}$).  The former setup
encompassed H$\alpha$, \ion{He}{i}, [\ion{O}{iii}], and H$\beta$ emission;
the latter setup encompassed [\ion{O}{iii}], H$\beta$, H$\gamma$,
H$\delta$, [\ion{O}{ii}]~$\lambda$3727, and [\ion{Ne}{v}]~$\lambda$3425. 
All RSS observations used the 2x2 spatial binning, the faint gain, and the slow readout mode.

For the SAAO/SpUpNIC spectrum, we used a 2.7$\arcsec$ slitwidth, and a
low-resolution grating to cover the entire optical range.

For the FORS2 observation on 27 December 2020 (\#4), we made use of three
gratings: G1400V (1000~s exposure), G300V + GG435+81 (450~s exposure),
and G300I + OG590+32 (450~s exposure). The 300V and 300I exposures
provide the full wavelength coverage and the G1400V exposure garnered
additional S/N near the H$\beta$--[\ion{O}{iii}] region. We
combined all exposures into a single high-quality spectrum. For the
FORS2 observation of 27 April 2022 (\#9), we used the G300V and G300I
gratings, with exposures of 1000~s each, and again combined the results
into a single spectrum covering 3520--9980 \AA.

All CCD data were reduced using standard bias corrections and
flat-fielding.  Wavelength calibration used arc-lamp spectra taken on
the night. Spectrophotometric calibrations were performed using data
for a standard star taken the same night in the case of the VLT-FORS2
and SAAO 1.9m spectra, and a standard star taken within the past year
in the case of the SALT-RSS spectra.


\subsection{Multiband variability overview}    

As an overview of $\jofour$'s X-ray variability, 
we list 0.2--5.0 and 0.5--2.0~keV fluxes in Table~\ref{tab:Xfluxes}.
For XM1--4, eR1--5, and Sw3, fluxes were derived directly from the best-fitting spectral
models, as detailed in Sect.~\ref{sec:Xspecfits}.
All other fluxes are based on count rates, and conversions using
the best-fitting models, also discussed in Sect.~\ref{sec:Xspecfits}.

\begin{figure*}[!h]
\includegraphics[width=1.98\columnwidth]{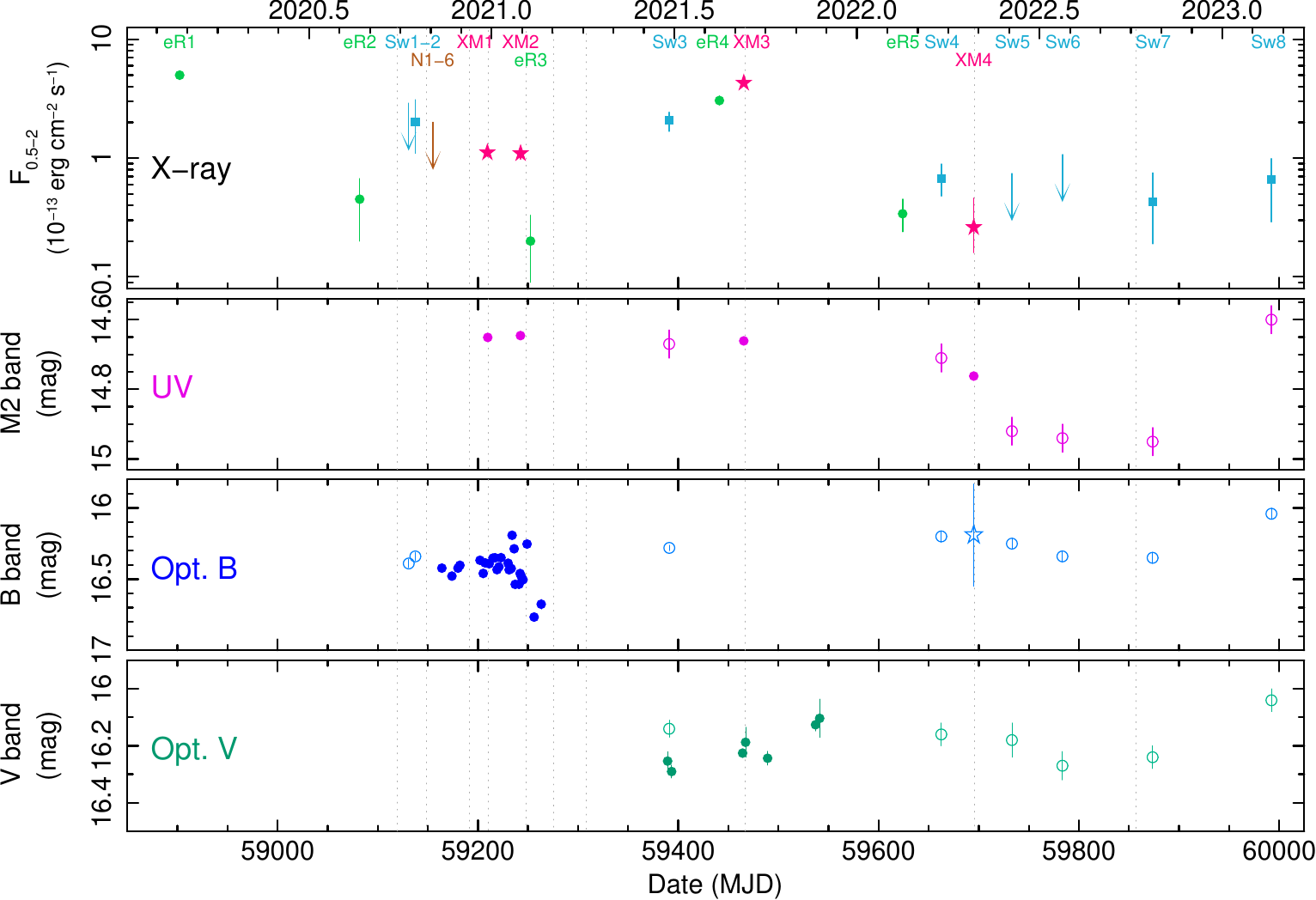}
\caption{Overview of multiband light curves. The top panel shows the
  soft X-ray (0.5--2.0~keV) fluxes across all observations. Green,
  blue, and red denote eRASS, \textit{Swift} XRT, and
  \textit{XMM-Newton} EPIC, respectively; 
the brown upper limit denotes the combined NICER observations.         
The corresponding observation abbreviation is written above each point.
The middle panel shows the UVM2
  photometric Vega magnitudes from \textit{XMM-Newton} OM (filled
  circles) and \textit{Swift} UVOT (open circles).  The bottom two panels
  show B- and V-band photometry from ground-based measurements (filled
  circles), \textit{Swift} UVOT (open circles), and \textit{XMM-Newton} OM 
(open star). The optical and UV magnitudes plotted here are not
corrected for Galactic absorption.  
\textit{XMM-Newton} M2 and B magnitudes were matched to those of \textit{Swift} UVOT
following the cross calibration procedure described in Appendix~\ref{sec:appdxoptmag}.
In each panel, there are some data points whose error bars are smaller than the data point symbol.    
Vertical dotted lines indicate the dates of optical spectroscopic measurements.}
\label{fig:LIGHTCURVE}
\end{figure*}

The 0.5--2.0~keV fluxes are plotted in the top panel of
Fig.~\ref{fig:LIGHTCURVE}.  Strong variations (roughly by factors of
5--10, as mentioned in Sect.~\ref{sec:intro}) are apparent.  eR1 represents a
relatively high soft X-ray flux state.  We refer to the first soft
X-ray flux dip as starting sometime after eR1 (February 2020) and before
eR2 (August 2020), and encompassing eR2, Sw1, Sw2, XM1, XM2, eR3, and
Sw3, and ending sometime between Sw3 (June 2021) and eR4
(August 2021).  Then, eR4 and XM3 represent another relatively high X-ray
flux state.  A second low soft-X-ray flux state started sometime between eR4 and
eR5 (February 2022), and encompassed eR5, Sw4, XM4, Sw5, Sw6, and Sw7, and was
still in-progress as of Sw8 (February 2023), our latest observation.

Meanwhile, the middle and lower panels of Fig.~\ref{fig:LIGHTCURVE}
display, respectively, the M2 and B-band photometry light curves.  As
discussed in Appendix~\ref{sec:appdxoptmag}, combining the \textit{XMM-Newton} OM and
\textit{Swift} UVOT M2 observed magnitudes into one light curve
required some magnitude corrections, given that the energy peaks of the two instruments'
effective areas  differ by $\sim$10 percent.  We
estimated spectral slopes from the SED and derived corrections of
$-0.070$ (XM1--3) and $-0.031$ (XM4) to apply to the OM data points,
as detailed in Appendix~\ref{sec:appdxoptmag}, with the final, corrected light curves
plotted in Fig.~\ref{fig:LIGHTCURVE}. Overall, the B and M2 light
curves display much lower levels of variability compared to the
X-rays.  The combined OM and UVOT M2 band light curve shows a
decrease in flux of only 
30~percent over $\sim$400--450 days through Sw7, followed by a recovery by Sw8. The
B-band continuum is overall characterized by mild variations on the order of
$\sim$30 percent over the two-year campaign, although there is
additionally a sharp, temporary drop in flux by $\sim$40 percent over
$\sim$40 days, starting after roughly MJD 59210.
Overall, the results of Figs.~\ref{fig:lcs_sw38_J0458} and
\ref{fig:LIGHTCURVE} rule out any major (e.g., order of magnitude of
more) drop in optical/UV luminosity or accretion rate occurring during
the campaign.

\renewcommand{\arraystretch}{1.20}
\begin{table}[h]
\caption[]{X-ray fluxes of $\jofour$}
\centering
\label{tab:Xfluxes}
\begin{tabular}{llll} \hline\hline
Obs. & Date  & $F_{0.2-5}$ & $F_{0.5-2}$ \\ 
     & (MJD) & ($10^{-13}$  & ($10^{-13}$   \\
     &       & \ecgs) & \ecgs)  \\   \hline
eR1 & 58902.22&  $14.62^{+1.13}_{-1.06}$ & $5.30^{+0.39}_{-0.42}$   \\
eR2 & 59081.91&   $1.29^{+0.87}_{-1.05}$ & $0.45^{+0.22}_{-0.25}$  \\
Sw1 & 59130.85&   $1.02^{+0.78}_{-0.51}$ &  $<$2.90  \\
Sw2 & 59137.49&   $3.77^{+1.61}_{-1.26}$ &  $2.00^{+1.10}_{-0.90}$  \\
N1--6 &59155.09& $<\sim$3    & $<\sim$2 \\       
XM1 & 59209.43&   $3.79\pm0.10$          &  $1.13\pm0.03$ \\
XM2 & 59242.62&   $3.46\pm0.12$          &  $1.10\pm0.04$ \\
eR3 & 59252.53&   $0.56^{+0.30}_{-0.38}$ &  $0.20^{+0.13}_{-0.11}$  \\
Sw3 & 59391.00&   $6.36^{+1.22}_{-1.53}$ & $2.08^{+0.35}_{-0.40}$ \\
eR4 & 59441.16&   $7.36^{+0.70}_{-0.76}$ &  $3.02^{+0.28}_{-0.25}$  \\
XM3 & 59465.40&  $12.15^{+0.24}_{-0.29}$ &  $4.28^{+0.17}_{-0.18}$  \\
eR5 & 59623.99&   $0.86^{+0.39}_{-0.27}$ &  $0.34^{+0.11}_{-0.10}$ \\
Sw4 & 59662.65&   $0.64\pm0.23$          &  $0.67^{+0.22}_{-0.19}$ \\
XM4 & 59695.01&   $0.80\pm0.05$          &  $0.25\pm0.02$ \\
Sw5 & 59733.06& $<$1.05                  & $<$0.74 \\
Sw6 & 59783.50& $0.47^{+0.28}_{-0.19}$   &  $<$1.07  \\ 
Sw7 & 59873.53& $0.60^{+0.44}_{-0.32}$   & $0.43^{+0.32}_{-0.24}$ \\ 
Sw8 & 59992.20& $1.51\pm0.58$           & $0.92^{+0.35}_{-0.29}$ \\ \hline
\end{tabular}
\tablefoot{Observed 
0.2--5.0 and 0.5--2.0~keV fluxes, determined from model fits to all
spectra except for Sw1, 2, 4, 5, 6, 7, and 8, which were based on measured
count rates.  N1--6 denotes the estimated upper limit associated with
the non-detection in the summed NICER data.  } \end{table}

%
%

\section{X-ray spectral fits} \label{sec:Xspecfits}      


Our spectral fitting strategy is to start by modeling
the \textit{XMM-Newton} spectra (Sect.~\ref{sec:xmmfits}), as they had
the highest S/N.  We use those results to inform fitting of
the other (lower S/N)
X-ray spectra; eRASS and \textit{Swift} XRT spectra are described in
Sects.~\ref{sec:erassfits} and \ref{sec:swiftfits}, respectively.

All X-ray spectral fits were done in \texttt{Xspec} version 12.13.0c.
All parameter uncertainties are for one interesting parameter, and
were derived using a Markov chain Monte Carlo (MCMC) algorithm via the \texttt{chain}
routine in \texttt{Xspec}.  We used the Goodman-Weare 
sampler \citep{Goodman10}, chains of length 25000,
20 walkers, and a burn length of 5000.  Parameter errors are at the
$90$~percent confidence level, and are taken from the 5th and 95th percentile
values of the parameter distribution.  In all models, we included a
\textsc{TBabs} component to account for Galactic absorption by
\ion{H}{i} and H$_2$ totaling $1.04 \times 10^{20}$~cm$^{-2}$
\citep{Willingale13}.  We assumed the abundances of \citet{Anders89}.

\subsection{Spectral fits to \textit{XMM-Newton} data} \label{sec:xmmfits}  

Our strategy was to start with XM3, which 
sampled a high soft X-ray flux state and had the highest total counts
(19200; summing pn0+pn14+MOS1+MOS2), and then fit XM1, XM2, and XM4,
which sampled low soft X-ray flux states and had fewer total counts
(16600, 6700, and 4000, respectively).  For each observation, 
we fit pn0 (0.25--10~keV) + pn14 (0.5--10~keV) + MOS1 + MOS2 (both
0.2--10~keV) jointly.  We applied instrumental constant components for
cross-calibration purposes, keeping the constant for pn0 fixed at
unity; constants for the other spectra were usually within a few per
cent of unity for best-fitting models.  All spectra were grouped to 20
counts per bin to ensure use of $\chi^2$ statistics.

\subsubsection{XM3} 

We used a baseline spectral model consisting of the
following components, and with an eye toward developing a 
physically self-consistent model to apply to all four \textit{XMM-Newton} spectra:

\begin{itemize}

\item A hard X-ray power law to model emission from the hot ($T_{\rm e} \sim 10^9$~K),
optically thin corona. 

\item A soft X-ray excess, modeled via \textsc{CompTT} \citep{Titarchuk94}.
Warm comptonization of optical/UV thermal photons emitted by the
accretion disk by plasma with electron temperature $T_{\rm e} \sim$ 
0.1--1.0~keV and optical depth $\tau \sim$ 10--40 has been successful in
modeling the soft X-ray excesses of many nearby Seyferts,
\citep[e.g.,][]{Mehdipour11, DiGesu14, Porquet18, Petrucci18}.
We assumed a sphere geometry, and fixed the seed photon temperature
$T_{\rm seed}$ at 20 eV, though the fit results were insensitive to
the value of $T_{\rm seed}$.  In all four \textit{XMM-Newton}
observations, the value of $T_{\rm e}$ pegged at a lower limit of
0.1~keV, so we froze $T_{\rm e}$ to this value.\footnote{As an aside,
we note that omission of a soft excess component from the model caused $\chi^2/dof$
to increase sharply to 2.05; modeling of the soft excess with a simple
power law yielded poor data/model residuals, and $\chi^2/dof$ near 1.2.
We also note that replacing the \textsc{CompTT} component with a  
phenomenological blackbody component yielded a fit with nearly identical quality, with
$\chi^2/dof$=699.57/650, and with data/model residuals virtually identical to that
obtained using \textsc{CompTT}. The best-fitting temperature was $k_{\rm B}T = 90\pm3$~eV.
However, such a model has been known to be physically
implausible in describing Seyferts' soft excesses \citep{Gierlinski06}.} 

\item A Compton reflection component (Compton hump and Fe K emission), 
assuming a clumpy medium lying
out of the line of sight. We used \textsc{UxClumpy} \citep{Buchner19},
which follows \citet{Nenkova08} in assuming a distribution of clouds
that is preferentially concentrated toward the equatorial plane, with
a Gaussian angular height distribution and a uniform radial
distribution. There is also a Compton-thick inner ring with variable
covering factor. However, data above 10 keV (observed frame) does not
exist for $\jofour$, and the insensitivity of each \textit{XMM-Newton}
spectrum to the exact shape of the Compton hump means we could not
attain reliable constraints on any \textsc{UxClumpy} geometry or
viewing parameters. We froze the system inclination to $30^{\circ}$,
the cloud angular Gaussian distribution $\sigma_{\rm TOR}$ to
$20^{\circ}$ and the Compton-thick inner ring's covering fraction to
0. We froze the input power-law normalization to $3.5\times10^{-4}$ ph
cm$^{-2}$ s$^{-1}$ keV$^{-1}$, the best-fitting value obtained from a
joint fit to all XM1--4 spectra
 -- likely constrained via the marginal detection of the Fe~K line in XM4, as discussed
below in Sect.~\ref{sec:xmmcompare} -- and representing the average flux
response from a cloud distribution extending up to a few light years or more away.  
The input photon index was frozen to 2.  

\end{itemize}

\begin{figure}
  \includegraphics[width=0.93\columnwidth]{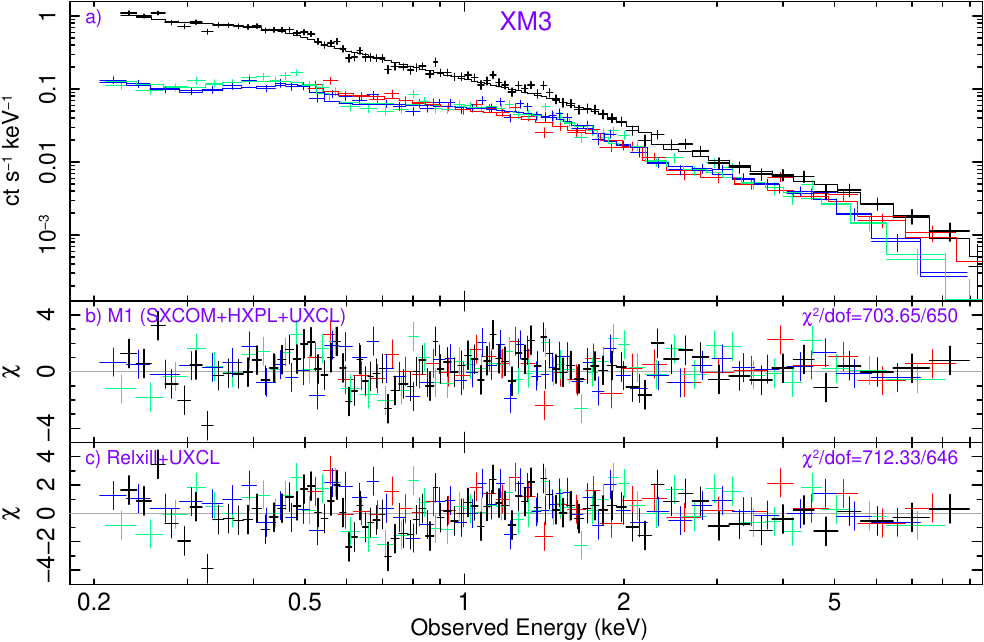}
  \caption{Spectral data and fits for XM3. Black, red, green, and blue denote pn0, pn14, MOS1, and
  MOS2, respectively.  Data have been rebinned by a factor of 3 for
  clarity. In panel a), crosses denote the counts spectra, and the
  histograms denote the best-fitting M1 ("SXCOM+HXPL+UXCL") model.
  The corresponding $\chi$ residuals are plotted in panel b).
  In this model, reflected emission is modeled with \textsc{UxClumpy}. 
  In panel c), we plot the $\chi$ residuals for the best-fitting model wherein the
   soft X-ray excess and Compton reflected component are modeled as blurred, ionized
   reflection via \textsc{relxill}.    } 
  \label{fig:XM3resids}
\end{figure}

We obtained a good fit, with $\chi^2/dof=704.26/650=1.083$. The
warm Comptonization optical depth was $\tau_{\rm warm}=45^{+9}_{-6}$.
The hard X-ray power-law photon index was $\Gamma_{\rm HX}=1.90\pm0.04$.
The data and model residuals are plotted in Fig.~\ref{fig:XM3resids}.
Other best-fitting parameters as listed in Table~\ref{tab:XM3parms}. 
We refer to this model as M1 or "SXCOM+HXPL+UXCL."

As an alternate model, we replaced the phenomenological
hard power law with a more physical, second \textsc{CompTT}
component.  We again used a spherical geometry and fixed $T_{\rm
seed}$ to 20 eV. We kept $T_{\rm e}$ fixed at 100~keV
\citep[e.g.,][]{Tortosa18}, due to the lack of constraints on any
high-energy power-law cutoff in $\jofour$.
The fit remained virtually unchanged, with $\chi^2/dof=703.65/650=1.083$, 
and with parameters for the warm Comptonization component unchanged.
The hot Comptonization depth was $\tau_{\rm hot}=0.27\pm0.03$.
Other best-fitting parameters as listed in Table~\ref{tab:XM3parms}. 
We refer to this model as M1-alt or "SXCOM+HXCOM+UXCL." In XM3 as well as to fits to all other
spectra, replacing the hard X-ray power law with a hot Comptonized component
made negligible impact on data/model residuals or fit statistics.   

Next, we tested if adding a full-covering neutral obscuration
component, modeled with \textsc{zTBabs} \citep{Wilms00}, improved the
fit; it did not, with an upper limit to column density $N_{\rm H}$ of
$8.6\times 10^{20}$~cm$^{-2}$. We thus do not include this component
further.

Finally, we tested if the soft excess could be modeled via reflection
off an ionized, relativistic accretion disk illuminated by a hard
X-ray power law.  We used \textsc{relxill} v.1.4.3 \citep{Garcia14, Dauser14},
keeping the outer radius fixed at
$400~R_{\rm g}$, the Fe abundance fixed to the solar value, and
power-law cutoff energy fixed at 100~keV (though the fit was
insensitive to thawing these parameters). We initially set the inner
radius to the inner stable circular orbit, but left this parameter
free.  We kept the black hole spin parameter, disk inclination, disk
ionization parameter, and power-law emissivity index all free.
We again include Compton-scattered components as in the previous fits
to model emission from a distant, neutral torus.  We refer to this
model as "RelXill+HXPL+UXCL."  
However, the best fit had $\chi^2/dof =
712.33/646=1.103$, with data/model residuals near 0.6--0.8 keV a bit
worse compared to the warm Comptonization model, as plotted in
Fig.~\ref{fig:XM3resids}.
In addition, the Akaike information criterion \citep[AIC;][]{Akaike73}
with finite sample correction by \citet{Sugiura78} yields that
${\Delta}$AIC going from the "SXCOM+HXCOM+UXCL" model to the
"RelXill+HXPL+UXCL" model is positive (+16.74, driven by both the
difference in fit statistic and the four additional free parameters),
indicating a preference for the warm Comptonization model.
However, we caution that the choice of soft excess model does not
significantly impact our conclusions in this paper concerning the
obscuring components.

\subsubsection{XM1}  \label{sec:xmm1fits} 

\begin{figure}
   \includegraphics[width=0.93\columnwidth]{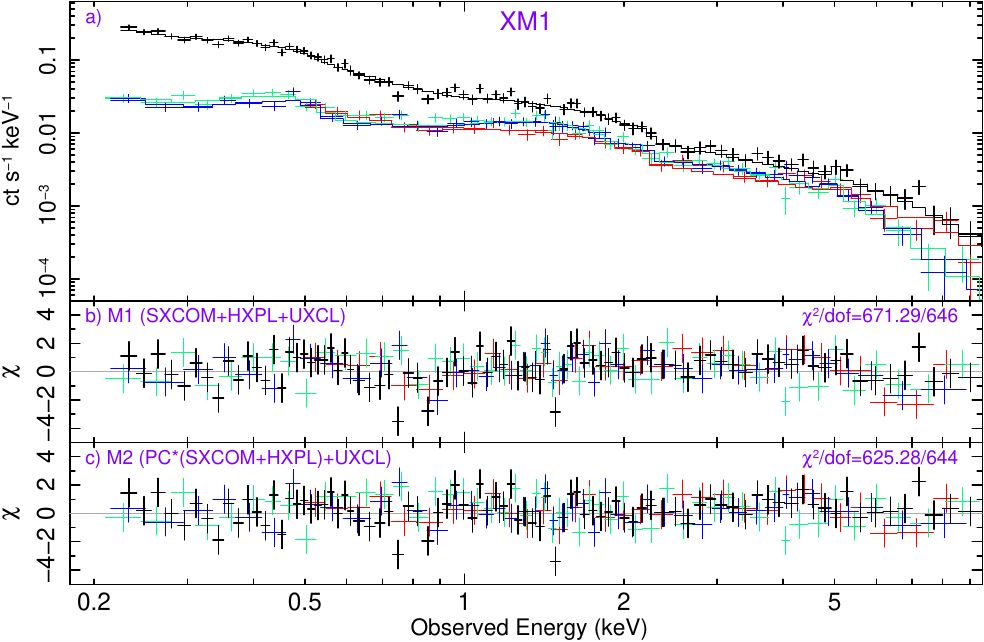}
    \caption{Same as Fig.~\ref{fig:XM3resids}, but for XM1. 
  In panel a), crosses denote the counts spectra, and the
  histograms denote the best-fitting M2 ("PC*(SXCOM+HXPL)+UXCL") model.
  In panels b) and c), respectively, we plot the
  $\chi$ residuals for the best-fitting
  M1 ("SXCOM+HXPL+UXCL") model (no obscuration),
  and for our preferred model, M2 ("PC*(SXCOM+HXPL)+UXCL"; with obscuration).}
    \label{fig:XM1resids}
\end{figure}

We now turn our attention to fitting XM1, which was obtained in December
2020, during the first low soft X-ray flux state.  We
first tried the unobscured phenomenological M1 ("SXCOM+HXPL+UXCL")
model.  We obtain a moderately reasonable fit, with
$\chi^2/dof=671.29/646=1.039$, and with data-model residuals plotted
in Fig.~\ref{fig:XM1resids}.  However, the hard X-ray power-law photon
index $\Gamma_{\rm HX}$ is quite flat, at $1.32\pm0.04$.

We then tested if the flat hard X-ray spectral slope could be
attributed to partial-covering obscuration masking a steeper power-law slope.
We applied a partial-covering component using
\textsc{zpcfabs}.\footnote{We also tried a full-covering component,
but obtained only an upper limit to $N_{\rm H}$, in this case $N_{\rm
H} < 5 \times 10^{20}$~cm$^{-2}$.}
We applied the partial-covering component only to the SXCOM and HXPL 
components, and not the \textsc{UxClumpy} component.
Our best-fitting "M2" model, or "PC*(SXCOM+HXPL)+UXCL" model, had $\chi^2/dof =
625.28/644=0.973$, with improved data/model residuals as 
plotted in Fig.~\ref{fig:XM1resids}, 
and a steeper value for $\Gamma_{\rm HX}$, $1.68^{+0.11}_{-0.10}$.
The partial-covering obscurer had column
density $N_{\rm H,PC} = 9.9^{+6.2}_{-3.3} \times 10^{21}$~cm$^{-2}$
and covering fraction $CF = 62^{+11}_{-10}$ percent. Other
best-fitting model parameters are listed in Table~\ref{tab:XM124parms}.
A set of 500 Monte Carlo simulations using the \texttt{simftest}
command in \texttt{Xspec} indicates that the addition of the
partial-covering obscurer improves the fit at the $>$99.8 percent
confidence level.  In addition, ${\Delta}$AIC when comparing the model
lacking the obscurer to the model containing the obscurer is $-$41.9,
confirming the necessity of the obscurer in the model.

Finally, we replaced the HXPL component with a hot Comptonization
component to form our M2-alt ("PC*(SXCOM+HXCOM)+UXCL") model, which
yielded $\chi^2/dof=626.61/644=0.973$, data-model residuals virtually
identical to those from the M2 fit, and best-fitting parameters as
listed in Table~\ref{tab:XM124parms}.

\subsubsection{XM2}   \label{sec:xmm2fits} 

Observation XM2 occurred 33 days after XM1 and we found similar soft
and hard X-ray flux levels and a very similar X-ray spectral shape to
XM1. Spectral fits to XM2 thus proceeded in a manner identical to
those for XM1, although the lower exposure time for XM2 yielded total
spectral counts a factor of 2.5 lower than for XM1.  Again, we first fit
an unobscured M1 ("SXCOM+HXPL+UXCL") model, whose best fit
again required a very flat power-law photon index ($\Gamma_{\rm HX} =
1.44^{+0.08}_{-0.05}$); $\chi^2/dof$ was 305.07/294 = 1.038.  

\begin{figure}
   \includegraphics[width=0.93\columnwidth]{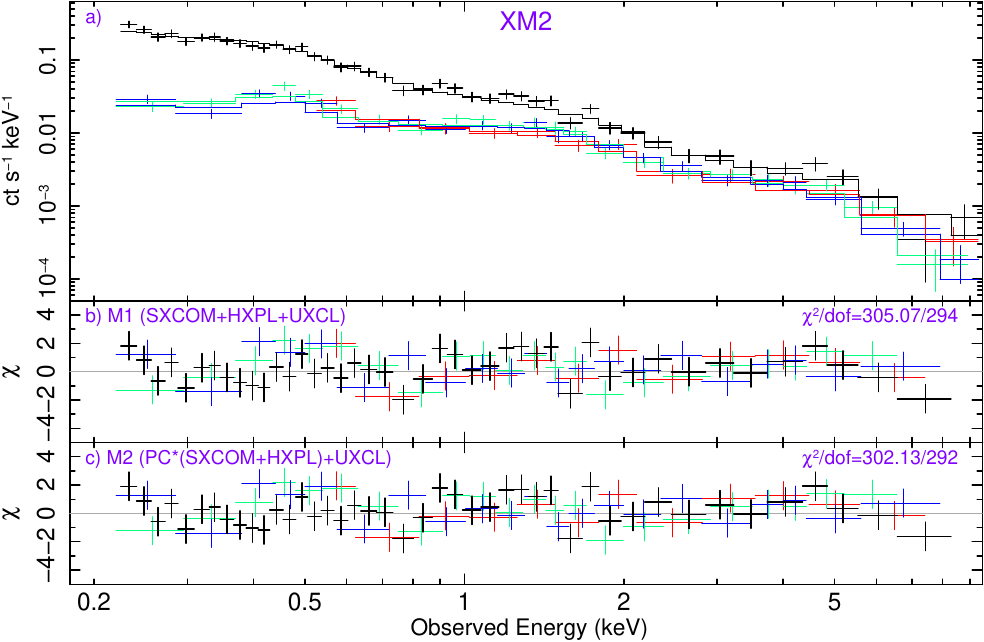}
    \caption{Same as Fig.~\ref{fig:XM1resids}, but for XM2.}
    \label{fig:XM2resids}
\end{figure}

When we added a partial-covering obscurer (M2; "PC*(SXCOM+HXPL)+UXCL"),
the improvement in fit was negligible, with $\chi^2/dof$ falling by
only 2.94/2 and with $\Gamma_{\rm HX}$ increasing to $1.64^{+0.11}_{-0.07}$.
In addition, Monte Carlo simulations indicate that the addition of the
obscurer improves the fit at merely the 80~percent confidence level. 
Other best-fitting model parameters are
listed in Table~\ref{tab:XM124parms}, and data/model residuals are
plotted in Fig.~\ref{fig:XM2resids}. 
The similarity in flux levels and spectral shape to XM1 suggest that
XM2 was similarly obscured.  However, we caution that detection of the partial-covering
obscurer in XM2 is not statistically robust.  Under the assumption
that during both XM1 and XM2, the continuum is obscured by the same
cloud of gas, XM1 provides better insight into the cloud properties.
Finally, the parameters for our best-fitting M2-alt
("PC*(SXCOM+HXCOM)+UXCL") model, which yielded data-model residuals
that are virtually identical to those for M2, are listed in
Table~\ref{tab:XM124parms}.

\subsubsection{XM4}   \label{sec:xmm4fits} 

\begin{figure}
   \includegraphics[width=0.93\columnwidth]{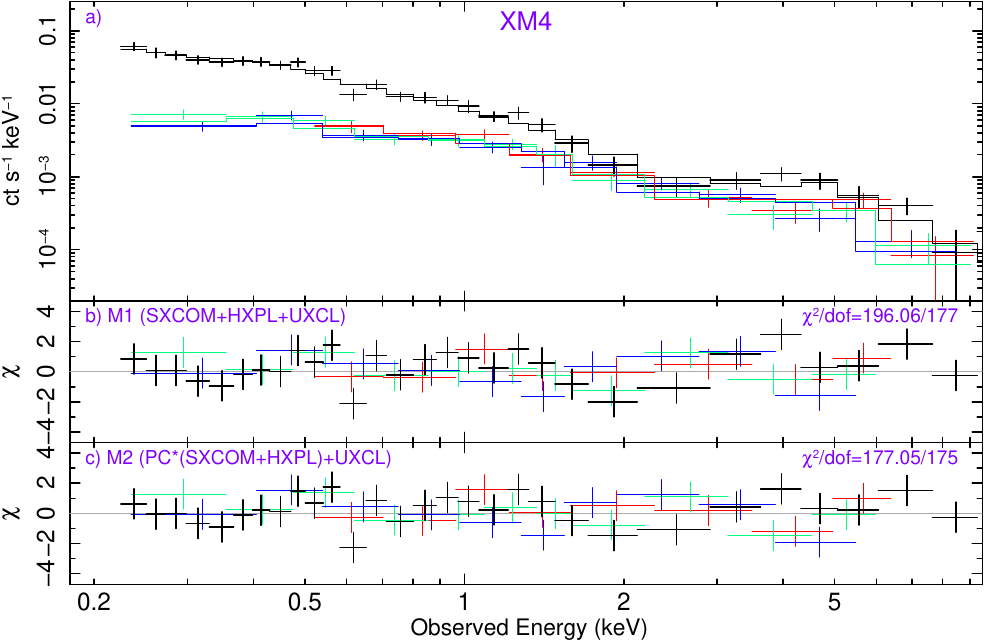}
    \caption{Same as Fig.~\ref{fig:XM1resids}, but for XM4.}
    \label{fig:XM4resids}
\end{figure}

Observation XM4 occurred in April 2022, during the second major soft
X-ray flux drop.  The counts spectra are plotted in the top panel of
Fig.~\ref{fig:XM4resids}, and indicate a hard X-ray spectrum that is
both very flat and has a strong convex shape. 
Again, we first tested an unobscured phenomenological M1 ("SXCOM+HXPL+UXCL") model,
which yielded $\chi^2/dof$=196.06/177=1.108, $\Gamma_{\rm HX} = 1.69\pm0.16$, and
moderate data-model residuals as plotted in Fig.~\ref{fig:XM4resids}.

Adding a full-covering neutral obscurer (\textsc{zTBabs}) to the model
failed to bring any improvement (upper limit to $N_{\rm H}$ of
$8\times10^{20}$~cm$^{-2}$), as model corrections impacting the hard
band were instead needed.  
Again, adding a partial-covering component (M2, or
"PC*(SXCOM+HXPL)+UXCL") yielded a good fit, with
$\chi^2/dof=177.05/175=1.012$ and good data/model residuals, as
plotted in Fig.~\ref{fig:XM4resids}.  A set of 500 Monte Carlo
simulations indicates that the addition of the partial-covering
obscurer improves the fit to XM4 at the $>$99.8 percent confidence
level.  In addition, ${\Delta}$AIC going from a model lacking the
obscurer to the model containing the obscurer is $-$14.62, confirming
the necessity of the obscurer in the fit to XM4.  The partial-coverer
had $N_{\rm H,PC}=2.84^{+3.98}_{-1.29}\times10^{23}$~cm$^{-2}$ and $CF
= 0.79^{+0.07}_{-0.13}$, with other best-fitting model parameters
listed in Table~\ref{tab:XM124parms}.
For completeness, we once again fit an M2-alt
("PC*(SXCOM+HXCOM)+UXCL") model, achieving a fit nearly identical to
that for M2, with $\chi^2/dof=176.50/175=1.009$ and best-fitting
parameters listed in Table~\ref{tab:XM124parms}.

\subsubsection{Comparison of \textit{XMM-Newton} spectral fit results} \label{sec:xmmcompare}

\begin{figure}
\includegraphics[width=0.93\columnwidth]{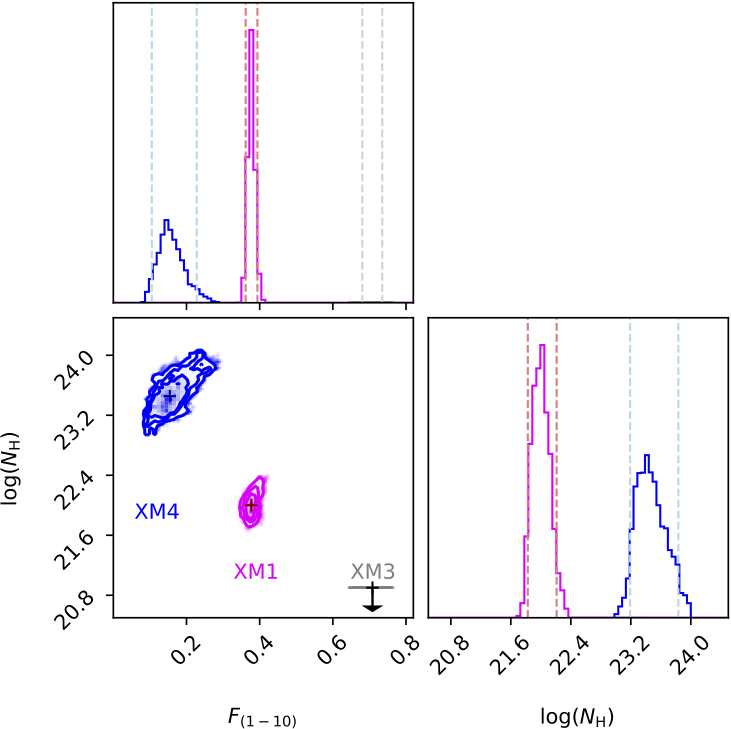}  
\includegraphics[width=0.93\columnwidth]{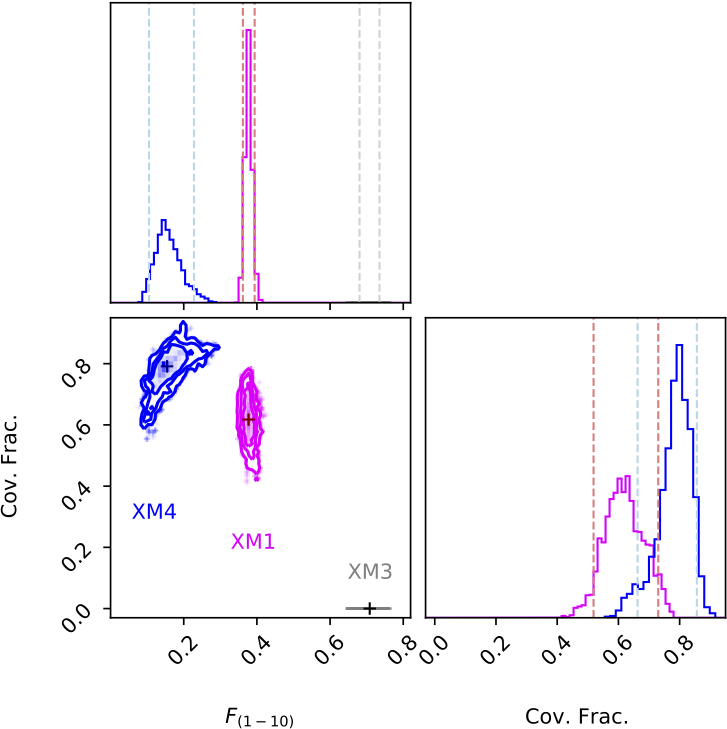}  
\caption{MCMC uncertainty analysis for column density $N_{\rm H,PC}$ (top panels)
and covering fraction $CF$ (bottom panels) as a function of $F_{1-10}$, the unabsorbed 1--10 keV
power-law best-fitting M1 (unobscured) model for XM3 (gray), and
M2 (obscured) models for XM1 (magenta), and for XM4 (blue). 
Black crosses represent the best-fitting parameter values;
dashed lines indicate the 90 percent confidence limits.
$F_{1-10}$ is plotted in units of $10^{-12}$ erg cm$^{-2}$ s$^{-1}$.
$N_{\rm H,PC}$ is in units of cm$^{-2}$.
XM2 is omitted for clarity, given its
relatively poor model constraints.
XM3 did not have a partial-covering obscurer. However, to be able to include it
in these panels for the purpose of comparing $F_{1-10}$ to XM1 and XM4, we include
XM3 here at an arbitrary limit of log($N_{\rm H,PC}$)$<$20.9 to represent the  zero value. }
\label{fig:MCMC_F0110}
\end{figure}

\begin{figure}
\includegraphics[width=0.93\columnwidth]{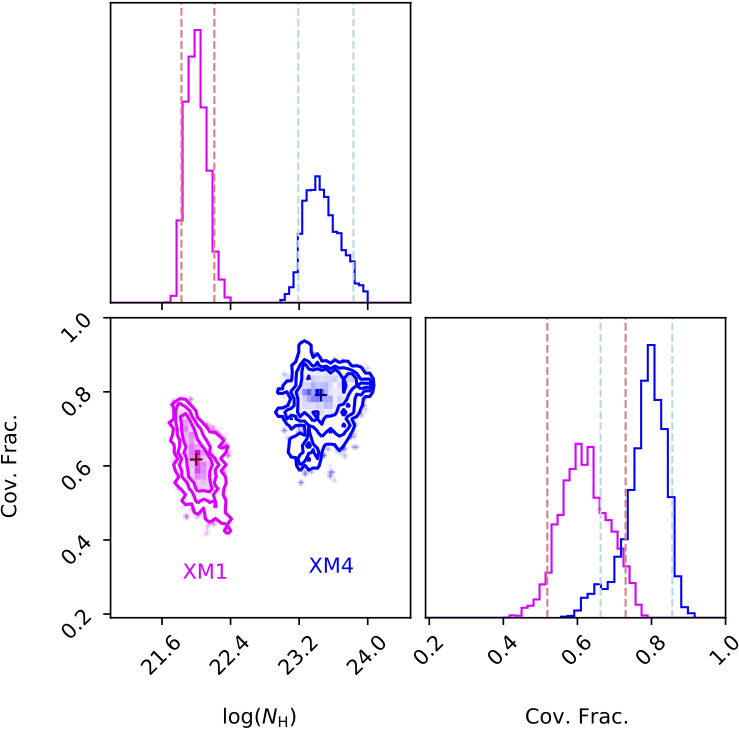} 
\caption{MCMC uncertainty analysis for covering fraction $CF$ as  
a function of column density $N_{\rm H,PC}$ for the best-fitting
M2 (obscured) models for XM1 (magenta) and XM4 (blue).
Black crosses represent the best-fitting parameter values;
dashed lines indicate the 90 percent confidence limits.
$N_{\rm H,PC}$ is in units of cm$^{-2}$.
}
\label{fig:MCMC_NHCF}
\end{figure}

In Fig.~\ref{fig:MCMC_F0110}, we plot the results of MCMC analysis for
column density $N_{\rm H,PC}$ and covering fraction $CF$ as a function
of the unabsorbed $1-10$~keV flux of the hard X-ray coronal
component flux $F_{1-10}$ for XM3,
XM1, and XM4. The covering fractions of the two clouds do not show any
trend with either $F_{1-10}$ or, as shown in Fig.~\ref{fig:MCMC_NHCF},
$N_{\rm H,PC}$.  Intriguingly, though, Fig.~\ref{fig:MCMC_F0110} seems
to suggest that periods of relatively higher values of $N_{\rm H,PC}$
occur during periods of relatively lower values of $F_{1-10}$ across
the unobscured state and both obscuration events. Such
anticorrelations have been reported previously for a tiny number of
Seyferts undergoing variable obscuration, for example
\citet{Couto16} or \citet{Marchesi22}.  However, given that we lack energy
coverage above 10~keV, and given that we have observed only two
obscuration events, such a conclusion for $\jofour$ must be treated as
tentative at best.

\begin{figure}
\includegraphics[width=0.93\columnwidth]{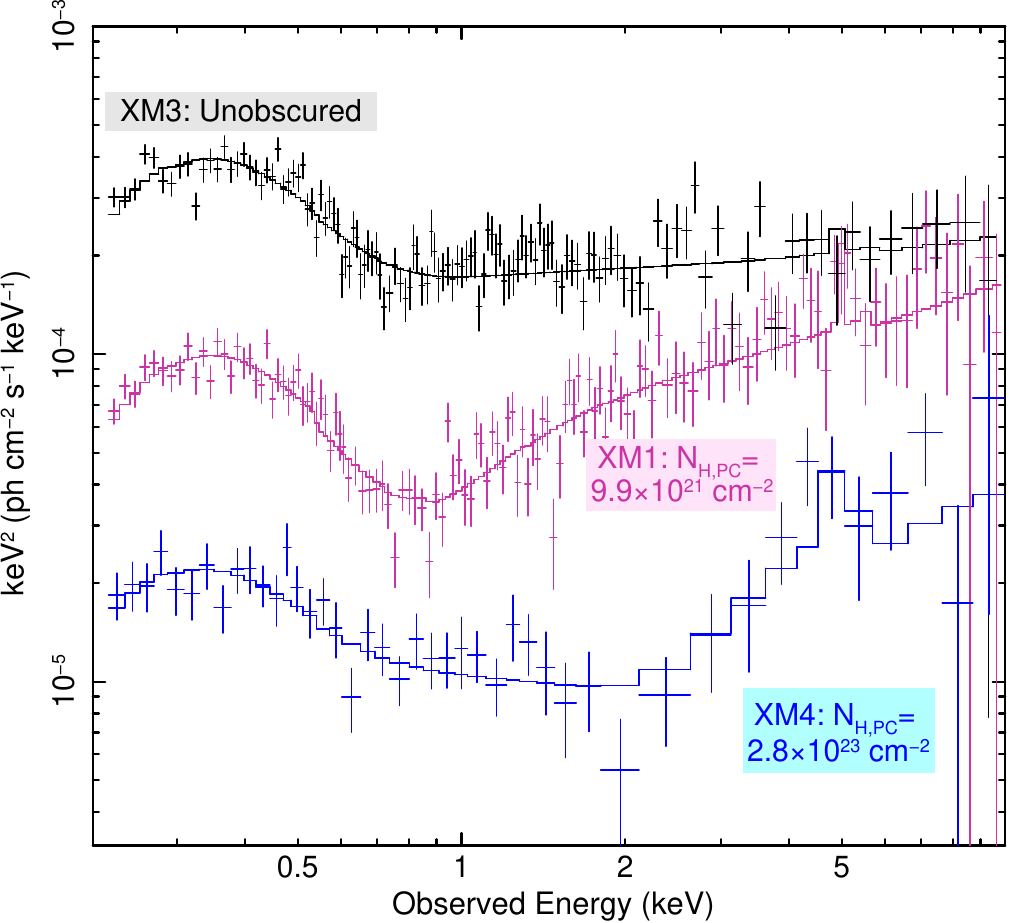} 
\caption{Unfolded spectra for the best fits to XM3, XM1, and XM4.
Data and the best-fitting M1 (unobscured) model for XM3 are shown in black.
For XM1 and XM4, we plot the data and the best-fitting
M2 (obscured) models in magenta and blue, respectively.}
\label{fig:PLOTEEUF}
\end{figure}

Finally, in Fig.~\ref{fig:PLOTEEUF}, as a summary of our best-fit
models, we present the spectra and best-fit models for XM3, XM1, and
XM4 in "model space" (i.e., "unfolded" and deconvolved from the
instrument responses).  One can see the hardening of the hard X-ray
spectrum as the source transitions between the unobscured state (XM3),
the first obscured state (XM1), and the second, more strongly obscured
state (XM4).  

One can also see the increasing relative prominence of the
Fe~K$\alpha$ line in the hard X-ray spectrum as the transmitted
continuum drops to the level in XM4, suggesting that the Fe~K line
originates in distant material.  A set of 500 Monte Carlo simulations
for a model lacking the \textsc{UxClumpy} component and instead
containing a Gaussian with rest-frame energy centroid fixed at 6.4 keV
and width fixed at 1 eV indicate that inclusion of the Gaussian
improves the fit at the 93.6 percent confidence levels for XM3, 97.2
(90.0) percent for XM1 (XM2), and 99.4 percent (2.8$\sigma)$  for
XM4.  In addition, in XM4, the 90 percent confidence uncertainty on
the equivalent width relative to the local continuum is quite high --
$339^{+147}_{-133}$~eV. We thus consider detection of the Fe~K line in
$\jofour$ to be tentative at best and only in XM4.


\renewcommand{\arraystretch}{1.18}
\begin{table*}
\caption[]{    Best-fitting models to XM3    }
        \centering
\label{tab:XM3parms}
\begin{tabular}{l|ll} \hline\hline

\textbf{M1 (SXCOM+HXPL+UXCL)} (preferred) & Soft Comptonization                               & Hard Comptonization         \\  \cline{2-3} 
$\chi^2/dof= 704.26/650=1.084$  & $T_{\rm seed}=20$~eV*                             & $\Gamma_{\rm HX}=1.90\pm0.04$  \\ 
$F_{0.5-2} =  4.28^{+0.17}_{-0.18}\times10^{-13}$ \ecgs & $T_{\rm e} = 0.1$~keV*    & $F_{1-10}^{(a)}= 7.11^{+0.24}_{-0.19}\times10^{-13}$ \ecgs\\  
$F_{0.2-5} = 12.15^{+0.24}_{-0.29}\times10^{-13}$ \ecgs & $\tau_{\rm warm} = 45^{+9}_{-6}$   \\
$F_{2-10}  =  5.29^{+0.23}_{-0.25}\times10^{-13}$ \ecgs & $N_{\rm SC}^{(b)}=8.51^{+3.22}_{-2.27}\times10^{-2}$   \\  
\textbf{M1-alt (SXCOM+HXCOM+UXCL):}       &    &   $T_{\rm seed}=20$~eV*,  $T_{\rm e} =100$~keV* \\     
$\chi^2/dof=703.65/650=1.083$             &    &   $\tau_{\rm hot} = 0.27\pm0.03$, $N_{\rm HC}^{(b)} =4.69^{+0.52}_{-0.47}\times10^{-5}$  \\    \hline\hline 
\textbf{RelXill+HXPL+UXCL (unpreferred)}        & Ionized Reflection   &  Hard X-ray power law  \\  \cline{2-3}
$\chi^2/dof= 712.33/646=1.103$ & Emissivity index = $6.7^{+3.0}_{-1.6}$  &  $\Gamma_{\rm HX}=2.33^{+0.06}_{-0.07}$  \\
                               & Spin $a^* > 0.89$    &   Relxill Norm.\ = $2.64^{+1.53}_{-1.68}\times10^{-6}$   \\ 
                               & Inner radius $< 2 R_{\rm ISCO}$ &   \\ 
                               & Incl.\ = $55^{+16}_{-11}$ deg. \\     
                               & log($\xi$, [erg~cm~s$^{-1}$]) = $0.70^{+0.61}_{-0.46}$ \\
                               & Refl.\ Frac.\ = $13.4^{+5.5}_{-4.0}$ \\  \hline\hline

\end{tabular}
\tablefoot{
An asterisk (*) denotes a fixed parameter.\\
\tablefoottext{a}{Unobscured 1--10 keV observed-frame flux of the hard power-law component.} \\
\tablefoottext{b}{$N_{\rm HC}$ and $N_{\rm SC}$ denote, respectively, 
the normalizations of the hard and soft X-ray \textsc{CompTT} components.}
}
\end{table*}


\begin{sidewaystable*}
\caption[]{   Best-fitting partial-covering obscuration models to XM1, XM2, and XM4    }
        \centering
\label{tab:XM124parms}
\begin{tabular}{l|llll} \hline\hline

\multicolumn{4}{c}{\large \bf XM1 (2020 Dec.\ 26)} \\ \hline\hline

\textbf{M2 (PC*(SXCOM+HXPL)+UXCL):}     &     Obscurer     & Soft Comptonization     & Hard Comptonization         \\  \cline{2-4}
$\chi^2/dof= 625.28/644=0.971$         &  $N_{\rm H,PC}=9.9^{+6.2}_{-3.3}\times10^{21}$~cm$^{-2}$  &  $T_{\rm seed}=20$~eV* &  $\Gamma_{\rm HX}=1.68^{+0.11}_{-0.10}$  \\
$F_{0.5-2} = (1.13\pm0.03)\times10^{-13}$  \ecgs &  $CF = 0.62^{+0.11}_{-0.10}$   &  $T_{\rm e}=0.1$~keV* &  $F_{1-10}^{(a)}= (3.77\pm0.16)\times10^{-13}$ \ecgs \\
$F_{0.2-5} = (3.79\pm0.10)\times10^{-13}$  \ecgs  & & $\tau_{\rm warm} = 63^{+23}_{-14}$ &  \\
$F_{2-10} = 2.97^{+0.19}_{-0.20}\times10^{-13}$ \ecgs  &  &  $N_{\rm SC}^{(b)}=3.97^{+2.07}_{-1.14}\times10^{-2}$ & \\
\textbf{M2-alt (PC*(SXCOM+HXCOM)+UXCL):}     & &    &   $T_{\rm seed}=20$~eV*,  $T_{\rm e} =100$~keV* \\
$\chi^2/dof=626.61/644=0.973$                & &    &   $\tau_{\rm hot} = 0.48^{+0.14}_{-0.12}$, $N_{\rm HC}^{(b)} =1.12^{+0.62}_{-0.34}\times10^{-5}$  \\    \hline\hline

\multicolumn{4}{c}{\large \bf XM2 (2021 Jan.\ 28)} \\ \hline\hline

\textbf{M2 (PC*(SXCOM+HXPL)+UXCL):}     &     Obscurer     & Soft Comptonization     & Hard Comptonization         \\  \cline{2-4}
$\chi^2/dof= 302.13/292=1.035$  &  $N_{\rm H,PC}=7.0^{+3.8}_{-0.8}\times10^{21}$~cm$^{-2}$ & $T_{\rm seed}=20$~eV* & $\Gamma_{\rm HX}=1.64^{+0.11}_{-0.07}$  \\
$F_{0.5-2} = (1.10\pm0.04)\times10^{-13}$  \ecgs &  $CF = 0.40^{+0.39}_{-0.15}$   &  $T_{\rm e}=0.1$~keV*  &  $F_{1-10}^{(a)} = (2.93\pm0.22)\times10^{-13}$ \ecgs \\
$F_{0.2-5} = (3.46\pm0.12)\times10^{-13}$  \ecgs  & & $\tau_{\rm warm} = 109^{+34}_{-51}$  & \\
$F_{2-10}  = 2.57^{+0.25}_{-0.19}\times10^{-13}$ \ecgs  & & $N_{\rm SC}^{(b)}=1.56^{+0.75}_{-0.12}\times10^{-2}$ & \\
\textbf{M2-alt (PC*(SXCOM+HXCOM)+UXCL):}     & &    &   $T_{\rm seed}=20$~eV*,  $T_{\rm e} =100$~keV* \\
$\chi^2/dof=302.96/292=1.038$                & &    & $\tau_{\rm hot} = 0.61^{+0.28}_{-0.19}$, $N_{\rm HC}^{(b)} = 6.92^{+3.24}_{-1.64}\times10^{-6}$   \\   \hline\hline\hline

\multicolumn{4}{c}{\large \bf XM4 (2022 Apr.\ 25)} \\ \hline\hline

\textbf{M2 (PC*(SXCOM+HXPL)+UXCL):}     &     Obscurer     & Soft Comptonization     & Hard Comptonization         \\  \cline{2-4}
$\chi^2/dof= 177.07/175=1.012$   &  $N_{\rm H,PC}=2.84^{+3.98}_{-1.29}\times10^{23}$~cm$^{-2}$ & $T_{\rm seed}=20$~eV* & $\Gamma_{\rm HX}=2.21^{+0.27}_{-0.26}$ \\

$F_{0.5-2} = (2.5\pm0.2)\times10^{-14}$  \ecgs &  $CF = 0.79^{+0.09}_{-0.13}$   &  $T_{\rm e}=0.1$~keV* & $F_{1-10}^{(a)} = 1.54^{+0.74}_{-0.49}\times10^{-13}$ \ecgs \\
$F_{0.2-5} = (8.0\pm0.5)\times10^{-14}$  \ecgs  & & $\tau_{\rm warm} = 32^{+15}_{-9}$ &  \\
$F_{2-10} = 6.6^{+1.3}_{-0.9}\times10^{-14}$ \ecgs  & & $N_{\rm SC}^{(b)}= 3.68^{+3.79}_{-1.79}\times10^{-2}$   &  \\
\textbf{M2-alt (PC*(SXCOM+HXCOM)+UXCL):}       &  &  &   $T_{\rm seed}=20$~eV*,  $T_{\rm e} =100$~keV* \\
$\chi^2/dof=176.50/175=1.009$   &  & & $\tau_{\rm hot} = 0.14^{+0.06}_{-0.05}$, $N_{\rm HC}^{(b)} = 2.46^{+1.25}_{-0.94}\times10^{-5}$  \\  \hline\hline

\end{tabular}
\tablefoot{
An asterisk (*) denotes a fixed parameter.\\
\tablefoottext{a}{Unobscured 1--10 keV observed-frame flux of the hard power-law component.} \\
\tablefoottext{b}{$N_{\rm HC}$ and $N_{\rm SC}$ denote, respectively,
the normalizations of the hard and soft X-ray \textsc{CompTT} components.}
}

\end{sidewaystable*}


\subsection{Spectral fits to eRASS data} \label{sec:erassfits}

\begin{figure}
\includegraphics[width=0.93\columnwidth]{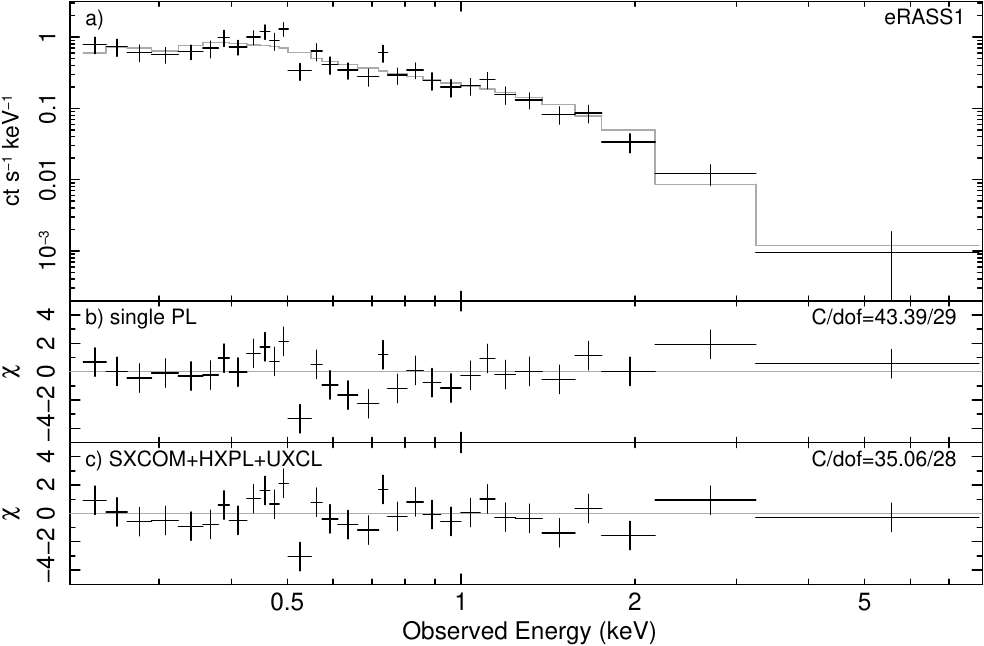}
 \caption{Spectral data for observation eR1. 
 In the top panel, crosses denote the counts spectra, and the
 histograms denote the best-fitting M1 (``SXCOM+HXPL+UXCL'') model, with
 data-model residuals plotted in panel c). Panel b) shows the residuals
 to a simple power-law fit.}
\label{fig:residser1}
\end{figure}

We first fit the spectrum for eR1, which had 531 counts in 0.2--5.0 keV.
We binned it to a minimum of 15 counts per bin, and fit using the C-statistic.
When we fit a simple power law, we get a modest fit, with $C/dof =
43.39/29=1.50$, and with large data-model residuals are plotted in
Fig.~\ref{fig:residser1}.  We then applied a 
"SXCOM+HXPL+UXCL" (M1)\footnote{The \textsc{UxClumpy} component is not detected but is
included in the model for consistency with models fit in
Sect.~\ref{sec:xmmfits}.} model.  $\tau_{\rm warm}$ was not constrained when left free,
so we froze it at 45, the best-fitting value from XM3. We obtained an
improved fit, with $C/dof = 35.06/28 = 1.25$, $\Gamma_{\rm HX} =
2.03^{+0.40}_{-0.36}$, power-law $F_{1-10} =
7.65^{+0.37}_{-0.24}\times10^{-13}$ {\ecgs}, and observed, absorbed flux
values as listed in Table~\ref{tab:Xfluxes}.  Data-model residuals are
plotted in Fig.~\ref{fig:residser1}.  A set of 500 MC simulations
indicate that adding the soft component to the model improves the fit
at the $>$99.8 percent confidence level; the model with the soft component
corresponds to a value of AIC that is lower by 5.93 compared to the simple power-law model.
The addition of a full-covering component does not improve the fit;
the upper limit is $3\times10^{20}$~cm$^{-2}$. A partial-covering
component does not improve the fit either. Upper limits to $N_{\rm
H,PC}$ are $(3-5)\times10^{20}$~cm$^{-2}$ when values of $CF$ above
0.75 are assumed, and peg at $1\times10^{24}$~cm$^{-2}$ for values of
$CF$ 0.7 or lower. 
These limits, combined with similarity in broadband flux level to XM3,
support the notion that eR1 sampled $\jofour$ in an unobscured state.


The spectrum of the other relatively high-flux eRASS observation, eR4,
had 297 counts in 0.2--5.0 keV.  Again, we binned to 15 counts per 
bin, and fit using the C-statistic.  A simple power-law yielded
an adequate fit, with $C/dof=8.41/15 = 0.56$, and unobscured flux
$F_{1-10}=2.31^{+0.85}_{-0.66} \times 10^{-13}$~erg cm$^{-2}$
s$^{-1}$.  In order to avoid overparametrizing the data
given the limited number of observed counts, we do not
explore more complex models\footnote{Upper limits to column density of
a partial- or full-covering component added to the model were not
attainable.}.  The resulting 0.2--5.0 and 0.5--2.0 keV observed,
absorbed fluxes are listed in Table~\ref{tab:Xfluxes}; these flux
values are closer to that of eR1 and XM3 (unobscured state) than to
those of XM1, XM2, or XM4. In addition, eR4 occurred only 24 days
before XM3. We thus infer that eR4 has also sampled $\jofour$ in
its unobscured state.

Finally, we turn our attention to eR2, eR3, and eR5, which had only
36, 19, and 31 counts in 0.2--5.0 keV, respectively.  We binned to five counts per
bin, and could only fit up to $\sim$2~keV. All fits used the
C-statistic.  The data quality was low enough that a simple power law
with $\Gamma$ frozen to 2, and obscured only by the Galactic column,
yielded satisfactory fits, with no significant improvement to the fits
when thawing $\Gamma$. The best fits had $C/dof$ = 6.5/5, 1.2/1, and
14.9/4, respectively. The resulting X-ray fluxes for all eRASS spectra
are listed in Table~\ref{tab:Xfluxes}; they are close to fluxes for the obscured
\textit{XMM-Newton} spectra.

\subsection{Spectral fits to \textit{Swift} XRT data}  \label{sec:swiftfits}

We performed spectral fitting only for Sw3 (90 total counts in 0.2--10 keV); 
all other spectra had too few counts (30 or fewer) for broadband spectral fitting.

Sw3 occurred 50 and 74 days, respectively, before eR4 and XM3, which
were consistent with lack of obscuration. We thus started with the
hypothesis that the source was also unobscured during Sw3.  
We binned Sw3 to 10 counts per bin, and fit the 0.4--4.5~keV band using the C-statistic.
A simple power law yielded a poor fit ($C/dof = 17.74/7 = 2.52$);
We thus added a soft excess via warm Comptonization, and fit using a
``SXCOM+HXPL+UXCL'' model following the M1 fit to XM3, keeping $\tau_{\rm
warm}$ fixed to 45.  Formally, the best fit was good, with $C/dof =
6.07/6 = 1.01$, and good data/model residuals.
A set of 500 MC simulations indicate that adding the soft component
to the model improves the fit at the $>$99.8 percent confidence
level; the model with the soft component corresponds to a value of AIC
that is lower (by 6.87) compared to the simple power-law model.  
However, it is suspicious that the best-fitting value of $\Gamma_{\rm HX}$ was very flat, at 
$0.7\pm0.5$.  
The addition of a partial-covering obscurer yields $C/dof = 5.55/4 =
1.39$; $\Gamma_{\rm HX}$ remains very poorly constrained, and there is
strong degeneracy in both the $N_{\rm H,PC}$--$N_{\rm WC}$ and
$CF$--$N_{\rm WC}$ planes.
Moreover, even with $\Gamma_{\rm HX}$ frozen in this model, the value
of AIC has actually increased from 16.87 in the model lacking the
obscurer to 23.55 here, as there is a heavy penalty for the correction
for the small sample size.
We thus cannot draw any firm conclusions about whether the first
obscuration event is still ongoing during Sw3. However, the closeness
in time to eR4 and XM3 as well as the similarity in soft X-ray flux
values (see Table~\ref{tab:Xfluxes}) would argue that it is more likely that the event had
ended (or was ending) by this time.

For Sw1, 2, 4, 5, 6, 7, and 8, we used the UKSSDC
Swift-XRT products website\footnote{\url{https://www.swift.ac.uk/};
  \citep{Evans07,Evans09}} to derive count rates and errors via the
Bayesian method of \citet{Kraft91}.  We converted to fluxes assuming a
power-law model with a photon index of 2.0.
Fluxes are listed in Table~\ref{tab:Xfluxes}.

%
%

\section{Checking for reddening in the optical/UV continuum SED}   \label{sec:OUVSED}

We now discuss fits to the time-variable optical/UV SED, using the
six-filter \textit{Swift} UVOT observations Sw3--8.  As demonstrated
by the light curves in Fig.~\ref{fig:lcs_sw38_J0458}, the optical/UV
SED becomes relatively redder as overall flux decreases from
observations Sw3 and Sw4 to Sw5--7, and bluer as flux increases
to Sw8. As Sw4--8 occur during the
second X-ray obscuration event, we thus fit their six-channel
SEDs to test if there is associated reddening impacting the optical/UV
SED.  We perform all fits in \texttt{Xspec}.  To model extinction
associated with the Galactic column, we include a \textsc{redden}
component with $E(B-V)$ fixed to 0.009 mag in all fits,
based on the dust maps of \citet{Schlegel98}.  We also
include a Sb host galaxy template from the SWIRE galaxy
template library \citep{Polletta07}
in all fits. However, due to the vast difference in spectral shape
between the Seyfert component and the galaxy component, the galaxy
component was not required in the fits; the AGN continuum clearly dominates.
We obtained small upper limits to the host template's
normalization: 2.0--7.5~eV (1653--6199 $\AA$) flux $<1.34\times10^{-14}$ 
erg~cm$^{-2}$~s$^{-1}$, compared to $1.22\times10^{-11}$ 
erg~cm$^{-2}$~s$^{-1}$ for $\jofour$ and for the same bandpass during Sw7.

\begin{figure*}[h]
\includegraphics[width=1.99\columnwidth]{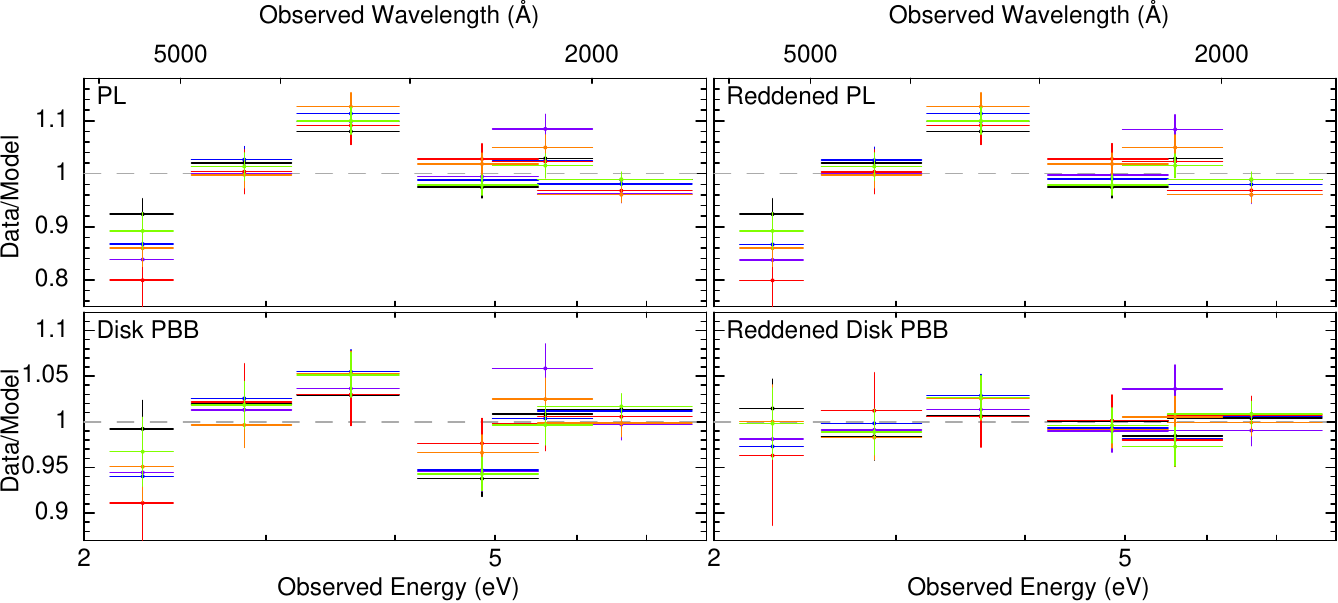}   
\caption{ Data-model residuals to fits to the optical/UV SED using the \textit{Swift} UVOT data from observations Sw3--8.
  Data for Sw3, 4, 5, 6, 7, and 8 are plotted in black, blue, purple, red, orange, and green, respectively.
  In the top panels, the continuum is modeled as a simple power law.
  In the bottom panels, it is modeled as a multitemperature disk blackbody.
  In the panels in the left column, the models contain no reddening, whereas
  in the right column, the models do contain reddening. } 
\label{fig:uvotsed}
\end{figure*}

We first attempted to model the optical/UV SEDs with simple power laws, with no reddening
beyond Galactic, and with all photon indices and normalizations
untied.  This is a relatively poor fit, with the sum of $\chi^2/dof$
across all six spectra equal to 203.30/23, and with strong convex
curvature across the UVOT band. However, the fits did support a
general trend of steeper spectral slope at lower 2.0--7.5~eV flux, with $\Gamma =
1.30\pm0.04$ and $1.39\pm0.04$ for Sw3 and Sw4, steepening to
$1.66\pm0.07$ and $1.58^{+0.04}_{-0.05}$ by Sw6 and Sw7, respectively,
and flattening to $1.33\pm0.05$ by Sw8,
and consistent with the picture of spectral variability in
Fig.~\ref{fig:lcs_sw38_J0458}. Applying a reddening component at the
system redshift with \textsc{zredden} did not yield any improvement in
fit, with upper limits to $E(B-V)$ ranging from 0.017 to
0.056. Data/model residuals are plotted in Fig.~\ref{fig:uvotsed}.

We substituted the power-law components in both the reddened and
unreddened fits with multitemperature blackbody disk components,
modeled with \textsc{diskpbb}. The index describing the power-law
dependence of temperature was tied across all fits.  We allowed the
temperature at the inner radius, $T_1$, to be untied across all SEDs,
since tying these parameters together in a joint fit caused
$\chi^2/dof$ to increase by over 70/6.  
Without reddening, the fit was still poor, with $\chi^2/dof=77.24/23$.

Allowing for reddening improved the fit considerably, to
$\chi^2/dof=13.56/17$.  The data-model residuals for both cases are
plotted in Fig.~\ref{fig:uvotsed}, and best-fitting parameters for the
reddened case are displayed in Table~\ref{tab:red_diskpbb}.  Compared
to the unreddened case, most of the improvement is in the W1 band bin.
The W1 bandpass's effective wavelength is 2630 $\AA$, which for $\jofour$, probes
roughly 2060 $\AA$ in the rest frame.
This wavelength is very close to 2175~$\AA$, the wavelength at which
there is a strong absorption bump associated with absorption in many
spiral galaxies, including the Milky Way, for instance as seen in
$A_{\lambda}(\lambda)$ plots of Galactic absorption
\citep[e.g.,][]{Cardelli89}.  That is, the \textsc{zredden} component
is impacting the W1 band significantly in this manner, and such
reddening is required for each UVOT spectrum: the resulting increase
in $\chi^2/dof$ when $E(B-V)$ is frozen at zero for any one spectrum
(when reddening is turned off) is always $\geq5/1$, with a
corresponding increase in the Akaike information criterion value
(${\Delta}$AIC) of at least +7.3.  We emphasize that the best-fit
values of $E(B-V)$ are always non-zero, but always less than 0.11, 
and that they do not increase as
optical/UV flux -- nor X-ray flux -- decreases. These observations suggest that the
decrease in observed optical/UV flux from Sw3--4 to Sw6--7 is not
attributable to an increase in reddening.

Assuming that the dust/gas ratio in the host galaxy of $\jofour$ is
identical to that for the Galaxy, $N_{\rm H} = A_{\rm V} \times 2.69
\times 10^{21}$ cm$^{-2}$ mag$^{-1}$ \citep{Nowak12}, and assuming
$R_{\rm V} \equiv A_{\rm V}/E(B-V) = 3.1$, then the average value of
$E(B-V)$ from the \textsc{diskpbb} fits, 0.08, corresponds to a column
of $6.7\times10^{20}$~cm$^{-2}$. This value is typical for
line-of-sight galactic absorption and smaller than the upper limits
from fitting a full-covering column to the \textit{XMM-Newton}
spectra.

\renewcommand{\arraystretch}{1.18}
\begin{table*}[h]
\caption[]{Reddened blackbody fit to \textit{Swift} UVOT spectra in Sw3--8}
        \centering
\label{tab:red_diskpbb}
        \begin{tabular}{lccc} \hline\hline
     &  $k_{\rm B}T_{\rm in}$  &  Unabs. BB flux                          & $E(B-V)$  \\
Obs. &  (eV)                   &  (2.0--7.5~eV, erg cm$^{-2}$ s$^{-1}$)   &  (mag.)  \\ \hline
Sw3  &  $2.62^{+0.22}_{-0.21}$ & $3.31^{+0.40}_{-0.38}\times10^{-11}$ &   $0.106^{+0.020}_{-0.022}$  \\
Sw4  &  $2.34^{+0.23}_{-0.17}$ & $2.88^{+0.51}_{-0.34}\times10^{-11}$ &   $0.085^{+0.030}_{-0.025}$ \\
Sw5  &  $2.07\pm0.16$          &  $2.24^{+0.41}_{-0.33}\times10^{-11}$ &   $0.080\pm0.032$  \\
Sw6  &  $1.92^{+0.21}_{-0.14}$ &  $2.09^{+0.56}_{-0.40}\times10^{-11}$ &    $0.056^{+0.043}_{-0.041}$  \\
Sw7  &  $1.98\pm0.13$          &  $2.00^{+0.37}_{-0.30}\times10^{-11}$ &   $0.059\pm0.032$\\
Sw8  &  $2.51^{+0.25}_{-0.18}$ &  $3.33^{+0.45}_{-0.54}\times10^{-11}$ &   $0.094^{+0.030}_{-0.028}$  \\
All:  & \multicolumn{3}{l}{$p > 1.1$ (pegged at 1.5)} \\
\multicolumn{4}{l}{$\chi^2/dof = 13.92/17$} \\ \hline
\hline \end{tabular}
        \tablefoot{The model used was \textsc{redden(zredden(cflux{$\ast$}diskpbb))}.
          $p$ is the \textsc{diskpbb} power-law index describing the radial temperature relation, $T(r)$.   }
\end{table*}

%
%

\section{Fits to optical emission-line spectra}\label{sec:optspec}     

All optical spectra were corrected for Galactic extinction. We used
$E(B-V)=0.009$ from the dust maps of \citet{Schlegel98}, obtained via
the \texttt{sfdmap} module in python.  We assumed $R \equiv A_{\rm
  V}/E(B-V) = 3.1$, and used the reddening curve of \citet{Fitzpatrick99},
via the \texttt{extinction} module in python.  Absolute
flux calibration was not available for the "red" setups for three SALT
spectra, \#5, 6 and 7, as well as for both SALT setups in spectrum
\#10. For spectra \#4, 5, 6, 7, and 9, fluxes were manually adjusted (gray
scaling) to match the H$\beta$--[\ion{O}{iii}] region from the
simultaneous "blue" setups. Both segments for spectrum \#10 were
adjusted to match the H$\beta$--[\ion{O}{iii}] region from spectrum \#9.

The spectra were taken across different instruments, spanning a
range of wavelength resolutions, slit widths, air masses,
and seeing conditions.  However, the integrated line flux of the
narrow [\ion{O}{iii}] emission line is a constant that can be used to
cross-calibrate the spectra \citep{vangroningen92,Fausnaugh17}
to account for differences in instrument resolutions,
seeing conditions, wavelength offsets, etc.

We flux-scaled all spectra using spectrum \#1 (SALT) as a reference spectrum, and 
following Saha et al.\ (submitted; their Sect.~4). To summarize our method,
for each spectrum (in \#2--10), we defined the continuum via bins on
either side of the [\ion{O}{iii}] $\lambda$5007 emission line,
interpolated, subtracted the continuum, and used a Goodman-Weare MCMC
sampler to fit the emission line
to determine wavelength and flux offsets.  Wavelength shifts for the
other spectra were typically $\sim + 1.6$~$\AA$ for the FORS2 spectra,
$\sim$ $-0.6$ $\AA$ for the SAAO spectrum, and $\sim -0.2$ to $-0.6$~$\AA$ for
the other SALT spectra.  Flux scaling values (applied as
gray-shifting) were in the range 0.95--1.71.

Prior to this campaign, the redshift of $\jofour$ was not known.  From
application of our best-fit model to the [\ion{O}{iii}] $\lambda$5007
emission line in spectrum \#1 (see below for fit details), we obtain
$z = 0.276$.

\begin{figure*}[!h]\centering
\includegraphics[width=1.98\columnwidth]{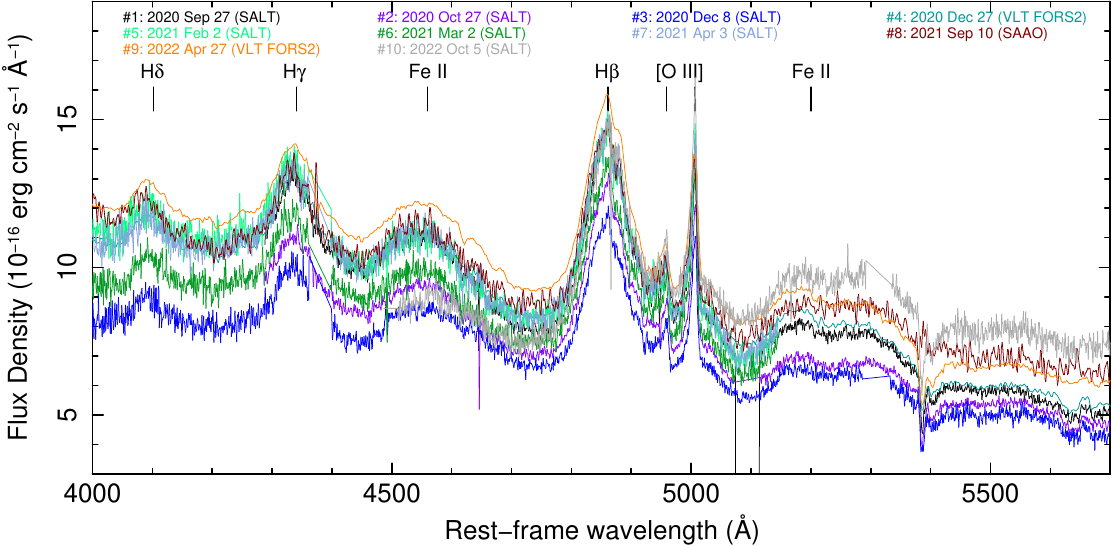}
\caption{Overplots of all ten optical spectra of $\jofour$, with the
most obvious emission features labeled, covering the region from
H$\delta$ through $\sim$5500~$\AA$.  Spectra have been re-scaled by their integrated [\ion{O}{iii}] fluxes,    
as described in the text, but otherwise have not been flux-shifted.}  
\label{fig:optspec_alspec2} 
\end{figure*}
\begin{figure*}[!h]\centering
\includegraphics[width=1.98\columnwidth]{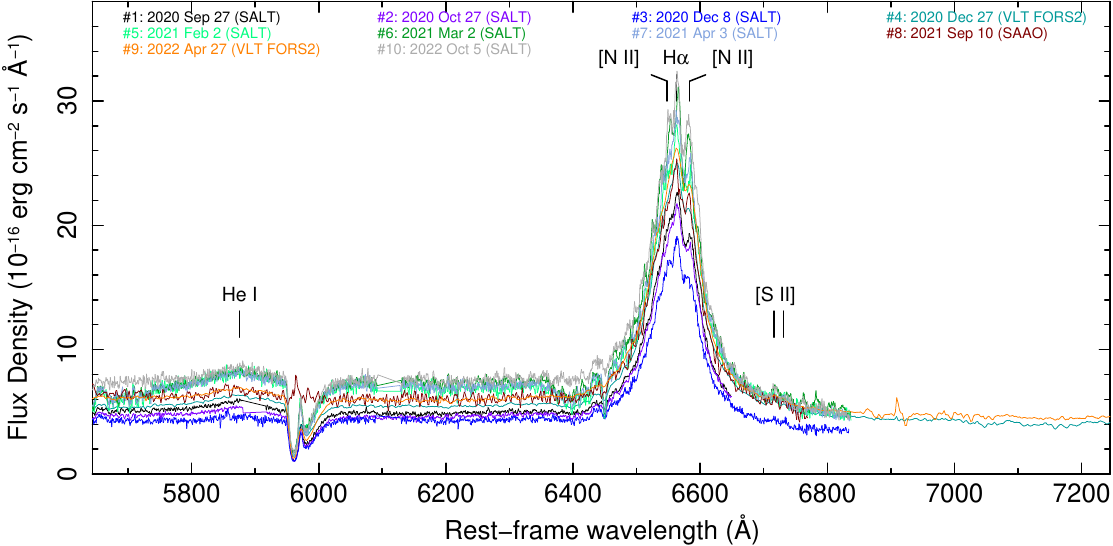}
\caption{Same as Fig.~\ref{fig:optspec_alspec2}, but covering the H$\alpha$ and \ion{He}{i} region.}   
\label{fig:optspec_alspec3}
\end{figure*}

The resulting spectra are presented in Figs.~\ref{fig:optspec_alspec2} and
\ref{fig:optspec_alspec3},
and they are all typical for a Sy~1.  The most obvious features by eye present in all spectra
are broad emission lines associated with H$\alpha$ $\lambda$6563,
H$\beta$ $\lambda$4861, H$\gamma$~$\lambda$4341,
H$\delta$~$\lambda$4102, 
and \ion{He}{i} $\lambda$5876; broad \ion{Fe}{ii} emission peaking near
4550~$\AA$; and narrow [\ion{O}{iii}]$\lambda$$\lambda$5007,4959.

Other narrow emission lines due to
[\ion{N}{ii}]~$\lambda$6584 (the accompanying
6548~$\AA$ line is blended with broad H$\alpha$ and not visually
obvious, but modeled in fits below), and
[\ion{S}{ii}]~$\lambda$$\lambda$6731,6716 are also noticeable in most,
if not all, spectra.
Figs.~\ref{fig:optspec_alspec2} and \ref{fig:optspec_alspec3} 
illustrate that there are no obvious, major changes (for example,
disappearance of broad lines) between the spectra.
The continuum levels across the spectra seem to vary, but such
variability is likely not directly representative of intrinsic
continuum variability. There can be artifacts associated with the
[\ion{O}{iii}] re-scaling process, or variations in seeing and sky
transparency during the observations.

Spectral modeling was done with the python \texttt{lmfit} package \citep{Newville14}.
We masked out regions of atmospheric telluric 
absorption (7590--7700, 7165--7300, and 6800--6900~$\AA$, observed frame).
We performed fitting for H$\beta$ and H$\alpha$ regimes separately.
For H$\beta$ fitting we fitted in the rest-frame window 4000--5200~$\AA$. 
For H$\alpha$, the fitting window extended from 5300~$\AA$ up to
7000~$\AA$ for the two FORS2 spectra (\#4 \& 9), 7000~$\AA$ for the SAAO
spectrum (\#8), 7000~$\AA$ for spectrum \#10, 6800~$\AA$ for spectrum \#3,
6585~$\AA$ (the limit of the data) for spectra \#1 \& 2, and 6566~$\AA$ (the
limit of the data) for spectra \#5, 6, \& 7.

First, we approximated the continuum, fitting by masking out all line
features.
For the H$\beta$ fitting, our continuum included a power-law
continuum; for the H$\alpha$ fitting, we required a broken power-law
continuum to obtain satisfactory fits.
We also included an \ion{Fe}{ii} template, convolved with a Gaussian to
include the effects of velocity broadening.  For the H$\beta$ fitting,
we used the templates of
\citet{Kovacevic10}\footnote{\url{http://servo.aob.rs/FeII_AGN/}},
which provide modeling for the F, S, G, P, and I~Zw~1 groups of
\ion{Fe}{ii} separately in the 4000--5500~$\AA$ range.  The F~group, whose
peaks blend to form a broad emission feature near 4550~$\AA$, was
especially prominent in $\jofour$.  For the H$\alpha$ fitting, we used
the template of \citet{Bruhweiler08}, which is based on 
observations of I~Zw~1.

We also included a host galaxy template from the SWIRE library
\citep{Polletta07}.  There are no host galaxy morphology
classifications available on NED for $\jofour$.  We used an Sb
template, and such a choice is admittedly arbitrary. However, 
our fits were insensitive to the sub-type (Sa--d) because a host
galaxy component was not required in the fits: fits for both the
H$\alpha$ and H$\beta$ regions preferred the power-law component to
strongly dominate the continuum and to fully describe the total
continuum.
Finally, we added Gaussians as needed for the emission lines.
The final model for the H$\beta$ region fitting contains:

\begin{itemize}

  \item A continuum dominated by a power-law with best-fitting values of 
  spectral index $\alpha$ spanning 1.51--1.95
  
  \item Narrow [\ion{O}{iii}] emission lines at both 5007 and 4959~$\AA$ (rest frame).
  Each line profile was modeled with the sum of 
  1) a narrow Gaussian with rest-frame width $\sigma \sim$ 3--5~$\AA$ (values between
  the two lines were kept tied) and energy centroid within 1--2~eV of
  the expected rest frame energy, plus 2)
  a wider Gaussian with rest-frame $\sigma \sim$ 10--12~$\AA$ (again, widths between
  the two lines were kept tied), and energy centroid 9--13~$\AA$ blue
  (5007~$\AA$ line) or 6--9~$\AA$ blue (4959~$\AA$ line) of the expected rest
  frame energy.  
  We tried fitting with the amplitude of the 4959~$\AA$ broad component tied to 1/3
  that of the 5007~$\AA$ component, but residuals became poor, so we had to leave
  the amplitudes untied. In the best fits, the mean and standard deviation of
  the ratio of amplitudes was $0.28\pm0.04$. 
  NLR outflows in AGN, manifested by blue-asymmetric
  [\ion{O}{iii}] profiles, are not uncommon in both 
nearby and high-redshift AGNs \citep[e.g.,][]{Schmidt18,Leung19,Mahoro23}.
  In $\jofour$, the energy centroid offsets correspond to
  line-of-sight velocities on the order of 400--800~km~s$^{-1}$ relative to
  systemic, and the line widths correspond to FWHM velocity
  dispersions of $\sim$600--700~km~s$^{-1}$, similar to typical values
  obtained by \citet{Schmidt18} and \citet{Leung19}.

  \item Broad emission components for H$\beta$, H$\gamma$, and H$\delta$;
  velocity widths were on the order of 4650--5050~km~s$^{-1}$,
  4900--6500~km~s$^{-1}$, and 3600--5200~km~s$^{-1}$, respectively.

  \item Narrow emission components for H$\beta$, H$\gamma$, and
    H$\delta$.  Their velocity widths were tied to that for the
    [\ion{O}{iii}]$\lambda$5007 line. It was necessary to keep their
    energy centroids frozen to expected rest-frame values, since the
    Balmer profiles were dominated by the broad components.

  \item There was no significant emission from broad or narrow \ion{He}{ii} $\lambda$4686 lines.

\end{itemize}


For the H$\alpha$ region fits, the final model for the contains:

\begin{itemize}

   \item A continuum modeled as a broken power law (an unbroken power law did not
   yield satisfactory fits).

   \item Narrow components for H$\alpha$, \ion{He}{i}~$\lambda$5876,
     and [\ion{N}{ii}]~$\lambda$$\lambda$6584,6548, with energy
     centroids frozen at expected rest-frame values.  In addition, we
     modeled [\ion{S}{ii}]~$\lambda$$\lambda$6731,6716 for spectra \#3,
     4, 8, 9, and 10, keeping intensities tied together.  All velocity
     widths were kept frozen at the best-fitting value from
     [\ion{O}{iii}]~$\lambda$5007 from the corresponding H$\beta$ fit.
   We also included narrow emission at rest-frame 6300~$\AA$ due to [\ion{O}{i}].

   \item Broad H$\alpha$ emission, with $\sigma$ typically 30--34~$\AA$ (FWHM
   velocity width of 3200--3650~km~s$^{-1}$.

   \item In addition to this broad component, all fits required an
   additional "ultra-broad" component to properly model the lower
   $\sim$1/3 of the H$\alpha$ profile.  Its average amplitude
   was 0.65 times that of the broad H$\alpha$ component; $\sigma$ was
   typically 46--100~$\AA$, corresponding to FWHM velocity widths
   4950--10500~km~s$^{-1}$.  Omission of this profile component caused
   $\chi^2$ to increase by typically 40--50 and the 
   AIC value to increase by over 600--700, and also
   produced very poor data-model residuals in all fits.  (As an aside,
   we subsequently re-fit the
   H$\beta$ profile to test for presence of this "ultra-broad"
   component in H$\beta$, but it was not required in those fits.
   Forcing an "ultra-broad" component with amplitude and $\sigma$ set
   to values 0.65 and 2.5 times those of the broad component caused
   fits to worse considerably, with $\chi^2$ typically increasing by
   +4 and AIC increasing by up to 100.)

   \item Broad emission from \ion{He}{i}, with $\sigma$ typically $\sim$ 30--45~$\AA$.

   \item Two absorption-like features modeled with Gaussians.  One was
     centered near 6400~$\AA$ (rest frame) = 8166~$\AA$ (observed
     frame), with an average observed-frame width of $\sigma= 21$~$\AA$.
     The other was centered at 6450~$\AA$ (rest frame) = 8224~$\AA$
     (observed frame), and was narrower, with $\sigma = 3$~$\AA$
     (observed frame).  These features were strongest in spectra
     \#5--7, moderately strong in \#1--4, and weak/absent in
     \#8--10. Their strengths correlate well with the strengths of
     atmospheric absorption features masked out above and we conclude
     they are also atmospheric in nature.  

\end{itemize}

\begin{figure*}[!h]\centering
\includegraphics[width=1.98\columnwidth]{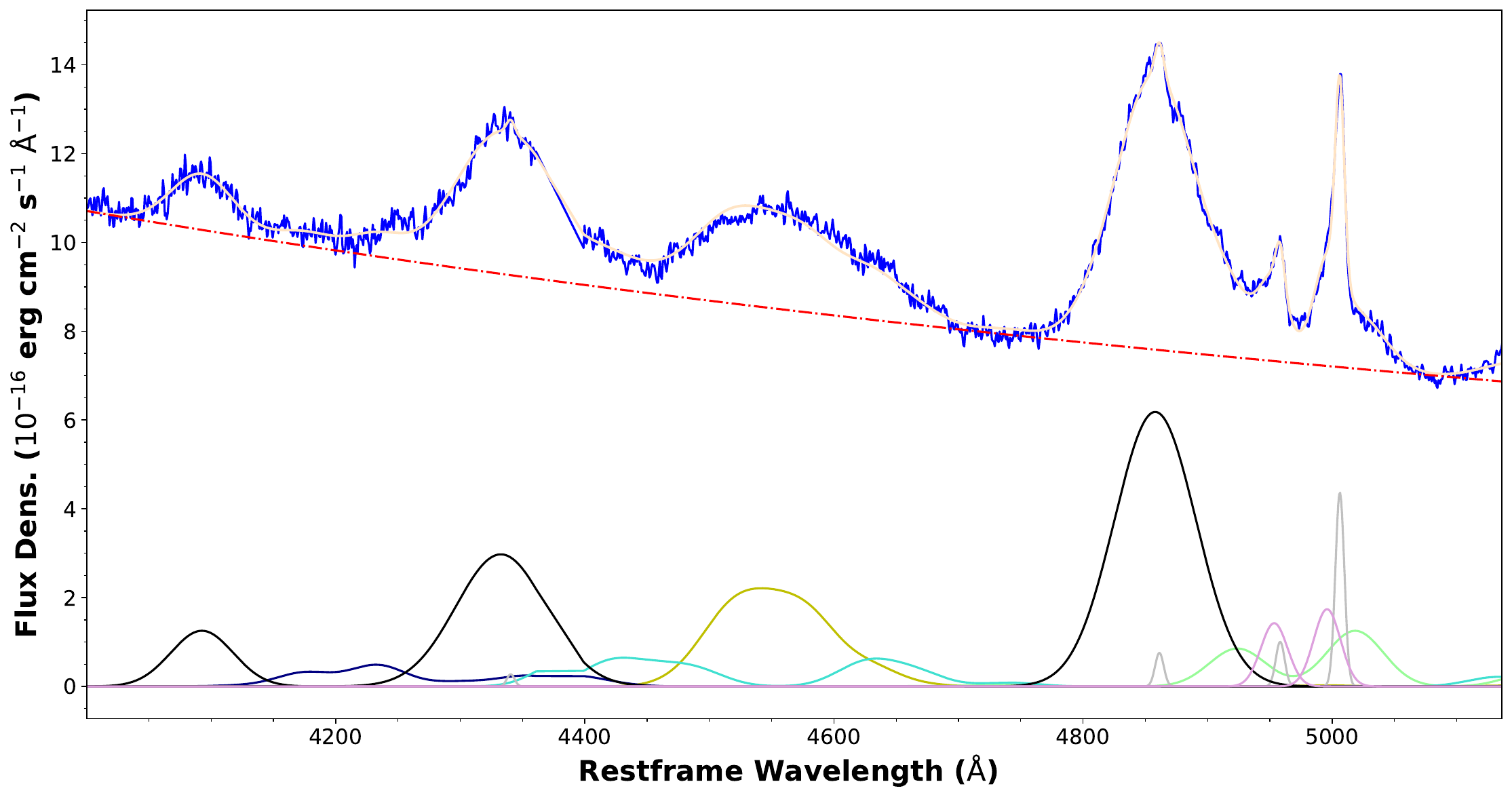}
\caption{Sample 
spectral deconvolution for the H$\beta$ fitting; shown here
is spectrum \#2.  The data are plotted in blue.  The red-dashed line
is the power-law continuum.  Black denotes broad Balmer components;
gray denotes narrow [\ion{O}{iii}] and Balmer components; pink denotes
the broad and blueshifted [\ion{O}{iii}] components; and navy blue,
yellow, turquoise, and green denote the P-, F-, I~Zw~1-, and S-groups
of \ion{Fe}{ii} emission.  The total model is shown in beige.  A host
galaxy template was included in the fit, but the best-fit value of its
amplitude is zero, as the AGN continuum dominates.}
\label{fig:optspec_fitspecHbeta}
\end{figure*}

\begin{figure*}[!h]\centering
\includegraphics[width=1.98\columnwidth]{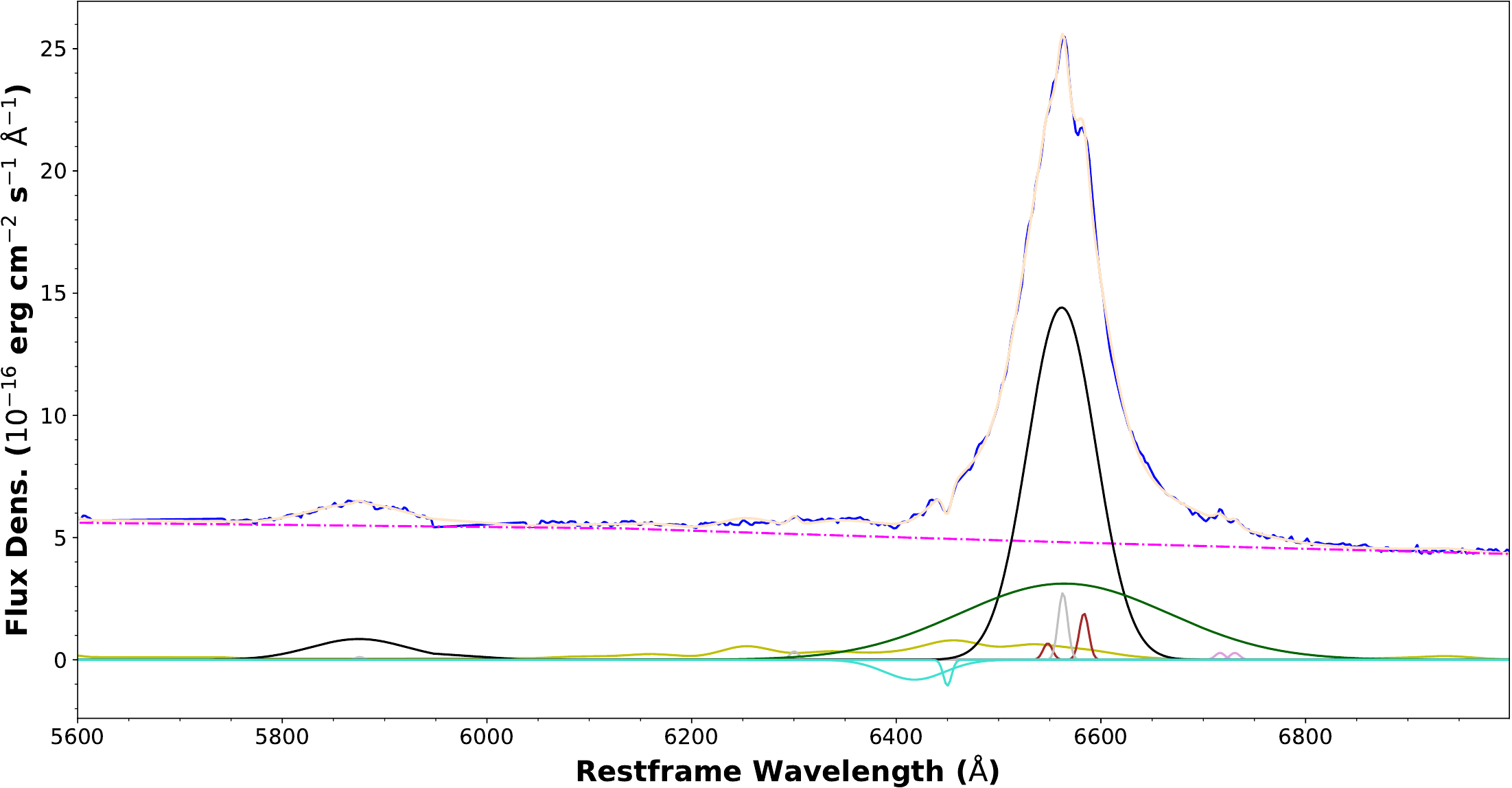}
\caption{Sample spectral deconvolution for the H$\alpha$ fitting; shown here
is spectrum \#4.  The data are plotted in blue.  The magenta-dashed
line is the broken power-law continuum.  Black denotes the broad
components of H$\alpha$ and \ion{He}{i}; gray denotes the narrow
components of H$\alpha$, \ion{He}{i}, and [\ion{O}{i}]$\lambda$6300;
dark green denotes the ``ultra-broad'' component of H$\alpha$; brown denotes       
the [\ion{N}{ii}]~$\lambda$$\lambda$6584,6548 emission lines; pink
denotes the [\ion{S}{ii}]~$\lambda$$\lambda$6731,6716 emission lines;
and yellow denotes \ion{Fe}{ii} emission.  Turquoise denotes the two
atmospheric absorption features near 6400 and 6450 $\AA$.
The total model is shown in
beige.  A host galaxy template was included in the fit, but the
best-fit value of its amplitude is zero.}
\label{fig:optspec_fitspecHalpha}
\end{figure*}

Gaussian components are adequate to fit all broad
and narrow emission lines. There is no strong evidence for asymmetry
in any of the broad Balmer components.
For brevity, all fit results are presented in Appendix~\ref{sec:appdxopt}, with
final best-fitting parameters listed in Tables~\ref{tab:fitspec_Hbeta4} and
\ref{tab:fitspec_Halpha1} for H$\beta$ and H$\alpha$, respectively.  Two typical
spectral decompositions are plotted in Figs.~\ref{fig:optspec_fitspecHbeta} and
\ref{fig:optspec_fitspecHalpha}, for H$\beta$ and H$\alpha$, respectively.


One main takeaway from the spectral fits is that there are no major
changes in continuum shape or in broad line profiles or heights across
spectra.
If either of the X-ray obscurers contained substantial dust,
and if they occulted substantial fractions of the BLR in
addition to occulting the X-ray corona, then we might expect the
H$\beta$ and H$\alpha$ broad component fluxes to drop appreciably due to the extinction,
accompanied by an increase in the observed ratio of H$\alpha$/H$\beta$
broad component fluxes, in spectra \#1--7 (event 1) and/or \#9--10
(event 2) compared to spectrum \#8 (unobscured).
However, the average H$\beta$ integrated broad component fluxes for
spectra \#1--7, \#8, and \#9--10 are $4.44\pm0.06 \times 10^{-12}$,
$4.13\pm0.06 \times 10^{-12}$, and $4.36\pm0.06 \times 10^{-12}$
ph~cm$^{-2}$ s$^{-1}$, respectively.
Similarly, the ratio of H$\alpha$/H$\beta$ broad component fluxes
remains roughly constant: Given the necessity of the "ultra-broad"
component to the H$\alpha$ profile, we consider the ratio of the maximum
heights\footnote{For amplitude, $A,$ in a Gaussian, the maximum height
is defined as $A$/($\sigma\sqrt{2\pi}$).}  of the wide components.
The average H$\alpha$/H$\beta$ ratio across spectra \#1--7 (event 1),
spectrum \#8 (unobscured), and spectra \#9--10 (event 2) are
$2.25\pm0.27$, $2.22\pm0.04$, and $2.67\pm0.03$, respectively.

We conclude that: 1) neither X-ray-obscuring cloud contains substantial dust and we can rule
out values of $E(B-V)$ above $\sim$0.1--2 (corresponding to values of
$A_{\rm V}\ \sim$ 0.3--0.7) in both clouds or 2) neither cloud significantly
occulted the line of sight to the BLR.

Furthermore, spectrum \#5 coincides with the dip in optical
photometric flux near MJD $\sim$ 59210--59250. However, there is no
change in the broad Balmer component fluxes or ratio here either.  We
conclude that the optical continuum dip is a luminosity variation
intrinsic to the disk, and not a product of dust obscuration within
cloud \#1.

%
%
\section{Discussion}  \label{sec:Discussion}

In this paper, we explore the first major obscuration events
detected with eROSITA and studied with triggered multiwavelength
followups.  We have tracked the durations and properties of the
obscurers through a series of X-ray flux and spectral monitoring
observations.  We summarize our main observations:

\begin{itemize}

\item eROSITA's all-sky X-ray surveys revealed in August 2020 (eR2)
  that $\jofour$'s 0.5--2.0~keV X-ray flux had dropped by a factor of
  $\sim$11 compared to February 2020 (eR1).  Follow-up
  \textit{XMM-Newton} EPIC spectra (XM1 and XM2) revealed obscuration
  by a Compton-thin cloud with $N_{\rm H,PC} \sim 1 
  \times10^{22}$~cm$^{-2}$ and covering fraction $CF \sim$ 60~percent
  (using values from XM1).  The soft X-ray flux remained low through
  eR3 (February 2021), though it had recovered by Sw3
  (June 2021).  Assuming that eR1 represented an
  unobscured state, this obscuration event lasted somewhere between
  309 and 539 days (observed frame).

\item By August 2021 (eR4), $\jofour$ had returned to an unobscured
  state, in which it remained through at least September 2021 (XM3).  It
  displayed relatively higher periods of soft X-ray flux ($F_{0.5-2}  
  \sim$ (3.0--4.3) $\times10^{-13}$~\ecgs) during this state, and the
  \textit{XMM-Newton} EPIC spectrum taken during XM3 was very typical
  for unobscured Seyferts.

\item By February 2022 (eR5), the soft X-ray flux had fallen back to
  $F_{0.5-2} = 3.4^{+1.1}_{-1.0} \times10^{-14}$~\ecgs, and remained
  near this level through at least February 2023 (Sw4, XM4, Sw5--8).
  As revealed by the \textit{XMM-Newton} EPIC spectrum taken
  during XM4, the source was partially obscured by a new transiting
  cloud, with $N_{\rm H,PC} \sim$ 2.8 $\times
  10^{23}$~cm$^{-2}$ and $CF \sim 79$~percent.  The duration of
  obscuration is a minimum of 527 days in the observed frame; 
  the eclipse had not ended as of the most recent
  \textit{Swift} observation (Sw8, in February 2023).

\item Optical and UV continuum photometry obtained via monitoring with
  ground-based telescopes, \textit{Swift} UVOT and \text{XMM-Newton}
  OM revealed minimal variations in optical/UV accretion disk luminosity, with
  long-term variations being less than $\sim$ 0.3 mag in both
  the B and M2 bands.  An exception is a temporary drop in B-band flux
  by $\sim$0.4 mag over 40 days in early 2021.

\item We obtained a series of optical spectra of $\jofour$ at SALT,
  VLT, and SAAO to establish the source's redshift and to
  monitor Balmer profiles.  We deconvolved the H$\alpha$, H$\beta$,
  and H$\gamma$ profiles into narrow and broad components, with the
  broad H$\beta$ component correspond to an average FWHM velocity of
  4870 km s$^{-1}$. We observed no significant changes in profile or
  intensity, and no significant changes in Balmer decrement.  In
  addition, the [\ion{O}{iii}] $\lambda$5007 and $\lambda$4959 emission
  lines were each modeled by a narrow component (FWHM velocity width
  $\sim$ 400--700~km~s$^{-1}$) and a blue-skewed broad component
  (offset by $\sim$ 400--800~km~s$^{-1}$), consistent with a bulk 
  outflow.

\end{itemize}

These observations argue strongly against the soft X-ray dimming being
due to temporary drops in accretion luminosity. 
We observed no major variations in the
Balmer intensities or profiles, suggesting that the $>$13.6~eV
continuum held steady.  The fact that we did not observe any major
decrease in the optical/UV thermal emission emanating from the inner
accretion disk ($<$ 18 lt-dy, or $<$ 500$R_{\rm g}$, as
estimated in Sect.~\ref{sec:discsub2}) strongly argues against any major
decreases in accretion rate or in disk temperature, for instance, due
to propagating cold fronts \citep{Ross18}.  Importantly, the minor
variations in optical/UV flux we did observe were not correlated with
the onset of X-ray obscuration.  Finally, we note the improvement in
X-ray spectral fits to XM1 and XM4 when obscuration components were
added.

As noted above, the high soft X-ray flux states, probed by eR1, eR4,
and XM3, support the absence of eclipsing by clouds along the line of
sight.  The limited data quantity means we cannot distinguish between
a scenario wherein obscuration in $\jofour$  
is due to discrete X-ray-obscuring clouds versus a scenario wherein
obscuration is caused by a contiguous but
patchy structure containing some ``holes.''  
In this scenario, the ratio of obscured/unobscured lines of sight can
be estimated very crudely via the fraction of time having observed the
source in obscured/unobscured states. During X-ray monitoring spanning
1090 days, $\jofour$ spent at least 846 days, or 78 percent, in an
obscured state.  However, we must note the limited number of observed
transitions between obscured and unobscured states during our
campaign, so this value is to be treated with extreme caution pending
significant amounts of additional long-term monitoring.

In addition, the limited data quantity, particularly with regard to
obtaining a large number of high-quality spectra, means we are not
highly sensitive to tracking the transverse column density profile of
the obscuring matter, as has been done by, for example, \citet{Maiolino10},
\citet{Rivers11,Rivers15}, and \citet{MKN14}.
For the purposes of this Discussion, we assume for simplicity that the
two obscuration events correspond to two separate X-ray-obscuring clouds
transiting the line of sight, and henceforth referred to as ``cloud 1''
and ``cloud 2.''

\subsection{Basic system parameters}  \label{sec:discsub2}

We first obtain virial estimates for the radial location of the BLR
and for the black hole mass, $M_{\rm BH}$, using the average value of
H$\beta$ FWHM velocity across all spectral fits, 4870 km s$^{-1}$ (rest frame),
and the empirical relation
between optical luminosity and BLR radius $R_{\rm BLR}$ in Seyferts
\citep{Bentz09,Bentz13}.  From the
optical spectral fits, we find an average flux density of
${\lambda}F_{5100\AA}$ to be $7.2\times10^{-16}$ \ecgsA, which for a
luminosity distance of 1400~Mpc, corresponds to a monochromatic
luminosity of ${\lambda}L_{5100\AA} = 1.1\times10^{45}$ erg~s$^{-1}$.
We apply the ${\lambda}L_{5100\AA}-R_{\rm BLR}$ relation of \citet{Bentz13},
log($R_{\rm BLR}$, lt-dy) = $K$ + 
${\alpha}$log(${\lambda}L_{5100\AA}$/ ($10^{44}$~\ecgs)), with their best-fitting
values of $K = 1.56$ and $\alpha=0.55$.  We obtain $R_{\rm BLR} =
2.8\times10^{17}$cm = 130 lt-days.  We do not have information on the inclination
of the system, so we assume an arbitrary virial factor $f$ of
1.0; we obtain $M_{\rm BH} = f R_{\rm BLR} v_{\rm FWHM}^2 G^{-1} =
6.2\times10^8
\Msun$. The corresponding Eddington luminosity is thus $L_{\rm Edd} =
7.8\times10^{46}$ erg~s$^{-1}$.  Considering the scatter in the
${\lambda}L_{5100\AA}-R_{\rm BLR}$ relation of \citet{Bentz13},
we assign an uncertainty of $\sim0.13$ dex to $R_{\rm BLR}$,
$M_{\rm BH}$, and $L_{\rm Edd}$.

We now estimate the bolometric luminosity, $L_{\rm Bol}$, via two methods:
First, we estimate $L_{\rm Bol}$ from the optical luminosity,
using the bolometric correction of \citet{Duras20}.  From the
\textit{Swift} UVOT observations, the observed B-band flux density is
typically $1.8-2.1 \times 10^{-15}$ \ecgsA.  These values include
contamination from the host galaxy, but the spectral decomposition of the optical
spectra suggest that by 4400\AA, such contamination is negligible, likely less than
5~percent. The corresponding monochromatic luminosity,
$\lambda$$L_{\rm 4400\AA}$, is $2\times10^{45}$~\ecgs.  From Eq.~4.2 of
\cite{Duras20}, the correction factor is $5$, and we
obtain $L_{\rm Bol} = 1 \times10^{46}$~\ecgs, and \lboledd = 0.13.
Given uncertainties on flux density, correction factor, and on $L_{\rm Edd}$ of
0.013 dex, 0.27 dex, and 0.13 dex, respectively,
we assign a total uncertainty of 0.41 dex to this estimate of $L_{\rm Bol}$.

Finally, we estimate $L_{\rm Bol}$ via a combination of X-ray
luminosity and {\aox}, the optical-to-X-ray spectral index: \aox =
$-$log($L_{\rm 2 keV}$/$L_{2500\AA}$)/2.605 \citep{Tananbaum79}.
We use Eq.~4.4 of \citet{Lusso10},
log($L_{\rm Bol}/L_{2-10}$) = 1.561 $-$ 1.853{\aox} + 1.226{\aox}$^2$. 
From XM3, the unabsorbed 2--10~keV flux of $5.2\times
10^{-13}$~\ecgs corresponds to a 2--10 keV luminosity
$L_{2-10} = 1.3\times10^{44}$~erg~cm$^{-2}$.
$L_{\rm 2 keV}$ and $L_{2500\AA}$ are the monochromatic rest-frame luminosities
at 2~keV and 2500\AA, respectively.  From the best fitting   
M1 model fit to XM3, the 2~keV flux density is 0.061$\pm$0.002~$\mu$Jy.
For the flux density at 2500\AA, we interpolate from the average of
the W1-band \textit{Swift} UVOT and the UVOT and \textit{XMM-Newton}
OM M2-band flux densities. We obtain a 2500~$\AA$ flux density of  
1.35$\pm$0.05~mJy, and a subsequent value of \aox=$1.67\pm0.08$.
Our best estimate of $L_{\rm Bol}$ is thus 
$1 \times10^{46}$~\ecgs, and \lboledd = 0.13, in excellent agreement with 
the value estimated from optical luminosity above.
Given uncertainties on correction factor based on \aox (0.18 dex), the
scatter in the $L_{\rm Bol} - L_{2-10}$ relation of \citet{Lusso10}
(0.08 dex), and uncertainty on $L_{\rm Edd}$ (0.13 dex), we assign a
total uncertainty of 0.39 dex to this estimate of $L_{\rm Bol}$.

Using this estimate of $L_{\rm Bol}$, we estimate the radial location
of the dust sublimation region, assuming isotropic luminosity.  We
follow Sect.~2.1 of \citet[][also discussed in \citealt{Barvainis87}]{Nenkova08}:
$R_{\rm dust}$ = 0.4 ($T_{\rm sub}$/(1500~K))$^{-2.6}$ ($L_{\rm
Bol}$/$10^{45}$~egs)$^{0.5}$.  For a sublimation temperature $T_{\rm
sub}$ of 1500~K, our estimate of $R_{\rm dust}$ is 1.3~pc
(1500~lt-day).  However, as discussed in \citet{Nenkova08}, the
transition between dusty and non-dusty regions is likely gradual, as
relatively larger grains are likely able to survive at radii factors
of up to $\sim$2--3 smaller than predicted from this equation. In
addition, different dust species sublimate at different temperatures,
e.g., sublimation of graphite grains and silicate grains at $T=1500$
and $T=1000$~K, respectively \citep{Schartmann05}. We therefore treat
this estimate of $R_{\rm dust}$ as an estimate of the  
dust sublimation radial "zone."

Finally, we estimate the radial location of the optical and UV thermal
emission in the inner accretion disk.  We follow \citet[][their
Eq.~1]{Edelson19} for a standard optically thick, geometrically thin
disk.  
\footnote{We note that the origin of the optical/UV continuum lags in Seyferts
is not fully settled at this time; an alternate explanation is the
optical/UV lag indicate diffuse emission from the BLR \citep{Netzer22}
instead of the inner accretion disk.  However, we note that even if
radii of peak optical/UV emission in the disk are factors of a few
smaller than estimated here, our conclusions are not significantly
impacted.}
We set the multiplicative scaling factor in that equation to $X
= 2.49$ for the flux-weighted radius.  For the M2 band (2246~$\AA$
observed frame $\sim$ 1760 $\AA$ rest frame), peak emission is at 
$4.6^{+2.9}_{-1.8}$ lt-dy, which corresponds to $130^{+90}_{-50}$ gravitational radii 
$R_{\rm g}$ for a $6.2\times10^8\Msun$ black hole\footnote{$ 1 R_{\rm
g} \equiv GM_{\rm BH}c^{-2}$}.  For the B band (4400 $\AA$ observed
frame = 3448 $\AA$ rest frame), we obtain radii of $11.3^{+6.9}_{-4.4}$ lt-dy, or
$310^{+190}_{-120}~R_{\rm g}$).

\subsection{Constraints on clouds' locations and sizes}  \label{sec:radloc}

 We can use constraints on ionization parameter, eclipse duration, and
column density to derive the distance $R_{\rm cl}$ from the X-ray
continuum source to each cloud, assuming for simplicity
that clouds' kinetics are dominated by Keplerian
motion. We follow \citet[][their Sect.~4]{Lamer03}: $R_{\rm cl} =
4\times10^{16}$~$M_7^{1/5}$~$L_{\rm ion,42}^{2/5}$~$t_{\rm
d}^{2/5}$~$N_{H,22}^{-2/5}$~$\xi^{-2/5}$~cm. Here, $M_7$ is the black hole
mass in units of $10^7\Msun$, $t_{\rm d}$ is the eclipse duration in
units of days, $N_{H,22}$ = $N_{\rm H}$/(10$^{22}$~cm$^{-2}$), and
$L_{\rm ion,42}$ is the ionizing ($>$13.6~eV) continuum luminosity in
units of $10^{42}$ \ecgs. To estimate $L_{\rm ion}$, we follow
\citet{VF09} and adopt $L_{\rm ion} = 0.2 \times L_{\rm
Bol} = 2\times10^{45}$~\ecgs.
The ionization parameter $\xi$ is defined as $L_{\rm
ion}/(n_{\rm e}r^2)$, where $n_{\rm e}$ is the gas electron density.
To estimate $\xi$, we revisited the best-fit ``PC*(SXCOM+HXPL)+UXCL''
models for XM1 and XM4, and replaced the partial-covering neutral
obscuring component with an ionized one, modeled with \textsc{zxipcf}
in \texttt{Xspec}. We obtain upper limits to log($\xi$,
erg~cm~s$^{-1}$) of +1.5 for cloud~1 and +2.5 for cloud~2.
For cloud 1, the eclipse duration is constrained to be in the range
309--539~days observed frame, or 242--422~days rest frame.  
The upper limit on $\xi$ translates into a lower limit on cloud location: 
$R_{\rm cl,1} > 3.0\times10^{18}$~cm =
1100~lt-dy, which would place it at radii well outside the H$\beta$-emitting
region of the BLR, and approaching the dust sublimation zone.
For cloud 2, which was still ongoing as of the most recent
observations, the eclipse duration is a minimum of 527~days
observed frame = 413~days rest frame.  For this
cloud, we obtain $R_{\rm cl,2} > 3.3\times10^{17}$~cm = 130~lt-dy. This
lower limit on radial location is commensurate with
the H$\beta$ region of the BLR. 
We emphasize the notion of "commensurate," as X-ray absorbers lie along the line of
sight, but BLR clouds likely contain components out of the line of
sight.

Still working with the simplifying assumption that the clouds are in strictly
Keplerian motion, we can derive upper limits on their transverse
velocity $v_{\rm cl} = ( G M_{\rm BH} R_{\rm cl}^{-1})^{1/2}$.  For
clouds 1 and 2, respectively, we obtain $v_{\rm cl,1} < 1700$
km~s$^{-1}$ and $v_{\rm cl,2} < 5000$ km~s$^{-1}$.
The clouds' sizes in the transverse direction, $D_{\rm cl}$ can be
estimated via $D_{\rm cl} = v_{\rm cl} t_{\rm D}$.  For cloud 1,
adopting $t_{\rm D} = 400$~days, $D_{\rm cl,1} < 6 \times10^{15}$~cm
$\sim$ 2 lt-dy $\sim$ 65~$R_{\rm g}$.
For cloud 2, we have a situation where transverse velocity is an upper
limit and duration is a lower limit.  Nonetheless, adopting $t_{\rm D}
\sim 400$~days, we obtain 
$D_{\rm cl,2} \sim 2 \times 10^{16}$~cm $\sim$ 6.5~lt-dy $\sim 180$~$R_{\rm g}$. 
Such inferred diameters are typically an order of magnitude more than those
found by \citet{MKN14}, and two orders of magnitude greater
than clouds inferred to exist in the BLR of NGC~1365 \citep{Maiolino10}.

We can estimate crude lower limits to the cloud density $n_{\rm H}$,
via $N_{\rm H}/D_{\rm cl}$. For cloud 1, we obtain $n_{\rm H,1} >
2\times10^6$~cm$^{-3}$.  For cloud 2, we obtain $n_{\rm H,2} >
2\times10^7$~cm$^{-3}$.
Such density limits are consistent with typical values for the BLR.

As a caveat, we recall that if the clouds' true motion is
strongly non-Keplerian (for instance, if the velocities contain a
strong radial component, such as that associated with a wind flowing
vertically upward from a disk), then the accuracy of these estimates
of $R_{\rm cl}$, $D_{\rm cl}$, and $n_{\rm H}$ can be compromised.

Finally, we can consider the results in the context of the
\textsc{clumpy} torus model of \citet{Nenkova08}, wherein all
obscuration is accounted for by a large number (on the order of $10^{4-5}$)
discrete clouds, and obscuration for any given source is
a probability depending on source inclination and cloud distribution
parameters.  Clouds are distributed preferentially toward the
equatorial plane, but with an angular distribution that is Gaussian.
The average number of clouds along our line of sight to the central
source is {$\mathcal{N_{\rm i}}$}($\sigma$, $i$, {$\mathcal{N_{\rm
0}}$}) = {$\mathcal{N_{\rm 0}}$}exp($-( (90-i)/\sigma)^2$). Here,
{$\mathcal{N_{\rm 0}}$} denotes the average number of clouds along an
equatorial ray, typically on the order of 5--15; $\sigma$ denotes the angular
width of the distribution.  The inclination of the system is denoted by $i$: $0^{\circ}$
for face-on, and $90^{\circ}$ for edge-on.  The likelihood to observe
a source in an obscured state is dependent on {$\mathcal{N_{\rm i}}$}
and $i$, \citep[e.g.,][]{Nikutta09} $P_{\rm obsc} = 1 -
$exp($-{\mathcal{N_{\rm i}}}$).  The \textsc{clumpy} model was originally used
to model the distribution of clouds residing in the dusty torus, that is,
outside the dust sublimation radius, but for the purpose of this
discussion, we can consider that the radial distribution of clouds may
extend to inside the dust sublimation radius.

In the case of $\jofour$, we have observed transitions between zero
and one cloud along the line of sight, such that ${\mathcal{N_{\rm
i}}}$ is always either 0 or 1.  If we assume for academic purposes
that ${\mathcal{N_{\rm 0}}}$ is, say, 5 (10), then having
${\mathcal{N_{\rm i}}} = 1$ means that $((90-i)/\sigma)^2$ must be
1.26 (1.52).  Though we lack information on the inclination of
$\jofour$, we can assume for simplicity that the system is relatively
face-on, with an arbitrarily chosen inclination angle of
$30^{\circ}$. In this case, the above value of ${\mathcal{N_{\rm 0}}}
= 5$ (10) yields a value of $\sigma = 48^{\circ}$ ($39^{\circ}$).
Such values are roughly consistent with values for the tori of type~1 Seyferts
obtained by \citet{RamosAlmeida11} in their SED model fits.

\subsection{Exploring whether the clouds dusty or non-dusty}

We did not observe drastic changes in optical or UV continuum
luminosity between the X-ray obscured and unobscured periods.  The
optical/UV thermal emission emanates from a relatively compact region,  
roughly $5 - 11$ lt-dy, as discussed in Sect.~\ref{sec:discsub2}. The
cloud sizes are crudely estimated to be on the order of $\lesssim$2--7 lt-day
(Sect.~\ref{sec:radloc}).   
If it were the case that the clouds somehow
partially obscured the X-ray-emitting region while simultaneously
covering all or most of the optical/UV continuum-emitting region of
the disk, then we might expect to observe evidence for extinction,
particularly in the UV band, if the clouds contained significant
amounts of embedded dust.

As a reminder, our best-fit values of column density were $N_{\rm
H,PC}$ = 1.0$\times10^{22}$~cm$^{-2}$ (cloud~1) and 2.8$\times10^{23}$~cm$^{-2}$
(cloud 2). To estimate the corresponding dust content, we assume a
dust/gas ratio equivalent to the Galactic value, $N_{\rm H} = A_{\rm
V} \times 2.69 \times 10^{21}$ cm$^{-2}$ mag$^{-1}$ \citep{Nowak12}. We
obtain $A_{\rm V}$ $\sim$ 3.7 mag (cloud~1) and 105 mag (cloud~2).
Following \citet{Cardelli89}, we can translate these V-band extinctions into
estimates of extinction in each of the M2 and B bands; we use
$A_{\rm M2} \sim 2.7 A_{\rm V}$ and $A_{\rm B} \sim 1.3 A_{\rm
V}$, respectively.  Subsequently, for cloud 1, we expect extinction levels of 
$A_{\rm M2} \sim$ 10 mag and $A_{\rm B} \sim$ 4.8 mag. For cloud 2, we expect  
$A_{\rm M2} \sim$ 280~mag and $A_{\rm B} \sim$ 135~mag.

However, we do not observe such strong variations in the light curves
of these bands, with conservative upper limits on observed variations
of 0.4 mag for the M2 band and 0.6 mag for B band, consistent with
standard variability in persistently-accreting Seyferts.

We conclude that if either cloud fully covers the optical/UV-emitting
regions of the disk, they must contain negligible dust: we estimate a
limit of $A_{\rm V} <$ $\sim$0.15~mag, assuming $R_{\rm V} \equiv
A_{\rm V}/E(B-V)=3.1$, $E(B-V) <$ $\sim$0.04~mag.  Similarly, the
optical-UV continuum SED fitting (Sect.~\ref{sec:OUVSED}) indicated
upper limits to $E(B-V)$ of roughly 0.1 mag.  On the basis of the
stability of the H$\beta$ broad component flux and the ratio of broad
H$\alpha$/H$\beta$ emission, if the X-ray-obscuring gas covered
substantial fractions of the BLR, then they must be dust-free, too.
In addition, the extreme tension between the lack of observed
extinction and the high expected values of extinction argue against a
scenario wherein the clouds are dusty and they partially cover the
optical/UV-emitting annular region of the disk with the same covering
fractions impacting the X-ray band.

However, an alternative and simple scenario is based on geometry: the
clouds may partially intersect the line of sight to the X-ray corona,
but they do not intersect the bulk of the lines of sight to the
optical/UV continuum-emitting annular region in the disk. In such a
case, both dusty and dust-free clouds could be accommodated.
Similarly, if the X-ray obscurers were discrete clouds residing
in the BLR, then single clouds could partially obscure the
X-ray corona, but not the other BLR clouds, depending on the BLR's
precise geometry and the system's inclination angle.

\subsection{Nature and origin of the clouds}

The lower limits to 
radial locations of the clouds, estimated in Sect.~\ref{sec:radloc},
are commensurate with the far-outer BLR or dust sublimation zone
(cloud 1), or commensurate with the H$\beta$-emitting BLR (cloud~2). 
The limits on cloud density and 
cloud size estimated in Sect.~\ref{sec:radloc}
are very crude, but these limits can accommodate typical
values for BLR gas \citep[e.g.,][]{Netzer13}.
Several previous works, such as \citet[][NGC~4151]{Mushotzky78}, \citet[][NGC~1365]{Risaliti09},
\citet[][Mkn~766]{Risaliti11}, and \citet[][ESO323$-$G77]{Miniutti14},
found support for their observed X-ray-eclipsing clouds
to be commensurate with, and possibly associated with, the same clouds
that emit optical broad lines, due to inferred radial locations,
densities, and/or sizes.   
The two obscuration events observed in $\jofour$ could potentially also fall into
this category, though this statement is likely more applicable to
cloud~2, given the limit on its inferred radial location.
The events observed in $\jofour$ are orders of magnitude longer in
duration than the events in \citet{Risaliti09} and others, on timescales
on the order of about a day or shorter.  This discrepancy is likely an
artifact of $\jofour$'s having a black hole mass $\sim$1--2 orders of
magnitude larger than these objects and having size scales for
structures scale at least roughly linearly with black hole mass.

The clouds' ionization levels are poorly constrained, although
ionization levels above $\sim 10^{2.5}$ erg cm s$^{-1}$ are excluded;
it is possible (albeit highly speculative) that these clouds' volumes
are dominated by neutral gas and have not yet reached an ionization
equilibrium.

The obscuration events in $\jofour$ are generally
consistent with multiple models of cloud formation.  A wind launched
from the disk and accelerated outward is a strong possibility and
there are multiple models for such winds.  A dusty, turbulent wind
launched from the disk may form the low-ionization part of the BLR
\citep{CH11,Naddaf21}.  Such a wind can launch from the region of the
disk where the temperature is near 1000~K, which we estimate to be
$200^{+130}_{-75}$ lt-dy, consistent with the radial location of cloud
2.  
Another possibility is magnetohydrodynamic-driven winds
\cite[e.g.,][]{Fukumura10}, wherein gas is accelerated outward from
the accretion disk along magnetic field lines.  
The obscurers are also broadly consistent with a UV line-driven wind
launched from the accretion disk \citep{Proga00,Proga05}.
Following \citet{Giustini19}, the estimated values of $M_{\rm BH}$ and
$L_{\rm Bol}/L_{\rm Edd}$ place $\jofour$ in the range of [$M_{\rm
BH}$, $L_{\rm Bol}/L_{\rm Edd}$] space where UV-driven line winds can
be launched and there is sufficient UV radiation from the disk to
sustain their driving.


One can consider a single stream or filament that partially covers the X-ray corona.
Such a stream would be rotated out of the line of sight on
dynamical timescales, which for a black hole mass of
$6\times10^8\Msun$ and for radii of $\gtrsim$4~lt-dy is $\gtrsim$360 days,
consistent with the event durations in $\jofour$.
However, at such radial distances, it would be challenging to only
partially, not fully, cover the compact corona.

Another plausible scenario is a non-homogeneous disk wind containing both
localized embedded dense regions of X-ray-obscuring gas 
\citep[e.g.,][]{Waters22} and regions where column densities are
low enough to not obscure X-rays. 
Critically, if the wind is continually replenished from the disk and
if the number density of clumps remains roughly constant, then such a
scenario can explain the long durations of sustained partial-covering
obscuration. Support for outflowing winds as the source of 
sustained periods of partial-covering, Compton-thin X-ray obscuration
has been found in several local Seyferts \citep[e.g.,][]{Mehdipour17, Mehdipour22, Kang23},
and $\jofour$ may thus be in that same category.
The similarity in covering fractions during the two events in $\jofour$ 
could simply indicate similar levels of clumping during each wind event.

On the other hand, it is not clear what would case a wind to launch in
$\jofour$, shut off the supply of gas to end the wind, 
and launch a second time, given that disk luminosity
was minimally variable during the entire campaign.
One possible cause is that both obscuration events are part of the same
continuous wind outflow, with dense X-ray-obscuring clumps being
absent (or having non-detectable column densities) during the
unobscured states (sampled by eR1 and eR4/XM3).


If the X-ray obscurers in $\jofour$ are outflowing winds located
interior to the BLR \citep{Dehghanian19}, 
and if the optical/UV continuum from the disk
must pass through the wind before intersecting the BLR, and if the
wind's number density is sufficiently high enough to impact the $>$13.6~eV
continuum, then the observed lack of response in the Balmer lines to the
presence or absence of the obscurer along our line of sight would suggest
that the presence or absence of the wind has no immediate global impact
on the BLR (even considering the $>$100-day light-travel time from the optical-UV-emitting
disk to the
BLR).  In such a case, perhaps the wind's global covering factor as
seen by the X-ray corona is small and we, as observers, are "lucky" that
the wind crosses our line of sight to the X-ray corona. Alternately,
if the wind's global covering fraction is high, then the wind's number
density must be lower than an order of $5\times10^{10}$~cm$^{-3}$ as to
not impact transmission of $>$13.6~eV continuum photons that
ultimately power the Balmer lines
\citep{Dehghanian19}\footnote{Similarly, transmission of the $>$15 eV
photons that power \ion{Mg}{ii}] $\lambda\lambda$2796,2803 emission
would not be affected.  In our campaign, our optical spectra also
captured consistent broad emission lines of \ion{Mg}{ii}].
We omitted spectral plots of that region of
near-UV spectrum from the paper for brevity, though.}.


Finally, if the clouds originate at radii larger than the dust sublimation
zone, then filament-like, dusty winds driven by infrared radiation
\citep{Dorodnitsyn12} or by X-ray heating and radiation pressure
\citep{Wada12} can be relevant.  Scenarios incorporating fragmentation
of cooling clouds into mildly-ionized or neutral filaments such as
those of \citet{Gaspari13,Gaspari18} and \citet{Sparre19} are
also relevant at such radial locations.


\subsection{A brief comment on the anticorrelation between $N_{\rm H,
PC}$ and hard X-ray flux}

As noted in Sect.~\ref{sec:xmmfits}, we find evidence from the fits to
the \textit{XMM-Newton} EPIC spectra for an anticorrelation between
$N_{\rm H, PC}$ and unabsorbed power-law flux $F_{1-10}$.  We consider
this anticorrelation to be only tentative at best given that it is
based on only two obscuration events and given that the lack of $>10$
keV coverage induces some degeneracy for cloud~2.

Nonetheless, such anticorrelations have been reported previously, by
\citet{Couto16} in their monitoring study of the partial-covering
obscurer in NGC~4151, and by \citet{Marchesi22} in their study of
NGC~1358's transition between Compton-thick- and Compton-thin-obscured
states.\footnote{We note that \citet{Marchesi22} had energy
  coverage above 10~keV courtesy of \textit{NuSTAR} data; our study
  lacks such coverage.}
As discussed in \citet{Krolik01} and \citet{Couto16},
lower-flux states may see high columns of neutral or lowly-ionized gas
because more material condenses from higher-ionization states.
Meanwhile, at sufficiently high luminosities, an excess of ionizing radiation
may overionize the wind, causing it to disappear \citep[e.g.,][]{Giustini19}.
Alternatively, as also discussed in \citet{Couto16}, if radiation
pressure is responsible for pushing gas structures into the line of
sight, then it is possible that the lower continuum radiation during
the lower flux states may simply be relatively less efficient at
driving the gas.

%
%

\section{Conclusions}   \label{sec:Conclusions}   

eROSITA's all-sky X-ray scans have been opening a new window into
time-domain astronomy, thus providing an
opportunity to search for major changes in soft X-ray flux on
timescales of six months or longer in a given object.  By monitoring a
very large starting sample of AGNs
(on the order of a million), we can thus
amplify small numbers of events that occur rarely on a per-object
basis.  Such events include major changes in soft X-ray flux due to
obscuration by discrete clouds transiting the line of sight to the
X-ray corona.

In this paper, we present the first X-ray obscuration events
detected in a Seyfert galaxy using eROSITA, including the results of a
two-year multiwavelength follow-up campaign.  In $\jofour$ (EC
04570-5206, eRASSt J045815$-$520200), a Sy~1 AGN at $z$=0.276, we
detected two major, separate dips in soft X-ray flux: the first in
2020, followed by a recovery in soft X-ray flux by August 2021, and a
second soft X-ray dip occurring by early 2022.  From October 2020
through February 2023, we complemented eRASS X-ray monitoring with X-ray
flux and spectral monitoring courtesy of \textit{XMM-Newton} EPIC,
\textit{Swift} XRT and NICER, optical and UV photometric monitoring
via \textit{XMM-Newton} OM, \textit{Swift} UVOT, and ground-based
photometry, and optical spectroscopy to track the BLR emission. These
followups confirmed each flux dip as being due to partial-covering
obscuration, with \textit{XMM-Newton} EPIC providing the constraints
on line-of-sight column density, $N_{\rm H,pc}$, and covering fraction,
$CF$. Meanwhile, the optical/UV continuum did not drop appreciably during
the campaign, with variations of less than 0.4 mag in any band. We can
thus exclude temporary major decreases in accretion or disk
temperature.

The obscuration events are consistent with transits by
discrete, Compton-thin clouds near the black hole.  
The first cloud obscuration event lasted between 309 and 539~d (observed frame), and is
attributed to a cloud with $N_{\rm H,PC} \sim$ 1
$\times10^{22}$~cm$^{-2}$ and $CF \sim$ 60 percent.  The second event
lasted at least 527 days (observed frame), and was still in-progress as of our most
recent observation in February 2023.  The responsible cloud 
had $N_{\rm H,PC} \sim 3 \times 10^{23}$~cm$^{-2}$
and $CF \sim 80$~percent.  Both clouds are neutral or lowly ionized,
with log ($\xi$, [erg cm s$^{-1}$]) $<$ 1.5--2.5.

The derived limit on cloud 1's radial distance
from the X-ray corona, assuming Keplerian motion, is $>$1100 lt-dys,
placing it at radii at least commensurate with the far-outer BLR or
the dust sublimation zone (which is on the order of 1500 lt-dy).  For cloud
2, the limit is $>$130 lt-dy, which means that a radial location
commensurate with the H$\beta$-emitting region of the BLR (also 130
lt-dy) cannot be excluded.  It is (speculatively) possible that
cloud 2 may even be a BLR cloud itself. In this case, the second
obscuration event may fall into the same category as previous events
that lasted up to roughly a day or shorter, and whose inferred cloud
properties place them at radii commensurate with the BLR
\citep{Risaliti09,Risaliti11}.  In the case of $\jofour$, perhaps
owing to its huge black mass, $M_{\rm BH} \sim 6.2\times10^8 \Msun$,
the resulting event timescales are on the order of several months to roughly a year.

Optical/UV continuum SED modeling did not yield any evidence for
reddening due to dust, and no change in Balmer decrement was observed
in the optical spectra.
One possibility is that the clouds obscure the X-ray emitting region
as well as the optical/UV continuum-emitting region and/or BLR, but
they are inherently non-dusty, thus yielding no reddening of the
optical/UV continuum or emission-line spectra.  However, a second
possibility is that the clouds are compact enough to partially obscure
only the X-ray-emitting corona while obscuring only a small fraction
of the inner accretion disk or BLR, in which case the clouds could be
either dusty or dust-free.

Outflowing winds, launched from the disk and radiatively accelerated
outward, are likely responsible for several long-duration
(months--years) partial-covering X-ray obscuration events observed in
nearby Seyfert galaxies. Such winds can contain numerous compact
embedded X-ray obscuring clumps with column densities on the order of
$10^{22-23}$ cm$^{-2}$, similar to those observed in $\jofour$.
Such a wind also seems to be a plausible explanation for the two
long-duration obscuration events observed in $\jofour$.  In this
scenario, the wind may be spatially extended (to yield the long event
durations). Consequently, the scenario where the wind covers large
swaths of the disk or even BLR (in addition to the X-ray corona) but
remains inherently dust-free may be preferred in this framework.

Although $\jofour$ has only been monitored for roughly three years, our
observations have captured the source to be in an obscured state 78
percent of the time, with three transitions between obscured and
unobscured states during that time (noting the obvious bias that our
observations were conducted because of the detection of the
obscuration events).
Nonetheless, it worth speculating on comparisons to NGC~1365, a source
with repeated transitions between Compton-thin and -thick-obscured
states, as well as with numerous observations of
discrete clumpy structures transiting the line of sight
to the corona \citep{Ricci23}. Thus,
$\jofour$ is worth keeping an eye on. 

We will continue to monitor this source to determine when the second
obscuration event ends, thereby improving estimates of that obscurer's
properties. In addition, it is a source worth monitoring in the coming
years in the pursuit of  potential additional obscuration events. Moreover,
additional tracking of the source and identification of new
transitions between obscured and unobscured states can solidify and
quantify any anticorrelation between column density and intrinsic
X-ray luminosity as in NGC~4151. Such observations can provide
constraints for modeling the obscurers in the context of outflowing,
radiatively driven winds.

\begin{acknowledgements}

The authors thank M.\ Mehdipour for useful discussions which helped guide
interpretation of results.

AM, TS, and SK acknowledge partial support from Narodowe Centrum Nauki
(NCN) grants 2016/23/B/ST9/03123 and 2018/31/G/ST9/03224.  AM also
acknowledges partial support from NCN grant 2019/35/B/ST9/03944.  DH
acknowledges support from DLR grant FKZ 50 OR 2003. MK is supported by
German Science Foundation (DFG) grant KR 3338/4-1.  AG and SH were
supported by DFG grant WI 1860/14-1.

This work is based on data from eROSITA, the soft X-ray instrument
aboard SRG, a joint Russian-German science mission supported by the
Russian Space Agency (Roskosmos), in the interests of the Russian
Academy of Sciences represented by its Space Research Institute (IKI),
and the Deutsches Zentrum für Luft- und Raumfahrt (DLR). The SRG
spacecraft was built by Lavochkin Association (NPOL) and its
subcontractors, and is operated by NPOL with support from the Max
Planck Institute for Extraterrestrial Physics (MPE). The development
and construction of the eROSITA X-ray instrument was led by MPE, with
contributions from the Dr. Karl Remeis Observatory Bamberg \& ECAP (FAU
Erlangen-Nuernberg), the University of Hamburg Observatory, the
Leibniz Institute for Astrophysics Potsdam (AIP), and the Institute
for Astronomy and Astrophysics of the University of Tübingen, with the
support of DLR and the Max Planck Society. The Argelander Institute
for Astronomy of the University of Bonn and the Ludwig Maximilians
Universität Munich also participated in the science preparation for
eROSITA.  

The eROSITA data shown here were processed using the eSASS/NRTA software
system developed by the German eROSITA consortium.  

Some of the observations reported in this paper were obtained with the
Southern African Large Telescope (SALT) under programme 2018-2-LSP-001
for transients (PI: Buckley), conducted within the eROSITA M.O.U.\ as
part of the eROSITA-SALT Transient collaboration, as well as under
programmes 2020-2-MLT-008 and 2021-2-MLT-003 (PI: Markowitz).  Polish
participation in SALT is funded by grant No.\ MEiN nr 2021/WK/01.  

This paper uses observations made using the South African Astronomical
Observatory (SAAO).

Based on observations obtained with XMM-Newton, an ESA science mission
with instruments and contributions directly funded by ESA Member
States and NASA.  The authors thank the \textit{XMM-Newton} director
for approving the DDT observations, and the \textit{XMM-Newton}
operations team for executing them.  This research has made use of
data and/or software provided by the High Energy Astrophysics Science
Archive Research Center (HEASARC), which is a service of the
Astrophysics Science Division at NASA/GSFC. 

This work made use of data supplied by the UK Swift Science Data
Centre at the University of Leicester \citep{Evans07,Evans09}.  Part
of this work is based on archival data, software or online services
provided by the Space Science Data Center - ASI.

This work was supported in part by NASA through the NICER mission and
the Astrophysics Explorers Program. NICER data used in this work were
gathered under a Guest Observer (GO) approved programme. 

This research has made use of the NASA/IPAC Extragalactic Database
(NED), which is funded by the National Aeronautics and Space
Administration and operated by the California Institute of Technology.

The Skynet Robotic Telescope Network is supported by the National
Science Foundation, the Department of Defense, the North Carolina
Space Grant Consortium, and the Mount Cuba Astronomical Foundation.

This publication makes use of data products from the Wide-field
Infrared Survey Explorer, which is a joint project of the University
of California, Los Angeles, and the Jet Propulsion
Laboratory/California Institute of Technology, funded by the National
Aeronautics and Space Administration.  This publication also makes use
of data products from NEOWISE, which is a project of the Jet
Propulsion Laboratory/California Institute of Technology, funded by
the Planetary Science Division of the National Aeronautics and Space
Administration.

This research has made use of ISIS functions (ISISscripts) provided by
ECAP/Remeis observatory and MIT
(http://www.sternwarte.uni-erlangen.de/isis/).

\end{acknowledgements}


\newcommand{\noop}[1]{}
\bibliographystyle{aa}
\bibliography{mybib}


\begin{appendix}

%
%

\section{Optical/UV continuum photometric data tables}        

\renewcommand{\arraystretch}{1.17}
\begin{table*}[b] \centering
\caption[]{Optical/UV continuum magnitudes from \textit{Swift}-UVOT and \textit{XMM-Newton}-OM}
\label{tab:OUVfluxes}
\begin{tabular}{lccccc} \hline\hline
Obs.\ & Date   & Filter & Flux Dens.\  & Flux Dens.\    & Flux Dens.\    \\
      & (MJD)  &        & (Vega mag)   & (mJy)          & (erg cm$^{-2}$ s$^{-1}$ {\AA}$^{-1}$) \\  \hline
Sw1 & 59130.95 & B  & $16.39\pm0.03$   & $1.16\pm0.03$ & $(1.82\pm0.05)\times10^{-15}$ \\
Sw2 & 59137.49 & B  & $16.34\pm0.03$   & $1.22\pm0.03$ & $(1.91\pm0.05)\times10^{-15}$ \\
XM1 & 59209.82 & M2 & $14.721\pm0.003$ & $1.108\pm0.003$ & $(6.229\pm0.014)\times10^{-15}$ \\
XM2 & 59242.45 & M2 & $14.644\pm0.003$ & $1.111\pm0.004$ & $(6.254\pm0.016)\times10^{-15}$ \\
Sw3 & 59391.00 & V  & $16.14\pm0.03$   & $1.31\pm0.04$ & $ (1.34\pm0.04)\times10^{-15}$ \\
    &          & B  & $16.24\pm0.04$   & $1.34\pm0.04$ & $ (2.14\pm0.07)\times10^{-15}$ \\
    &          & U  & $15.11\pm0.04$   & $1.36\pm0.05$ & $ (3.32\pm0.11)\times10^{-15}$ \\
    &          & W1 & $14.85\pm0.04$   & $1.08\pm0.04$ & $ (4.74\pm0.19)\times10^{-15}$ \\
    &          & M2 & $14.67\pm0.04$   & $1.12\pm0.04$ & $ (6.77\pm0.26)\times10^{-15}$ \\
    &          & W2 & $14.71\pm0.04$   & $1.03\pm0.03$ & $ (7.15\pm0.25)\times10^{-15}$ \\
XM3 & 59465.41 & M2 & $14.731\pm0.004$ & $1.097\pm0.004$ & $ (6.169\pm0.020)\times10^{-15}$ \\
Sw4 & 59662.59 & V  & $16.16\pm0.04$ & $1.28\pm0.05$ & $(1.32\pm0.05)\times10^{-15}$ \\
    &          & B  & $16.20\pm0.04$ & $1.39\pm0.05$ & $(2.22\pm0.07)\times10^{-15}$ \\
    &          & U  & $15.05\pm0.04$ & $1.43\pm0.05$ & $(3.50\pm0.12)\times10^{-15}$ \\
    &          & W1 & $14.86\pm0.04$ & $1.07\pm0.04$ & $(4.69\pm0.16)\times10^{-15}$ \\
    &          & M2 & $14.71\pm0.04$ & $1.08\pm0.04$ & $(6.58\pm0.26)\times10^{-15}$ \\
    &          & W2 & $14.76\pm0.04$ & $0.99\pm0.03$ & $(7.18\pm0.24)\times10^{-15}$ \\
XM4 & 59694.64 & B  & $16.38\pm0.36$ & $1.24\pm0.27$   & $(1.84\pm0.60)\times10^{-15}$ \\
    & 59695.04 & M2 & $14.793\pm0.004$ & $1.036\pm0.004$ & $ (5.827\pm0.020)\times10^{-15}$ \\
Sw5 & 59733.06 & V  & $16.18\pm0.06$ & $1.26\pm0.07$ & $(1.29\pm0.07)\times10^{-15}$ \\
    &          & B  & $16.25\pm0.04$ & $1.32\pm0.05$ & $(2.12\pm0.09)\times10^{-15}$ \\
    &          & U  & $15.17\pm0.04$ & $1.28\pm0.05$ & $(3.14\pm0.13)\times10^{-15}$ \\
    &          & W1 & $14.99\pm0.04$ & $0.95\pm0.04$ & $(4.16\pm0.16)\times10^{-15}$ \\
    &          & M2 & $14.92\pm0.04$ & $0.89\pm0.04$ & $(5.38\pm0.21)\times10^{-15}$ \\
    &          & W2 & $15.01\pm0.04$ & $0.79\pm0.03$ & $(5.70\pm0.22)\times10^{-15}$ \\
Sw6 & 59783.49 & V  & $16.27\pm0.05$ & $1.16\pm0.05$ & $(1.19\pm0.06)\times10^{-15}$ \\
    &          & B  & $16.34\pm0.04$ & $1.22\pm0.04$ & $(1.95\pm0.07)\times10^{-15}$ \\
    &          & U  & $15.27\pm0.04$ & $1.17\pm0.04$ & $(2.86\pm0.10)\times10^{-15}$ \\
    &          & W1 & $15.11\pm0.04$ & $0.85\pm0.03$ & $(3.73\pm0.15)\times10^{-15}$ \\
    &          & M2 & $14.94\pm0.04$ & $0.88\pm0.03$ & $(5.28\pm0.21)\times10^{-15}$ \\
    &          & W2 & $15.09\pm0.04$ & $0.73\pm0.03$ & $(5.30\pm0.21)\times10^{-15}$  \\
Sw7 & 59873.53 & V  & $16.24\pm0.04$ & $1.19\pm0.05$ & $(1.22\pm0.05)\times10^{-15}$ \\
    &          & B  & $16.35\pm0.04$ & $1.21\pm0.04$ & $(1.93\pm0.06)\times10^{-15}$ \\
    &          & U  & $15.24\pm0.04$ & $1.20\pm0.04$ & $(2.94\pm0.10)\times10^{-15}$ \\
    &          & W1 & $15.07\pm0.04$ & $0.88\pm0.03$ & $(3.87\pm0.15)\times10^{-15}$ \\
    &          & M2 & $14.95\pm0.04$ & $0.87\pm0.03$ & $(5.23\pm0.20)\times10^{-15}$ \\
    &          & W2 & $15.07\pm0.04$ & $0.74\pm0.03$ & $(5.39\pm0.18)\times10^{-15}$ \\
Sw8 & 59992.20 & V  & $16.04\pm0.04$ & $1.42\pm0.05$ & $(1.47\pm0.06)\times10^{-15}$ \\
    &          & B  & $16.14\pm0.04$ & $1.46\pm0.05$ & $(2.34\pm0.08)\times10^{-15}$ \\
    &          & U  & $14.99\pm0.04$ & $1.51\pm0.05$ & $(3.70\pm0.12)\times10^{-15}$ \\
    &          & W1 & $14.76\pm0.04$ & $1.17\pm0.04$ & $(5.15\pm0.17)\times10^{-15}$ \\
    &          & M2 & $14.60\pm0.04$ & $1.20\pm0.05$ & $(7.23\pm0.28)\times10^{-15}$ \\
    &          & W2 & $14.63\pm0.04$ & $1.11\pm0.04$ & $(8.09\pm0.27)\times10^{-15}$ \\
\hline  \end{tabular}
\tablefoot{All magnitudes are Vega magnitudes, and are observed magnitudes, i.e., not
corrected for Galactic absorption. In contrast, all flux densities are
  corrected for Galactic absorption. Uncertainties are statistical only (see Appendix~\ref{sec:appdxoptmag}).}
\end{table*}

\renewcommand{\arraystretch}{1.17}
\begin{table}
\caption[]{B- and V-band photometric observations at PROMPT6}
        \centering
\label{tab:Bphotlog}
        \begin{tabular}{lccc} \hline\hline
Date   & Filter & Expo.\  & Obsd.   \\
(MJD)  &        & (s)    & Mag.    \\
       &        &        & (Vega)  \\ \hline
59164.06 & B& 1030 &  $16.42\pm0.02$  \\  
59174.08 & B& 1224 &  $16.48\pm0.02$  \\
59180.16 & B& 1350 &  $16.42\pm0.02$  \\
59182.01 & B& 1350 &  $16.40\pm0.03$  \\
59202.04 & B& 1251 &  $16.37\pm0.02$  \\
59205.35 & B& 1251 &  $16.46\pm0.03$  \\
59207.14 & B& 1251 &  $16.39\pm0.02$  \\
59211.19 & B& 1251 &  $16.39\pm0.02$  \\
59215.04 & B& 1350 &  $16.35\pm0.02$  \\  
59217.10 & B& 1350 &  $16.35\pm0.02$  \\  
59219.11 & B& 1350 &  $16.43\pm0.02$  \\
59221.04 & B& 1350 &  $16.41\pm0.02$  \\
59223.10 & B& 1350 &  $16.35\pm0.02$  \\  
59230.09 & B& 1371 &  $16.39\pm0.01$  \\  
59231.14 & B& 1200 &  $16.43\pm0.02$  \\  
59233.18 & B& 1200 &  $16.43\pm0.02$  \\  
59234.15 & B& 1200 &  $16.19\pm0.01$  \\  
59236.19 & B& 1200 &  $16.29\pm0.02$  \\  
59237.18 & B& 1200 &  $16.54\pm0.02$  \\
59241.09 & B&  600 &  $16.54\pm0.03$  \\
59242.09 & B&  900 &  $16.46\pm0.02$  \\
59243.21 & B& 1200 &  $16.48\pm0.03$  \\  
59245.08 & B& 1200 &  $16.50\pm0.02$  \\
59249.08 & B& 1200 &  $16.25\pm0.01$  \\
59256.16 & B& 1200 &  $16.77\pm0.03$  \\
59263.16 & B& 1200 &  $16.68\pm0.03$  \\ 
59272.07 & B& 1800 &  $16.52\pm0.04$  \\ 
59274.08 & B&  720 &  $16.54\pm0.05$  \\
59279.02 & B&  720 &  $16.33\pm0.10$  \\
59283.99 & B&  692 &  $16.44\pm0.06$  \\
59291.03 & B&  720 &  $16.52\pm0.05$  \\
59294.00 & B&  900 &  $16.51\pm0.05$  \\
59308.97 & B&  764 &  $16.72\pm0.10$  \\  \hline 
59389.38 & V& 1440 &  $16.25\pm0.03$  \\  
59393.34 & V& 1440 &  $16.29\pm0.02$  \\
59464.23 & V& 1440 &  $16.23\pm0.01$  \\
59467.28 & V& 1603 &  $16.19\pm0.05$  \\
59489.25 & V& 1620 &  $16.24\pm0.03$  \\
59537.06 & V& 1746 &  $16.13\pm0.02$  \\
59541.19 & V& 1620 &  $16.10\pm0.07$ \\  \hline \end{tabular} 
\tablefoot{All magnitudes in Col.~4 are Vega magnitudes, and are
observed magnitudes, i.e., not corrected for Galactic absorption, and not
corrected for contamination by the host galaxy.}
\end{table}

\clearpage

%
%

\section{Cross-calibration of UVOT and OM M2 magnitudes} \label{sec:appdxoptmag}

In this work, we combine photometry data from both
\textit{XMM-Newton} OM and \textit{Swift} UVOT.  For purposes of cross
calibration between the two instruments, and for determining if the
variability observed in the B and M2 light curves is real, we
additionally examined the optical/UV data of two stars in the field of
view, each located 1{\farcm}3--1{\farcm}9 from $\jofour$. These stars are
TYC 8083-1113-1 ($\alpha = 04^{\rm h} 58^{\rm m} 08{\fs}3$, $\delta=-52^{\circ} 02\arcmin 12{\farcs}1$) and
TYC 8083-1608-1 ($\alpha = 04^{\rm h} 58^{\rm m} 14{\fs}7$, $\delta=-52^{\circ} 03\arcmin 20{\farcs}9$).

For the UVOT B-band, we converted UVOT B-band magnitudes to Johnson
B-band magnitudes following the "color correction" document available
at the Swift UVOT Calibration Documents
webpage\footnote{\url{https://heasarc.gsfc.nasa.gov/docs/heasarc/caldb/swift/docs/uvot/}}
and using observed $B-V$ and $U-B$ colors for each object.  This action
resulted in magnitude corrections of +0.017, +0.012, and +0.015 for
$\jofour$, TYC 8083-1113-1, and TYC 8083-1608-1, respectively.

We used star TYC 8083-1113-1 to estimate systematic uncertainties for
the \textit{Swift} UVOT data, under the assumption that the star is
not variable.  We estimated systematic uncertainties to be 0.015,
0.012, 0.013, 0.015, 0.029, and 0.023 mag for the V, B, U, W1, M2, and W2
bands, respectively.  Star TYC 8083-1608-1 seems to display some
energy-dependent variability during Sw7, so we did not use this star to
estimate systematics. The four OM observations each used the M2 band;
from TYC 8083-1113-1, we estimated the systematic uncertainty to be $\sim$0.015 mag.

A direct comparison between the UVOT and OM M2 magnitudes, however, was
still hindered by the fact that the energy peaks of the two instruments' effective areas
differ by $\sim$10 percent: OM M2 peaks near 2300~$\AA$ (5.4~eV), while UVOT
M2 peaks near 2140~$\AA$ (5.8~eV).  To extrapolate from the OM M2 peak
energy to the UVOT M2 peak energy, we took advantage of having
observed the field with all six UVOT filters in Sw3--8.  For the
stars, we used the time-averaged V, B, U, W1, M2, and W1 magnitudes to
construct SEDs, and fit simple power-law models via linear regression.
For $\jofour$, we used the Sw5 SED to estimate adjustments for XM4, as
Sw5 occurs closest in time to XM4; we used Sw3 to estimates adjustments
to XM1--3. We obtained spectral indices of 
$+0.97\pm0.05$ ($\jofour$, Sw3), 
$+0.42\pm0.10$ ($\jofour$, Sw5), 
$-3.75\pm0.07$ (TYC 8083-1113-1, average of Sw3--8), 
and $-2.61\pm0.07$ (TYC 8083-1608-1, average of Sw3--8). These
values correspond to magnitude corrections 
applied to the OM M2 light curves of $-0.070$ ($\jofour$, XM1--3), $-0.031$ ($\jofour$, XM4), +0.276 (TYC 8083-1113-1), and +0.192 (TYC 8083-1608-1).

\begin{figure*}
\includegraphics[width=1.99\columnwidth]{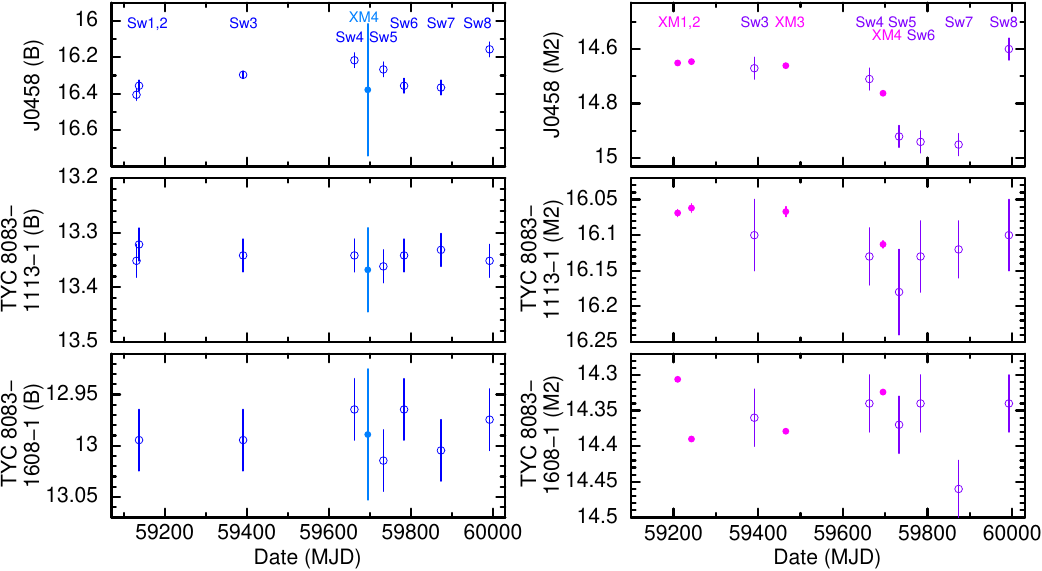}   
\caption{B-band and M2-band Vega magnitude light curves for $\jofour$
and for nearby stars TYC 8083-1113-1 and TYC 8083-1608-1.
\textit{XMM-Newton} OM points are denoted by filled circles;
\textit{Swift} UVOT points are denoted by open circles. Uncertainties for $\jofour$ are statistical only.
Some \textit{XMM-Newton} OM M2-band data points have error bars smaller than their data point symbol.}  
\label{fig:finalBM2}
\end{figure*}

The final combined M2 and B light curves are shown in
Fig.~\ref{fig:finalBM2}.  Magnitudes are not corrected for Galactic
extinction, just for the aforementioned magnitude shifts.  In the M2 band, there is agreement between the UVOT and
OM magnitudes to within 0.05--0.1 mag.  In addition, with the
exception of a 0.1 mag drop between Sw6 and Sw7 in object C, both
stars' light curves display minimal variability -- $\sim$0.1 mag or
less within any one instrument or between instruments.  We thus
conclude that the M2 flux drop by $>$0.2 mag in $\jofour$ from
Sw4/XM4 to Sw5--7 is real and not an artifact.

\begin{figure*}
\includegraphics[width=1.99\columnwidth]{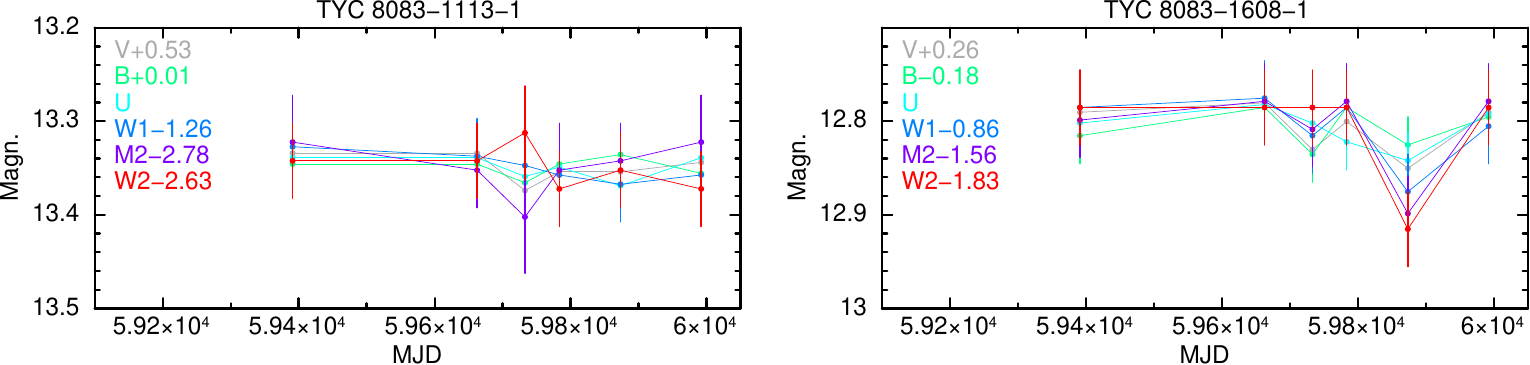}
\caption{Same as Fig.~\ref{fig:lcs_sw38_J0458}, but for two stars 
nearby in the field of view, TYC 8083-1113-1, and TYC 8083-1608-1. 
For plotting purposes, the magnitudes for the V, B, W1, M2, and W2 bands have been offset by the magnitude values
listed in the top left corner, such that their mean magnitudes match that for U-band.} 
\label{fig:lcs_sw38_BC}
\end{figure*}

Finally, for completeness, in Fig.~\ref{fig:lcs_sw38_BC}, we
display the two stars' V, B, U, W1, M2, and W2-band light curves,
from observations Sw3--8.

\onecolumn

%
%

\section{Optical spectral fit result tables}  \label{sec:appdxopt}

\renewcommand{\arraystretch}{1.18}
\begin{table}[h]   
\caption[]{Optical spectral fit parameters, H$\beta$ region: continuum}
        \centering
\label{tab:fitspec_Hbeta1}
\begin{tabular}{lcc} \hline\hline
Spec.\ & \multicolumn{2}{c}{Power-law continuum}  \\
 \#    &   ${\nu}F_{5100\AA}$ & Slope \\ \hline 
1 & $6.64\pm0.02$ & $-1.71\pm0.02$ \\ 
2 & $6.96\pm0.16$ & $-1.77\pm0.05$ \\ 
3 & $5.96\pm0.02$ & $-1.51\pm0.02$ \\ 
4 & $7.61\pm0.14$ & $-1.58\pm0.04$ \\ 
5 & $7.24\pm0.31$ & $-1.94\pm0.09$ \\ 
6 & $6.73\pm0.23$ & $-1.55\pm0.07$ \\ 
7 & $7.26\pm0.28$ & $-1.76\pm0.08$ \\ 
8 & $7.86\pm0.03$ & $-1.81\pm0.02$ \\ 
9 & $8.41\pm0.02$ & $-1.60\pm0.02$ \\ 
10& $7.51\pm0.02$ & $-1.60\pm0.05$ \\ \hline 
\end{tabular}
\tablefoot{${\nu}F_{5100\AA}$ is in units of $10^{-16}$~\ecgsA.   
}
\end{table}


\renewcommand{\arraystretch}{1.18}
\begin{table*}[h!]     
\caption[]{Optical spectral fit parameters, H$\beta$ region: [\ion{O}{iii}] narrow line emission}
        \centering
\label{tab:fitspec_Hbeta2}
\begin{tabular}{l|ccc|ccc} \hline\hline
Spec. &  \multicolumn{3}{c|}{Narrow [\ion{O}{iii}]$\lambda$5007}&  \multicolumn{3}{c}{Narrow [\ion{O}{iii}]$\lambda$4959}     \\
      &  Centroid  & Width $\sigma$  &   Flux ($10^{-14}$        &  Centroid  & Width $\sigma$ &  Flux ($10^{-14}$\\
      &   (\AA)    &(km s$^{-1}$)    &   erg cm$^{-2}$ s$^{-1}$)  &  (\AA)     &(km s$^{-1}$)    &   erg cm$^{-2}$ s$^{-1}$)\\ \hline
1 & $5006.1 \pm 0.1$ & $  214 \pm   7$ & $ 3.51 \pm  0.20$ &  $4958.8 \pm 0.3$  & (tied) & $ 1.04 \pm  0.12$ \\
2 & $5006.1 \pm 0.1$ & $  214 \pm   7$ & $ 3.53 \pm  0.20$ &  $4958.3 \pm 0.4$  & -- & $ 0.81 \pm  0.12$ \\
3 & $5006.0 \pm 0.1$ & $  220 \pm  11$ & $ 3.08 \pm  0.26$ &  $4958.9 \pm 0.4$  &  -- & $ 0.97 \pm  0.15$ \\
4 & $5005.7 \pm 0.1$ & $  275 \pm  13$ & $ 5.33 \pm  0.55$ &  $4958.8 \pm 0.5$  &  -- & $ 1.52 \pm  0.28$ \\
5 & $5006.3 \pm 0.1$ & $  195 \pm  10$ & $ 3.31 \pm  0.27$ &  $4959.4 \pm 0.4$  &  -- & $ 1.10 \pm  0.09$ \\
6 & $5006.0 \pm 0.1$ & $  215 \pm   9$ & $ 3.66 \pm  0.27$ &  $4958.6 \pm 0.4$  &  -- & $ 1.17 \pm  0.16$ \\
7 & $5006.1 \pm 0.1$ & $  202 \pm  10$ & $ 3.37 \pm  0.27$ &  $4958.3 \pm 0.5$  &  -- & $ 0.87 \pm  0.16$ \\
8 & $5004.7 \pm 0.2$ & $  238 \pm  16$ & $ 3.97 \pm  0.50$ &  $4959.5 \pm 0.9$  &  -- & $ 0.76 \pm  0.24$ \\
9 & $5004.5 \pm 0.2$ & $  378 \pm  12$ & $10.53 \pm  0.51$ &  $4955.0 \pm 0.5$  &  -- & $ 3.51 \pm  0.17$ \\
10& $5006.2 \pm 0.1$ & $  202 \pm   7$ & $ 3.27 \pm  0.18$ &  $4959.1 \pm 0.3$  &  -- & $ 1.09 \pm  0.06$ \\  \hline
\end{tabular}
\tablefoot{Line parameters are reported for the rest frame. 
The velocity width $\sigma$ for the 4959~$\AA$ line was tied to that for the 5007~$\AA$ line.}
\end{table*}

\renewcommand{\arraystretch}{1.18}
\begin{table*}
\caption[]{Optical spectral fit parameters, H$\beta$ region: [\ion{O}{iii}] broad, blueshifted line emission}
        \centering
\label{tab:fitspec_Hbeta3}
\begin{tabular}{l|ccc|ccc} \hline\hline
Spec. &  \multicolumn{3}{c|}{Broad [\ion{O}{iii}]$\lambda$5007}&  \multicolumn{3}{c}{Broad [\ion{O}{iii}]$\lambda$4959}     \\
      &  Centroid  & Width $\sigma$  &   Flux  ($10^{-14}$       &  Centroid  & Width $\sigma$ &  Flux ($10^{-14}$\\
      &   (\AA)    &(km s$^{-1}$)    &   erg cm$^{-2}$ s$^{-1}$)  &  (\AA)     &(km s$^{-1}$)    &   erg cm$^{-2}$ s$^{-1}$)\\ \hline
1 & $4996.2 \pm 0.9$ & $  653 \pm  24$ & $12.68 \pm  0.90$ &  $4951.0 \pm 0.8$  & (tied) & $ 9.27 \pm  0.59$ \\
2 & $4995.9 \pm 0.9$ & $  643 \pm  23$ & $12.60 \pm  0.86$ &  $4953.4 \pm 0.7$  & --& $10.32 \pm  0.58$ \\
3 & $4996.0 \pm 1.6$ & $  688 \pm  40$ & $10.82 \pm  1.26$ &  $4950.0 \pm 1.4$  & --& $ 8.14 \pm  0.79$ \\
4 & $4994.7 \pm 1.9$ & $  671 \pm  40$ & $12.01 \pm  1.60$ &  $4950.9 \pm 1.3$  & --& $ 9.55 \pm  0.98$ \\
5 & $4995.1 \pm 1.4$ & $  662 \pm  40$ & $14.74 \pm  1.45$ &  $4950.3 \pm 1.1$  & --& $ 9.70 \pm  1.57$ \\
6 & $4994.2 \pm 1.3$ & $  665 \pm  35$ & $13.62 \pm  1.21$ &  $4949.9 \pm 1.2$  & --& $ 9.74 \pm  0.57$ \\
7 & $4996.5 \pm 1.2$ & $  670 \pm  32$ & $15.54 \pm  1.43$ &  $4951.1 \pm 1.1$  & --& $11.64 \pm  0.94$ \\
8 & $4995.6 \pm 1.9$ & $  620 \pm  39$ & $10.67 \pm  1.64$ &  $4951.6 \pm 1.3$  & --& $ 9.14 \pm  0.94$ \\
9 & $5008.2 \pm 3.9$ & $ 1602 \pm 226$ & $41.60 \pm 33.57$ &  $4956.5 \pm 23.7$ & --& $24.23 \pm  5.11$ \\
10& $4996.9 \pm 0.8$ & $  665 \pm   0$\tablefootmark{a} & $15.46 \pm  0.80$ &  $4951.4 \pm 0.7$  & --& $ 9.72 \pm  0.49$ \\ \hline
\end{tabular}
\tablefoot{Line parameters are reported for the rest frame. The
velocity widths ($\sigma$) for both broad components were kept tied in fits.\\
\tablefoottext{a}{Width was kept frozen to enable reasonable constraints on other parameters.}}
\end{table*}

\renewcommand{\arraystretch}{1.18}
\begin{table*}
\caption[]{Optical spectral fit parameters, H$\beta$ region: broad and narrow H$\beta$~$\lambda$4861 emission}
        \centering
\label{tab:fitspec_Hbeta4}
\begin{tabular}{l|ccc|ccc} \hline\hline
Spec. &  \multicolumn{3}{c|}{Broad H$\beta$}&  \multicolumn{3}{c}{Narrow H$\beta$}     \\
\#    &  Centroid  & Width $\sigma$  &   Flux   ($10^{-14}$                 &  Centroid  & Width $\sigma$ &  Flux ($10^{-14}$  \\
      &   (\AA)    &(km s$^{-1}$)    &   erg cm$^{-2}$ s$^{-1}$)  &  (\AA)     &(km s$^{-1}$)    &   erg cm$^{-2}$ s$^{-1}$)\\ \hline
1 & $4858.5 \pm 0.2$ & $ 1996 \pm  13$ & $400.86 \pm  2.58$ &  (4861.3) & (tied) & $ 0.53 \pm  0.07$     \\  
2 & $4857.9 \pm 0.2$ & $ 2019 \pm  15$ & $416.12 \pm  2.83$ &  -- &-- & $ 0.58 \pm  0.07$     \\
3& $4857.2 \pm 0.3$ & $ 2124 \pm  21$ & $380.62 \pm  3.73$ &  -- &-- & $ 0.50 \pm  0.09$     \\
4 & $4856.5 \pm 0.2$ & $ 2081 \pm  17$ & $462.87 \pm  3.54$ &  -- &-- & $ 0.87 \pm  0.13$     \\
5 & $4857.1 \pm 0.4$ & $ 2111 \pm  25$ & $482.24 \pm  5.50$ &  -- &-- & $ 0.52 \pm  0.10$     \\
6 & $4857.1 \pm 0.3$ & $ 2151 \pm  22$ & $451.67 \pm  4.48$ &  -- &-- & $ 0.64 \pm  0.09$     \\
7 & $4856.3 \pm 0.3$ & $ 2167 \pm  23$ & $508.53 \pm  5.35$ &  -- &-- & $ 0.61 \pm  0.10$     \\
8 & $4856.5 \pm 0.3$ & $ 1988 \pm  20$ & $412.25 \pm  3.95$ &  -- &-- & $ 0.11 \pm  0.12$   \\  
9 & $4859.1 \pm 0.5$ & $ 2087 \pm  29$ & $488.98 \pm  6.91$ &  -- &-- & $ 0.69 \pm  0.26$ \\    
10& $4860.0 \pm 0.3$ & $ 2057 \pm  19$ & $405.33 \pm  3.62$ &  -- &-- & $ 0.33 \pm  0.08$\\ \hline
\end{tabular}
\tablefoot{Line parameters are reported for the rest frame. The energy centroid for narrow H$\beta$ was kept fixed at 4861.3~$\AA$.
The velocity width for the narrow H$\beta$ component was tied to that for the narrow [\ion{O}{iii}]$\lambda$5007 line.}
\end{table*}

\renewcommand{\arraystretch}{1.18}
\begin{table*}
\caption[]{Optical spectral fit parameters, H$\alpha$ region: broad, ultra-broad, and narrow H$\alpha$~$\lambda$6563 emission}
        \centering
\label{tab:fitspec_Halpha1}
\begin{tabular}{l|ccc|ccc|c} \hline\hline
Spec. &  \multicolumn{3}{c|}{Broad H$\alpha$}&  \multicolumn{3}{c|}{Ultra-Broad H$\alpha$}  & Narrow H$\alpha$   \\
 \#   &  Centroid  & Width $\sigma$ & Flux ($10^{-14}$      &  Centroid  & Width $\sigma$ &  Flux ($10^{-14}$    &  Flux ($10^{-14}$ \\
      &   (\AA)    &(km s$^{-1}$) & erg cm$^{-2}$ s$^{-1}$) &   (\AA)    & (km s$^{-1}$)  & erg cm$^{-2}$ s$^{-1}$) & erg cm$^{-2}$ s$^{-1}$)\\ \hline

1 & $6565.9\pm0.6$ & $ 1495\pm 49$ & $833.85\pm148.83$ & $6529.0\pm20.4$ & $ 2111\pm268$ & $396.63\pm225.80$ & $ 3.84 \pm  0.14$\\
2 & $6564.5\pm0.4$ & $ 1450\pm 24$ & $742.41\pm44.69$ & $6538.2\pm5.1$ & $ 2415\pm100$ & $521.70\pm84.30$ & $ 2.80 \pm  0.13$\\
3 & $6561.5\pm0.2$ & $ 1536\pm 14$ & $737.51\pm11.90$ & $6562.2\pm2.4$ & $ 4548\pm190$ & $1416.61\pm61.80$ & $ 3.13 \pm  0.15$\\
4 & $6561.7\pm0.2$ & $ 1524\pm 15$ & $1007.48\pm16.81$ & $6564.1\pm3.5$ & $ 4684\pm126$ & $2057.88\pm87.36$ & $ 3.63 \pm  0.21$\\
5 & $6563.2\pm0.2$ & $ 1548\pm 20$ & $1084.18\pm29.80$ & $6567.7\pm2.8$ & $ 4250\pm296$ & $1807.65\pm100.85$ & $ 5.08 \pm  0.28$\\
6 & $6560.7\pm0.3$ & $ 1371\pm 38$ & $751.19\pm60.11$ & $6559.8\pm1.2$ & $ 2811\pm138$ & $1603.46\pm116.41$ & $ 5.19 \pm  0.34$\\
7 & $6562.3\pm0.2$ & $ 1501\pm 23$ & $1044.80\pm39.05$ & $6565.0\pm2.0$ & $ 3658\pm224$ & $1606.93\pm 0.08$ & $ 5.23 \pm  0.27$\\
8 & $6560.9\pm0.3$ & $ 1477\pm 15$ & $918.87\pm14.03$ & $6566.3\pm2.7$ & $ 4571\pm  0$\tablefootmark{a} & $2145.13\pm101.36$ & $ 2.72 \pm  0.27$\\
9 & $6562.1\pm0.2$ & $ 1538\pm 11$ & $1126.17\pm13.26$ & $6563.8\pm2.5$ & $ 4571\pm  0$\tablefootmark{a} & $1888.62\pm97.62$ & $ 2.88 \pm  0.21$ \\
10 & $6563.7\pm0.2$ & $ 1528\pm 12$ & $1229.23\pm13.34$ & $6534.7\pm2.1$ & $ 4571\pm  0$\tablefootmark{a} & $2782.34\pm76.25$ & $ 5.16 \pm  0.28$ \\ \hline
\end{tabular}
\tablefoot{Line parameters are reported for the rest frame.
The velocity width for the narrow H$\alpha$ component was tied to that for the narrow [\ion{O}{iii}]$\lambda$5007 line;
the energy centroid was kept frozen at 6562.8~$\AA$.\\
\tablefoottext{a}{Width was kept frozen to enable reasonable constraints on other parameters.}}
\end{table*}

\end{appendix}

\end{document}